\documentclass[11pt]{article}
\pdfoutput=1
\usepackage{jcapmod}
\usepackage{shorthand}
\usepackage[scaled]{helvet}
\usepackage[toc,page]{appendix}
\usepackage{mathtools}
\usepackage{booktabs}
\usepackage[english]{babel}
\usepackage{amsmath,amssymb,amsbsy,amstext, amsthm, simplewick, amsfonts}
\usepackage{hyperref}
\usepackage{graphicx}
\usepackage[small]{caption}
\usepackage{siunitx}
\usepackage{upgreek}
\usepackage{framed}
\usepackage{wrapfig}
\usepackage{multirow}
\usepackage{bbm}
\usepackage[svgnames,dvipsnames,x11names]{xcolor}
\usepackage[utf8x]{inputenc}
\usepackage{selinput}
\usepackage{bm}
\usepackage{float}
\usepackage{geometry}
\usepackage{yfonts}
\usepackage{caption}
\usepackage{subcaption}
\usepackage{sidecap}
\usepackage{longtable}
\usepackage{anyfontsize}
\usepackage{dsfont}
\usepackage{tikz}
\usepackage{relsize}
\usepackage{shuffle}
\usetikzlibrary{matrix}
\usepackage{bigstrut}
\usepackage{fnpct}
\usetikzlibrary{arrows.meta}
\usetikzlibrary{decorations.pathreplacing,calligraphy}
\usepackage{tensor}
\usepackage{mleftright}
\usepackage[most]{tcolorbox}
\usepackage{fancybox} 
\usepackage{tikz}
\usepackage{tikz-3dplot}
\usepackage{ytableau}
\usepackage{braket}

\usetikzlibrary{decorations.markings, decorations.pathmorphing, decorations.pathreplacing, decorations.shapes}

\usepackage{xcolor}
\usepackage{xparse}
\definecolor{mplBlue}{rgb}{0,0,1}   
\definecolor{mplRed}{rgb}{1,0,0}    
\definecolor{mplGreen}{rgb}{0,0.5,0}   

\usepackage{fontawesome5}

\makeatletter
\newcommand{\github}[1]{%
   \href{#1}{\faGithub}%
}
\makeatother

\NewDocumentCommand{\colornucleus}{omme{_^}}{%
  \begingroup\colorlet{currcolor}{.}%
  \IfValueTF{#1}
   {\textcolor[#1]{#2}}
   {\textcolor{#2}}
    {%
     #3
     \IfValueT{#4}{_{\textcolor{currcolor}{#4}}}
     \IfValueT{#5}{^{\textcolor{currcolor}{#5}}}
    }%
  \endgroup
}

\usepackage{array}
\newcolumntype{L}[1]{>{\raggedright\let\newline\\\arraybackslash\hspace{0pt}}m{#1}}
\newcolumntype{C}[1]{>{\centering\let\newline\\\arraybackslash\hspace{0pt}}m{#1}}
\newcolumntype{R}[1]{>{\raggedleft\let\newline\\\arraybackslash\hspace{0pt}}m{#1}}

\usepackage[framemethod=default]{mdframed}
\newmdenv[skipabove=7pt,
skipbelow=7pt,
rightline=true,
leftline=true,
topline=true,
bottomline=true,
backgroundcolor=gray!10,
linecolor=black,
innerleftmargin=5pt,
innerrightmargin=5pt,
innertopmargin=5pt,
innerbottommargin=5pt,
leftmargin=0cm,
rightmargin=0cm,
linewidth=1pt]{eBox}

\definecolor{Red}{RGB}{214, 39, 40}
\definecolor{Blue}{RGB} {31, 119, 180}
\definecolor{Orange}{RGB}{255, 153, 51}
\definecolor{Purple}{RGB}{178, 102, 255}
\definecolor{Green}{RGB}{44, 160, 44}
\definecolor{regal}{RGB}{90,0,120}
\definecolor{darkblue}{rgb}{0.15,0.35,0.55}
\definecolor{reddish}{rgb}{0.65, 0.2, 0.2}
\definecolor{darkgreen}{RGB}{50,150,0}
\definecolor{greyish}{rgb}{.90,.90,.90}
\definecolor{greyish2}{rgb}{.96,.96,.96}
\definecolor{greyish3}{rgb}{.37,.37,.37}
\definecolor{darkblue2}{rgb}{0.3,0.4,0.9}
\definecolor{Blue3}{RGB}{31, 119, 180}

\definecolor{blue3}{RGB}{31, 119, 180}
\definecolor{red3}{RGB}{	214, 39, 40}
\definecolor{orange3}{RGB}{255, 127, 14}
\definecolor{green3}{RGB}{44, 160, 44}
\definecolor{repBlue}{RGB}{31, 119, 180}
\definecolor{repRed}{RGB}{	214, 39, 40}
\definecolor{repGreen}{RGB}{44, 160, 44}

\renewcommand\d{\mathrm{d}}

\definecolor{vio}{RGB}{19, 130, 164}
\definecolor{vioo}{RGB}{89, 2, 155}
\newcommand{\Comment}[1]{{}}
\usepackage{empheq}

\usepackage[linktocpage=true]{hyperref}
\hypersetup{
colorlinks=true,
citecolor=darkblue,
linkcolor=reddish,
urlcolor=darkblue,
pdfauthor={},
pdftitle={},
pdfsubject={}
}

\renewcommand{\d}{\text{d}}

\newcommand{\bubA}{
 {\mathchoice
  {\includegraphics[height=2ex]{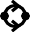}}
  {\includegraphics[height=2ex]{Figures/Bubble_Graphs/Bubble_Icon1.pdf}}
  {\includegraphics[height=1.6ex]{Figures/Bubble_Graphs/Bubble_Icon1.pdf}}
  {\includegraphics[height=1ex]{Figures/Bubble_Graphs/Bubble_Icon1.pdf}}
 }
}

\newcommand{\bubC}{
 {\mathchoice
  {\includegraphics[height=2ex]{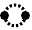}}
  {\includegraphics[height=2ex]{Figures/Bubble_Graphs/Bubble_Icon3.pdf}}
  {\includegraphics[height=1.6ex]{Figures/Bubble_Graphs/Bubble_Icon3.pdf}}
  {\includegraphics[height=1ex]{Figures/Bubble_Graphs/Bubble_Icon3.pdf}}
 }
}

\usepackage{colortbl}
\definecolor{lightgreen}{cmyk}{0.2, 0, 0.2, 0.2}
\definecolor{lightgray2}{cmyk}{0.1,0.1,0,0.1}
\definecolor{Red2}{RGB}{214, 39, 40}
\definecolor{Blue2}{RGB} {31, 119, 180}
\definecolor{Orange2}{RGB}{255, 127, 14}
\definecolor{Green2}{RGB}{44, 160, 44}
\definecolor{blue2}{rgb}{0.27451, 0.482353, 0.92549}
\definecolor{green2}{rgb}{0., 0.85098, 0.278431}
\definecolor{red2}{rgb}{0.94902, 0.168627, 0.262745}

\makeatletter
\newlength{\apb@width}
\newcommand{\autoparbox}[2][c]{\settowidth{\apb@width}{#2}\parbox[#1]{\apb@width}{#2}}

\makeatother


\def\beq{\begin{equation}}
\def\eeq{\end{equation}}
\def\be{\begin{equation}}
\def\ee{\end{equation}}

\allowdisplaybreaks[1]
\newcommand\sqmatrix[2][c]{%
  \fixTABwidth{T}%
  \setbox0=\hbox{$\tabbedCenterstack{#2}$}%
  \setstackgap{L}{\dimexpr\maxTAB@width+\tabbed@gap}%
  \tabbedCenterstack[#1]{#2}%
}

\usetikzlibrary{shapes.misc}

\tikzset{cross/.style={cross out, draw=black, minimum size=2*(#1-\pgflinewidth), inner sep=0pt, outer sep=0pt},
cross/.default={1pt}}

\usetikzlibrary{calc}

\tcbset{
	mybox/.style={
		colframe=black,
		colback=white,
		boxrule=0.5pt,
		arc=3pt,
		width=\textwidth-25pt,
		boxsep=8pt,
		left=10pt, right=5pt, top=-10pt, bottom=5pt
	}
}

\begin{document}


\newgeometry{top=2cm, bottom=2cm, left=2cm, right=2cm}

\begin{titlepage}
\setcounter{page}{1} \baselineskip=15.5pt 
\thispagestyle{empty}

\begin{center}
{\fontsize{18}{18} \bf On Cosmological Correlators at One Loop}
\end{center}

\vskip 30pt
\begin{center}
\noindent
{\fontsize{12}{18}\selectfont 
Guilherme L.~Pimentel and Tom Westerdijk}
\end{center}

\begin{center}
\vskip 4pt
\textit{Scuola Normale Superiore and INFN, Piazza dei Cavalieri 7, 56126, Pisa, Italy}
\end{center}

\vspace{0.4cm}
\begin{center}{\bf Abstract}\end{center}
We study equal-time in-in correlators of massless scalar fields in flat space at one loop. Using the time-ordered decomposition of correlators together with a cosmological analogue of the Baikov representation, we systematically construct relatively simple loop integrals and make manifest why, in this setting, loop corrections to correlators are simpler than those of wavefunction coefficients. As benchmark examples, we analyse the bubble and triangle diagrams. The bubble exhibits a UV divergence that can be removed by a local counterterm, while the triangle yields a finite result, which we evaluate explicitly in terms of dilogarithms using an integral transform for the Laplacian Green’s function. We classify the kinematic singularities of these diagrams using Landau analysis, identifying novel types of singular behaviour, and validate this analysis against the explicit results. Finally, we derive a factorisation property of one-loop cosmological correlators at singular kinematics, relating them to flat-space loop amplitudes and lower-point tree-level correlators.

\noindent 

\vskip 15 pt

\end{titlepage}
\restoregeometry

\newpage
\setcounter{tocdepth}{2}
\setcounter{page}{2}

\linespread{0.75}
\tableofcontents
\linespread{1.}

\newpage
\section{Introduction} \label{sec:intro}

Spatial correlation functions of cosmological observables such as the cosmic microwave background (CMB) and large-scale structure (LSS) can be traced back to the initial conditions at the hot Big Bang. In inflationary cosmology~\cite{Guth:1980zm,Linde:1981mu}, a period of accelerated expansion precedes the hot Big Bang, during which quantum fluctuations are generated and subsequently stretched to cosmological scales, thereby seeding these initial conditions~\cite{Mukhanov:1981xt,Guth:1982ec,Starobinsky:1982ee,Bardeen:1983qw,Hawking:1982cz}. The statistics of this primordial snapshot—taken at the end of inflation, marking the onset of the hot Big Bang—are encoded in \textit{cosmological correlators}.

The physical processes that occurred during inflation are imprinted in cosmological correlators in much the same way that scattering amplitudes encode the outcome of particle collisions measured far from the interaction region. Both objects are computed within the framework of perturbative quantum field theory, which has demonstrated remarkable predictive power in the context of scattering amplitudes. However, unlike in flat-space scattering, the relevant quantum field theory in cosmology is defined on a time-dependent background. This introduces significant conceptual and technical challenges, leaving open even the most basic question: is the theory itself well defined?

The study of cosmological correlators has developed rapidly in recent years, addressing both of these challenges. Broadly speaking, there are two main approaches. One line of research, known as the \textit{cosmological bootstrap}~\cite{Arkani-Hamed:2018kmz,Baumann:2022jpr}, has bypassed many conceptual difficulties by rephrasing the question. Rather than evolving a quantum field theory throughout inflation and extracting correlation functions at its end, the bootstrap approach asks a different question: what is the space of spatial correlation functions at the end of inflation that is consistent with symmetry, unitarity, and other fundamental physical principles? Substantial progress in understanding how these principles are encoded has carved out the landscape of cosmological correlators~\cite{Baumann:2019oyu,Sleight:2019hfp,Mata:2012bx,Bzowski:2011ab,Bzowski:2012ih,Bzowski:2013sza,Bzowski:2019kwd,Kundu:2014gxa,Kundu:2015xta,Shukla:2016bnu,Baumann:2021fxj,Meltzer:2021zin,Hogervorst:2021uvp,Goodhew:2021oqg,Melville:2021lst,Jazayeri:2021fvk,Goodhew:2020hob,Goodhew:2024eup,Thavanesan:2025kyc}, including scenarios with reduced symmetry~\cite{Pajer:2020wxk,Cabass:2021fnw,DuasoPueyo:2023kyh}.

On the other hand, a complementary approach is to assume that quantum field theory simply works and to compute a large variety of correlators in order to generate \textit{data}. From a phenomenological perspective, it is important to provide this data in the form of templates that can be compared with (future) observations. From a more theoretical viewpoint, having access to sufficiently rich data, even in simple toy models, may allow us to uncover deep structures paralleling developments in scattering amplitudes, such as the Amplituhedron~\cite{Arkani-Hamed:2013jha}. Recent works, including Kinematic Flow~\cite{Baumann:2025qjx,Baumann:2024mvm,Arkani-Hamed:2023bsv,Arkani-Hamed:2023kig,Glew:2025ypb,Hang:2024xas}, have provided hints in this direction by revealing geometric structure residing at the late-time boundary of inflation.

Despite substantial progress in both the data-generating approach and the cosmological bootstrap, this success has largely been confined to tree level. At loop level, results are considerably more scarce, and even a systematic understanding of one-loop cosmological correlators remains elusive.
The scarcity of results should not be attributed to a lack of importance. Loop corrections are an inevitable consequence once perturbative quantum field theory is assumed to describe very early-universe physics and, more generally, the validity of the theory relies on a controlled loop expansion. 

In addition to the ultraviolet divergences remedied by standard renormalisation, cosmological loops involving sufficiently light particles are also plagued by infrared (IR) divergences specific to cosmological backgrounds~\cite{Ford:1984hs,Antoniadis:1985pj,Tsamis:1994ca,Tsamis:1996qm,Tsamis:1997za,Polyakov:2007mm,Polyakov:2009nq,Polyakov:2012uc,Giddings:2010nc,Burgess:2010dd,Marolf:2010nz,Krotov:2010ma,Marolf:2010zp,Rajaraman:2010xd,Marolf:2011sh,Giddings:2011zd,Giddings:2011ze,Senatore:2012nq,Pimentel:2012tw,Senatore:2012ya,Beneke:2012kn,Akhmedov:2013vka,Anninos:2014lwa,Akhmedov:2017ooy,Hu:2018nxy,Akhmedov:2019cfd}%
\footnote{More precisely, IR divergences already appear at tree level, but become increasingly severe at higher loop order.}. 
These secular divergences reflect the build-up associated with continuous particle production in an expanding spacetime. They have the potential to spoil the perturbative expansion and may even trigger eternal inflation. Furthermore, they could play an important role in a putative holographic description of de~Sitter space.
While there has been progress towards a systematic treatment of these IR divergences~\cite{Gorbenko:2019rza,Baumgart:2019clc,Mirbabayi:2019qtx,Cohen:2020php,Mirbabayi:2020vyt,Baumgart:2020oby,Cohen:2021fzf,Bzowski:2023nef,Cespedes:2023aal}, a fully satisfactory framework—on an equal footing with renormalisation theory in flat space—is still missing. A more complete understanding of loop-level cosmological correlators would therefore naturally complement and inform this line of research.

Our interest in loop corrections goes beyond their divergent behaviour. A detailed understanding of their analytic dependence on external kinematics addresses more mathematically oriented questions, such as: what is the space of functions that appears at one loop? In flat-space scattering amplitudes, it is well known that a finite family of functions suffices to describe all one-loop amplitudes~\cite{Melrose:1965kb,vanNeerven:1983vr,Bern:1993kr,Binoth:1999sp,Fleischer:1999hq,Denner:2002ii,Duplancic:2003tv,Binoth:2005ff}. Although the four simplest topologies—from the tadpole up to the box—require distinct classes of functions, the pentagon can be expressed entirely within this same basis. An analogue of this statement in cosmology would require the computation, or at least a sufficiently detailed understanding, of higher-site loop diagrams.

Another structural question concerns the kinematic singularities of cosmological correlators and their behaviour near such singularities. Within the bootstrap approach, this information provides essential input for fixing the integration constants left undetermined by differential equations~\cite{Arkani-Hamed:2018kmz}. Similarly, in the context of Kinematic Flow, the set of kinematic `letters' forms the basis of the underlying geometric structure.

Beyond their conceptual relevance, loop corrections to cosmological correlators also have clear phenomenological applications. In the context of the \textit{cosmological collider}, for example, a neutral inflaton can couple to charged particles only through loops. As a consequence, a background of Standard-Model correlator loops appears unavoidable~\cite{Chen:2016uwp,Chen:2016hrz}. Even more tantalisingly, very massive charged particles that are inaccessible to terrestrial colliders could, in principle, be probed through loop effects imprinted in the sky. 
In most concrete models, however, large suppression factors render such signals unobservable. There exist mechanisms that can partially alleviate this issue by enhancing particle production in specific sectors, for instance through the presence of a chemical potential~\cite{Wang:2019gbi,Bodas:2025wuk,Chen:2018xck,Wang:2020ioa,Bodas:2020yho}. 

Loop effects also play an important role in the context of primordial black hole (PBH) formation and scalar-induced gravitational waves (SIGW). In these scenarios, the classes of inflationary models involved often push perturbation theory to its limits, making the issue of perturbative control particularly acute~\cite{Cheng:2021lif,Kawaguchi:2024rsv,Ballesteros:2024zdp,KristianoYokoyama2211,KristianoYokoyama2303,Riotto2303,Firouzjahi2303,FirouzjahiRiotto2304,Firouzjahi2311,ChoudhuryPandaSami2303,MotohashiTada2303,Tasinato2305,Fumagalli2305,ChengLee2305,Franciolini2305,Fumagalli2307,Sheikhahmadi:2024peu,Fumagalli:2024jzz}. More pragmatically, observables in this context frequently require explicit loop calculations. For example, the bispectrum of scalar-induced gravitational waves can be viewed as arising from a triangle loop diagram.

Despite these important motivations and objectives, progress has remained limited for the simple reason that the relevant integrals are hard to compute. Compared to flat-space scattering amplitudes, the technical hurdles for cosmological correlator loops are heightened by the presence of additional time integrals inherent to cosmological calculations. In addition, the integrands of loop integrals lose their Lorentz-invariant structure, obstructing many of the widely used parametrisations familiar from amplitudes.

For two-site loops, most notably the bubble diagram, these difficulties can be efficiently tamed by regarding the loop as a tree-level exchange of composite operators. This connection renders the computation of the loop integral amenable to a variety of (standard) techniques~\cite{Weinberg:2005vy,Senatore:2009cf,Westerdijk:2025ywh,Beneke:2023wmt,delRio:2018vrj,Chowdhury:2023khl,Chowdhury:2023arc,Jain:2025maa,Lee:2023jby,Salcedo:2022aal,Cacciatori:2023tzp,Cacciatori:2024zrv}, and makes it particularly well suited to spectral decompositions~\cite{Qin:2024gtr,Zhang:2025nzd,Loparco:2023rug,Qin:2023nhv,Qin:2023bjk}. Moving to the one-loop three-site diagram—the triangle—this powerful connection to tree-level physics is lost, and correspondingly far fewer results are currently available.

To gain analytic control, we consider a simple toy model of massless scalars. We further neglect spacetime expansion by working in flat space; nevertheless, by evaluating equal-time correlators, boosts and time translations remain broken, as in cosmology. Within this framework, \cite{Benincasa:2024ptf} derived differential equations for the triangle loop correction to the \textit{wavefunction}. The wavefunction and its associated coefficients constitute a distinct class of observables, which serve as an intermediate step towards physical correlation functions. 

In this article, we adopt the same setup but instead focus directly on correlator loops. By a judicious choice of parametrisation, we show explicitly why correlator loops are simpler than their wavefunction counterparts in line with the findings of~\cite{Chowdhury:2023arc,Arkani-Hamed:2025mce}. Furthermore, we fully characterise the singularity structure of the triangle diagram and, for the first time, derive its explicit expression in terms of dilogarithms.

\newpage
\noindent {\bf Roadmap and Summary of Results}

\noindent The computation of the triangle diagram requires a range of techniques that are not standard in the literature. To guide the reader through the remainder of the paper, we therefore provide a roadmap and overview of our main results below.
The final formula is given explicitly in a companion Mathematica notebook \cite{Notebook}, without its derivation.

\vskip 10pt
\noindent
\textbf{1. Constructing the integrals:}
Starting with the Feynman rules for in-in correlators~\cite{Weinberg:2005vy}, a one-loop integral for massless scalars in flat space takes the form
\begin{equation}
    \mathcal I = \int \frac{\d^d q_1}{(2\pi)^d} \, R(\vec q_1,X_1,...,X_V,k_1,...,k_V)
\end{equation}
where $\vec q_1$ is one of the loop three-momenta, and $R$ is a rational function depending on the loop momentum and external kinematics, parametrised by vertex energies $X_1,...,X_V$ and internal energies $k_1,...,k_V$. The integral cannot be performed directly, so it is essential to break it into smaller pieces. Finding such a set of simple integrals relies on a precise symbiosis between the measure and the rational function. 

The rational integrand is uniquely determined by a Feynman graph. Conventionally, this one-to-one relation is described by Feynman rules which express the rational function as a sum of time-integrals of propagators. In general, a propagator can link two time integrals in three different ways: time-ordered, anti-time-ordered or factorised. Each term in the sum of time-integrals can be represented by a Feynman graph with oriented edges or time-ordering. Interestingly, we can write down a rational function associated to a single time-ordered diagram by means of elementary graphical rules~\cite{Fevola:2024nzj,Glew:2025mry}. This function has the following schematic form:
\begin{equation}
    \text{diagram} \qquad \leftrightarrow \qquad \frac{1}{q_1 \cdots q_V} \frac{1}{L_1 \cdots L_V},
\end{equation}
where $q_i = |\vec q_i|$ is what we call a `loop energy,' $L_i$ is a linear factor depending on $q_i$ and external kinematics, and $V$ is the number of vertices in the graph. Taking the sum of a specific set of time-ordered pieces produces the complete integrand. This statement holds for both in-in correlators and wavefunction coefficients, although the specific sets differ. However, we will \textit{not} consider the sum, but the individual terms, since they turn out to be a natural companion for the loop measure. 

The natural set of integration variables from the perspective of the integrand are the loop energies, $q_i$. We need to change variables from the components of the $d$-dimensional vector $\vec q_1$ to the set of $V$ loop energies $q_1,...,q_V$\footnote{Setting $d=3$, this change of variables is, strictly speaking, only guaranteed to be valid for $V \leq 3$, which is sufficient for our purposes.}. The corresponding Jacobian measures the squared volume, $\mathcal V^2$, of the simplex created out of $\vec q_1,...,\vec q_V$ in terms of a polynomial of its side lengths, the loop energies. From the schematic formula~\cite{Benincasa:2024lxe}
\begin{equation}
    \d^d q_1 \propto \mathcal{V}(q_1,...,q_V)^{d-V-1} q_1 \cdots q_V \, \d q_1 \cdots \d q_V
\end{equation}
we see that the $q_1 \cdots q_V$ factor in the measure will cancel against the same factor in the denominator of a single time-ordered rational function. Formulating the integrals in terms of loop energies naturally selects the time-ordered pieces as the minimal partial fraction decomposition. 

At the same time, the change of variables maps the integration region from $\mathbb R^d$ to the positive volume region of loop energies, $\mathcal V^2\geq 0$. Assembling the measure and rational function yields an elegant formula, where the integration is in terms of only loop energies:
\begin{equation}
    \mathcal I = \int_{\mathcal V^2\geq 0} \mathcal V(q_1,...,q_V)^{d-V-1} \frac{1}{L_1 \cdots L_V} \, \d q_1 \cdots \d q_V.
\end{equation}

Integrands of in-in correlators possess a special property: the denominator, $L_1 \cdots L_V$, depends only on $V-1$ linear combinations of loop energies. This property distinguishes loop corrections to in-in correlators from those to wavefunction coefficients, for which $L_1 \cdots L_V$ is a function of the full space of loop energies. As a result, wavefunction coefficients always include more complex functions (higher transcendental weight, hyperelliptic sectors, etc.). A simple linear change of variables from $q_1,...,q_V$ to $u_1,...,u_V$ makes this property manifest so that only the polynomial $\mathcal V^2$ depends on $u_V$. We find that it is possible to consistently `integrate out' this last variable $u_V$ to arrive at \textit{reduced integrals}, which we define as
\begin{equation}
    \mathcal I_{\text{red}} \equiv \int_{\Sigma} \Omega(u_1,...,u_{V-1})^{(d-V-1)/2} \frac{1}{\tilde L_1 \cdots \tilde L_{V-1}} \, \d u_1 \cdots \d u_{V-1},
\end{equation}
where $\tilde L_1\cdots \tilde L_{V-1}$ denotes a new product of linear factors that is obtained by applying partial fractions to the original rational integrand. 
As a result of performing one of the integrals, the polynomial in the integrand changes from $\mathcal V^2$ to $\Omega$ which has, at least for the bubble and triangle diagrams, a lower degree in the $u_i$ variables. 
Finally, integrating out $u_V$ projects the original integration region $\mathcal V^2 \geq 0$ onto a codimension $1$ region $\Sigma$. For the triangle, this means the three-dimensional region $\mathcal V^2 \geq 0$ gets \textit{squashed} onto a surface $\Sigma$ which is bounded by four lines. 
\vskip 24pt
\noindent
\textbf{2. Singularity analysis:} the singularities of the reduced integrals have a physical interpretation in terms of the kinematic loci where time-ordered classical scattering of point particles occurs~\cite{Coleman:1965xm,Mizera:2023tfe}. These are regions in kinematic space, where every time-ordered edge in the loop diagram represents a classical object moving along the direction of its momentum, and which collide and scatter with other objects at a definite point in space and time. The classical constraints are captured in the Landau equations:
\begin{equation}
\begin{split}
    \sum_i \pm_i \alpha_i q_i & = 0 \\
    \sum_i \alpha_i \vec{q}_i & = 0,
\end{split}
\end{equation}
which resemble the Landau equations for amplitudes~\cite{Landau1965,Cutkosky:1960sp,Eden:1966dnq}, apart from specifying the time-ordering with $\pm_i$. In contrast to amplitudes, the quantum process which the in-in correlator probes, does not conserve energy. Classicality, therefore, implies additional energy conservation constraints. Roughly speaking, we need to conserve energy for all vertices that are part of the diagram we enforce to be classical (equivalently, which $\alpha_i \neq 0$). 

From the perspective of the reduced integrals, singularities occur when the integration contour $\Sigma$ gets trapped while the integrand blows up. To be more precise, we first extend the space that $u_1,...,u_{V-1}$ live in to $\mathbb{CP}^{V-1}-S$ where $S$ denotes the singular variety defined by $\Omega=0$ and $L_i=0$ for $i=1,...,V-1$. The linear factors, $L_i$ (in green below), contain the vertex energies, $X_1,...,X_V$, and we are interested in varying these parameters to see where the integral develops singularities. We consider the internal energies $k_1,...,k_V$ to be constant. The integration region is now a $(V-1)$-dimensional contour embedded in $2(V-1)$-dimensional complex space. Varying the position of a set of $L_i$ by tuning the vertex energies, we can position these hyperplanes in such a way that they intersect with $\Sigma$ (shown in gray below). 
\vspace{-.5em}
\begin{center}
\includegraphics{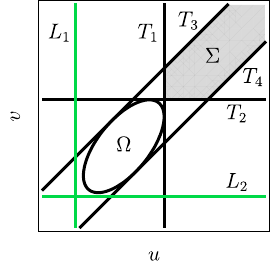}
\end{center}
\vspace{-.8em}
If at the same time, this arrangement prohibits small complex deformations of $\Sigma$, the values of $X_1,...,X_V$ for which this happens define a singular hypersurface in kinematic space. These types of singularities are called physical. In addition, there are configurations of the singular hypersurfaces that can potentially trap the contour but do not intersect it. These configurations can be accessed by letting $X_1,...,X_V$ go to non-physical regions in $\mathbb C^V$, thus accessing different branches of the integral. The values of $X_1,...,X_V$ for which these branches develop singularities are referred to as being on the \textit{second sheet}. 

After performing both types of analyses for a time-ordered diagram of the triangle,
\vspace{-.5em}
\begin{center}
\includegraphics{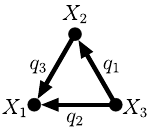}
\end{center}
\vspace{-1em}
we conclude that, in addition to the total energy pole, there are four physical singularities. In the physical Landau analysis, all four correspond to setting a subset of $\alpha_i$ to zero. This can be visualised by pinching the corresponding edges by bringing pairs of vertices together. For the specific time-ordered diagram at hand, we list the physical singularities together with their pinched graph below:

\noindent
\begin{tabular}{>{\centering\arraybackslash} m{.22\textwidth} >{\centering\arraybackslash} m{.22\textwidth}>{\centering\arraybackslash} m{.22\textwidth} >{\centering\arraybackslash} m{.22\textwidth}}
    \includegraphics{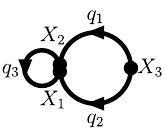} & $X_3+k_3=0$ & \includegraphics{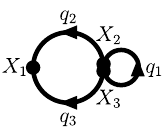} & $X_2+X_3+k_1=0$ \\
    \includegraphics{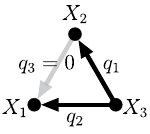} & $X_3+k_1+k_2=0$ & \includegraphics{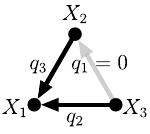} & $X_2+X_3+k_2+k_3=0$
\end{tabular}

When approaching one of the partial energy singularities, $X_2+X_3+k_1=0$, the correlator shows an interesting leading order behaviour reminiscent of tree-level factorisation:
\begin{equation}
    \mathcal I_{\text{triangle}} \sim \Big( \psi(X_1,k_1) + \psi(X_1,-k_1) \Big) \times \mathcal A_{\text{triangle}}(P_1,P_2,P_3)  \qquad \text{for} \qquad X_2+X_3+k_1 \sim 0
\end{equation}
where $\psi(x,y)$ describes a contact diagram with energies $x$ and $y$ on its external legs and $\mathcal A_{\text{triangle}}$ is the triangle amplitude for massless scalars with off-shell external momenta, $P_1=(-X_2-X_3,\vec k_1),P_2=(X_2,\vec k_2)$ and $P_3=(X_3,\vec k_3)$. Notice that $X_2+X_3+k_1=0$ corresponds to $P_1^2=0$, so one vertex is on-shell. The physics behind this factorisation property resembles that of tree level: the limit $X_2+X_3+k_1 \rightarrow 0$ pulls the two vertices $X_2$ and $X_3$ to the infinite past. The two internal edges connecting to vertex $X_1$ form an infinitely long `tube' that goes on-shell. Only far in the past, it `sees' the other two vertices and splits up, thus forming a triangle in the infinite past. In other words, a triangle amplitude. 
\vspace{-0em}
\begin{center}
\includegraphics{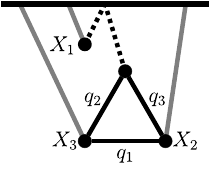}
\end{center}
\vspace{-1.8em} 
\noindent
\vskip 24pt
\noindent
\textbf{3. Explicit evaluation:} for the examples of the bubble and the triangle, we express the reduced integrals in terms of logarithmic and dilogarithmic functions. In our formulation, the reduced integral for the bubble is a hypergeometric function in $d$ dimensions. On the other hand, the triangle does not present itself directly as a standard integral, being given by
\begin{equation}
    \mathcal I_{\text{triangle}}^{\text{(reduced)}} = \int_\Sigma  \frac{1}{\sqrt{\Omega(u,v)}} \ \frac{1}{(X_3+u)(X_2+X_3+v)} \ \d u \, \d v \ ,
\end{equation}
where the irreducible quadratic polynomial in the square root,
\begin{equation}
    \Omega(u,v)=k_1^4+k_2^4+k_3^4-2 k_1^2k_2^2-2k_1^2k_3^2-2k_2^2k_3^2-2uv(k_1^2-k_2^2+k_3^2)+4k_1^2u^2+4 k_3^2 v^2,
\end{equation}
remains the only obstacle to writing the integral directly in terms of dilogarithms since the integration region $\Sigma$ is bounded by hyperplanes.

We use the integral representation for the Green's function for the three-dimensional Klein-Gordon operator due to Whittaker~\cite{Whittaker1902},
\begin{equation}
    \frac{1}{\sqrt{x^2+y^2-z^2}}
    =
    \frac{1}{-2\pi i}\int_0^{2\pi}\frac{\d \theta}{x \cos\theta + y \sin\theta + z} ,
\end{equation}
where $x,y \in \mathbb R$ and $z$ is real with a small imaginary component. After complexifying $\theta = -i\log(w)$, the integral on the right hand-side becomes a contour integral over the unit circle in the complex plane of $w$. It can be computed by picking the residue of one of the two poles of $w$ in the integrand.  

By expressing $\Omega(u,v)$ as a Whittaker integral, the integrals over $u$ and $v$ become standard iterated integrals  
\begin{align}
    \mathcal{I}_{\text{triangle}}^{\text{(reduced)}} &= \int\frac{\d\theta}{-2\pi i}\int_\Sigma \ \frac{\d u \, \d v}{(Uu+Vv+W)(X_3+u)(X_2+X_3+v)} = \nonumber\\&=
    \int\frac{\d\theta}{-2\pi i} \frac{\mathcal  L(\theta)}{-UX_3-V(X_2+X_3)+W}
     ,
\end{align}
where $U, V$ and $W$ are functions of $\theta$ and the parameters $k_1,k_2,k_3$. The function $\mathcal L(\theta)$ expands into a sum of dilogarithms with arguments that depend on the external kinematics and $\theta$. After analytically continuing this function, $\bar{\mathcal{L}}(w)=\mathcal{L}(-i\log(\theta))$, we encounter a problem: $\bar{\mathcal{L}}$ has branch cuts inside and outside the unit circle. Interestingly, the function can be organised in such a way that it becomes a sum of two pieces, $ \bar{\mathcal{L}}= \bar F_I+\bar F_O$,  which have a `big' branch cut inside and outside the unit circle respectively. They also possess `small' branch cuts, outside for $\bar F_I$ and inside for $\bar F_O$, but they can be shrunk to zero in a controlled fashion. Therefore, the $\theta$ integral can be performed by taking residues,
and the result is a sum of dilogarithms with complicated kinematic dependence confirming the naive expectation of \cite{Chowdhury:2025ohm}.

Having found a single time-ordered loop integral, $\mathcal I$, we can generate the results for all other diagrams by simple permutation- and replacement rules. Summing the integrals of all contributing time-ordered graphs, we obtain the full correlator. By fixing all external parameters but one, $X_3$, and choosing appropriate $i\epsilon$ prescriptions, we plot the real and imaginary parts of the triangle correlator below specifying the singularities.
\begin{figure}
\centering
\includegraphics{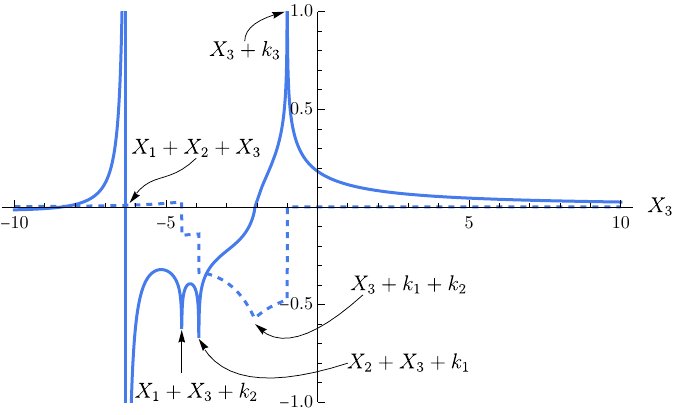}
\caption{Real (solid line) and imaginary (dashed line) parts of the triangle integral $I_\triangle$ with $X_1=3.55$, $X_2=2.76$, $k_1=1.15$, $k_2=0.93$, and $k_3=1.00$. Two distinct $i\varepsilon$ prescriptions have been implemented, as discussed in detail in Sec.~\ref{sec:ExplicitResults}.
}
\vspace{-10pt}
\label{fig:Triangle_Correlator_Result_Intro}
\end{figure} 
\vskip 10pt

\noindent {\bf Outline}  In Sec.~\ref{sec:setup}, we set up the loop computation and break it into various pieces, leading to ``reduced integrals." Before proceeding to direct computation, in Sec.~\ref{sec:Sing} we perform a detailed Landau analysis of the integrals, and show that the singularities are related to classical scattering in a similar fashion as in amplitudes. In Sec.~\ref{sec:ExplicitResults} we perform explicit computations of the in-in correlators, for the bubble and triangle. We state our conclusions in Sec.~\ref{sec:Outlook}.

\newpage
\section{Constructing the Integrals} \label{sec:setup}

In flat space, loop integrands of correlators of massless scalars are rational functions of kinematics. Their derivation relies on performing a set of time-integrals of a product of propagators. The expansion of this product results in a sum of iterated time-integrals that is efficiently organized by a set of \textit{time-ordered} diagrams. These graphs, and an associated set of tubing rules, allow a shortcut to the derivation of an optimal partial fraction decomposition for the loop integrals. 

The decomposition into time-ordered graphs works for both wavefunction coefficients and in-in correlators; the difference between the two observables lies solely in the specific diagrams we include. Interestingly, one-loop wavefunction coefficients require a larger set of time-ordered graphs. The integrals of the rational functions corresponding to these diagrams are described by more complicated functions. 

After changing the integration variables appropriately, it becomes clear why loop integrals of individual time-ordered pieces are simpler. The resulting Jacobian consists of a polynomial raised to a dimension dependent power multiplied by a product of integration variables. After the cancellation of this product between the measure and rational function, the simplicity of correlators emerges in its most concrete form: the possibility to `reduce' the loop integral. The reduced loop integrals will form the basis of the singularity analysis and explicit computations.

\subsection{The Rational Function}\label{sec:Rational}

We will introduce a graph-based method to derive the partial fraction decomposition without performing iterated time-integrals. After highlighting the crucial difference between correlators and wavefunctions, we will compute the in-in rational functions for the bubble and triangle. 

\subsubsection{Rational functions from Time-Ordered Diagrams}

The integrands in flat space of the different observables are associated to unique sets of time-ordered diagrams. Consider a Feynman graph, $\mathcal G$, consisting of $V$ vertices, $\nu_i$, and $E$ edges, $e_i$. 
Let us consider three different assignments for the edges: they can either be time-ordered or anti-time-ordered which we denote with the two different orientations we can draw on an edge. 
An edge can also be non-time-ordered which amounts to `cutting' or `disconnecting' it. 
We represent non-time-ordered edges with dashed lines. We denote by $G$ a diagram of which every edge is time-ordered, anti-time-ordered or disconnected.  

\begin{figure}[h!]
\centering
\begin{subfigure}[b]{.1\textwidth}
    $r=0$ \vspace{.25em}
\end{subfigure}
\begin{subfigure}[b]{.2\textwidth}
    \centering
    \includegraphics{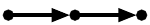}
\end{subfigure}
\begin{subfigure}[b]{.2\textwidth}
    \centering
    \includegraphics{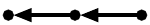}
\end{subfigure}
\begin{subfigure}[b]{.2\textwidth}
    \centering
    \includegraphics{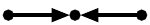}
\end{subfigure}
\begin{subfigure}[b]{.2\textwidth}
    \centering
    \includegraphics{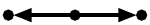}
\end{subfigure}
\vspace{1em}
\hrule
\vspace{1em}
\begin{subfigure}[b]{.1\textwidth}
    $r=1$ \vspace{.25em}
\end{subfigure}
\begin{subfigure}[b]{.2\textwidth}
    \centering
    \includegraphics{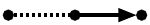}
\end{subfigure}
\begin{subfigure}[b]{.2\textwidth}
    \centering
    \includegraphics{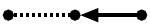}
\end{subfigure}
\begin{subfigure}[b]{.2\textwidth}
    \centering
    \includegraphics{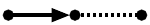}
\end{subfigure}
\begin{subfigure}[b]{.2\textwidth}
    \centering
    \includegraphics{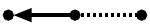}
\end{subfigure}
\vspace{1em}
\hrule
\vspace{1em}
\begin{subfigure}[b]{.1\textwidth}
    $r=2$
\end{subfigure}
\hspace{.3em}
\begin{subfigure}[b]{.8\textwidth}
    \centering
    \includegraphics{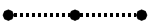}
\end{subfigure}
\hspace{.4em}
\caption{The nine different time-orderings for a three-site graph grouped into the three different sectors.}
\label{fig:ThreeSite_Sectors}
\end{figure}

The number of disconnected edges, $r$, naturally grades the set of possible time-ordered diagrams. Every sector, $r$, consists of the $\left(\begin{smallmatrix} r \\ E \end{smallmatrix}\right)$ different ways of cutting an edge. The diagram for each unique way of disconnecting $r$ edges can be assigned $2^{e-r}$ different time-orderings. In Fig.~\ref{fig:ThreeSite_Sectors} we have displayed the full set of time-ordered three-site graphs. The $r=0,1$ and $r=2$ respectively contain $4$, $4$ and $1$ graphs. 
The difference between wavefunction coefficients and correlators lies solely in which $r$-sectors we include. We will come back to this soon.

\begin{figure}[h!]
\centering
\begin{subfigure}[c]{.24\textwidth}
    \centering
    \includegraphics{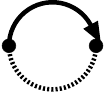}
\end{subfigure}
\begin{subfigure}[c]{.24\textwidth}
    \centering
    \includegraphics{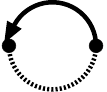}
\end{subfigure}
\begin{subfigure}[c]{.24\textwidth}
    \centering
    \includegraphics{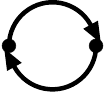}
\end{subfigure}
\caption{The first two diagrams contribute to the wavefunction but not to the in-in correlator. The rightmost diagram is excluded from both due to an ill-defined time ordering.
}
\label{fig:Bubble_OneTimOrd}
\end{figure}

The time-ordered diagrams are in one-to-one correspondence with consistent sets of tubings. A tubing is a connected subgraph of the parent Feynman graph. We define a tubing on a time-ordered diagram to be allowed if all edges that cross the tubing are either positively oriented or non-time-ordered. In other words, edges can never flow into the tubing.
A consistent or admissible set of tubings, $S(G)$, consists of all the subsets of $V$ tubings drawn from the set of allowed tubings for the specific $G$. Furthermore, each tubing in a subset has to satisfy one of the following conditions: it has zero intersection with any other tubing in the subset, is fully contained in another tubing or contains all the other tubings in the subset. This definition automatically removes graphs that contain cycles such as the one illustrated in Fig.~\ref{fig:Bubble_OneTimOrd} (see also~\cite{Baumann:2025qjx}).
\begin{figure}[h!]
\centering
\begin{subfigure}[c]{.24\textwidth}
    \centering
    \includegraphics{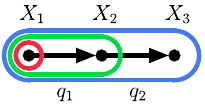}
\end{subfigure}
\begin{subfigure}[c]{.24\textwidth}
    \centering
    \includegraphics{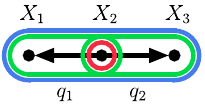}
\end{subfigure}
\begin{subfigure}[c]{.24\textwidth}
    \centering
    \includegraphics{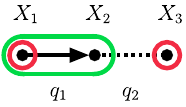}
\end{subfigure}
\begin{subfigure}[c]{.24\textwidth}
    \centering
    \includegraphics{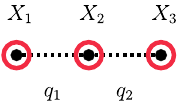}
\end{subfigure}
\caption{Four examples of consistent tubings for the time-ordered three-site graphs.}
\label{fig:ThreeSite_Tubings}
\end{figure}

The rational function associated to a subset of tubings can be read off using a simple rule. First of all, associated to each vertex, $\nu_i$, is a variable, $X(\nu_i)\equiv X_i$, and to each edge connecting $\nu_i$ and $\nu_j$, a variable, $Y(e_{ij})\equiv Y_{ij}$. From now on, we will refer to edges and vertices with the variables we associated to them.
The map $L$ takes in an individual tubing and returns the following sum of vertex- and edge-variables
\begin{equation}
    L(H)=\sum_{\nu_i \in H} X_i + \sum_{\begin{matrix}
        \nu_i,\ \nu_j\in H \\
        e_{ij} \notin H
    \end{matrix}} Y_{ij}.
\end{equation}
Using the $L$ map, we can build the rational function for any time-ordering:
\begin{equation}
    R(G)=\sum_{S_i \in S(G)} \frac{1}{\prod_{H_j\in S_i} L(H_j)}
\end{equation}

If we draw a Feynman graph, $\mathcal{G}$, there are two important observables we can derive from it: wavefunction coefficients and in-in correlators. They differ only in the way we disconnect edges in the diagram. The rule for the wavefunction is simple: every edge can be non-time-ordered. Therefore, the number of graphs, $G_i$, we need to build the wavefunction is simply $3^e$. 

\begin{figure}[h!]
\centering
\begin{subfigure}[c]{.24\textwidth}
    \centering
    \includegraphics{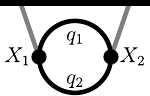}
    \caption{$++$}
    \label{fig:Bubble_InIn_++}
\end{subfigure}
\begin{subfigure}[c]{.24\textwidth}
    \centering
    \includegraphics{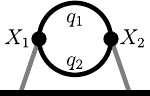}
    \caption{$--$}
    \label{fig:/Bubble_InIn_--}
\end{subfigure}
\begin{subfigure}[c]{.24\textwidth}
    \centering
    \includegraphics{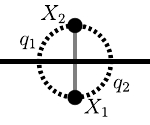}
    \caption{$+-$}
    \label{fig:Bubble_InIn_+-}
\end{subfigure}
\begin{subfigure}[c]{.24\textwidth}
    \centering
    \includegraphics{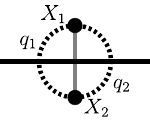}
    \caption{$-+$}
    \label{fig:/Bubble_InIn_-+}
\end{subfigure}
\caption{The nature of the in-in time contour is such that not all graph topologies appear in the correlator. In the example of the bubble, there is no diagram involving one time-ordered and one disconnected propagator, they always come in pairs.}
\label{fig:Bubble_InIn}
\end{figure}
The situation is markedly different for the correlator. The set of $G_i$'s is constructed by only considering \textit{full} cuts: if we split up the set of vertices of $\mathcal{G}$ into two disjoint subsets, a full cut disconnects all edges that connect the two vertex subsets. The origin of these \textit{full} cuts lies in the $+$ and $-$ branches of the in-in-formalism. The different contributions to a correlation function are characterised by the locations of the interaction vertices on the two different branches. 

To make this more explicit, we can draw a thick horizontal line representing the late-time surface and then call the plane below this line, the $+$ branch and the plane above this line the $-$ branch. Taking any Feynman graph $\mathcal G$, the different ways of non-time-ordering the edges are classified by all the different ways of positioning the vertices on the $+$ and $-$ branches. 
We illustrate this with the bubble diagram in Fig.~\ref{fig:Bubble_InIn_++}. Importantly, if an edge crosses the final time surface, it is non-time-ordered. For loops, this leads to a crucial difference between wavefunction coefficients and correlators in the sectors that must be included, since each vertex is necessarily connected to more than one edge.
This becomes abundantly clear when considering the bubble. Whereas graphs that have one non-time-ordered edge as in Fig.~\ref{fig:Bubble_OneTimOrd} contribute to the wavefunction coefficient, they play no role in the correlator. 

This basic difference has a profound implication on the structure of the integrals. In the next section we will show in two examples, the bubble and the triangle, that the loop \textit{integrands} depend on one less linear combination of loop energies than the total number of edges. For the bubble this means that there is a single linear combination that parametrises the integrand. The triangle depends on two linear combinations. We will discuss in Sec.~\ref{sec:ReducedIntegrals} how this allows us to write so-called `reduced' loop integrals that greatly simplify the analysis and computations. 

\subsubsection{The Bubble Correlator}

The rational function for the bubble is remarkably simple. It only consists of two sectors: the fully connected part, $r=0$, and the fully disconnected part with $r=2$. From the perspective of the in-in-formalism, the $r=0$ sector is generated by the $++$ and $--$ contributions which we have shown in Figs.~\ref{fig:Bubble_InIn_++} and~\ref{fig:/Bubble_InIn_--}. The other two in-in terms, the $+-$ and $-+$ graphs as in Figs.~\ref{fig:Bubble_InIn_+-} and~\ref{fig:/Bubble_InIn_-+}, belong to the non-time-ordered $r=2$ sector.

Traditionally, the formulae for the correlators are derived using the Feynman rules for the in-in formalism~\cite{Weinberg:2005vy}. Every in-in graph contributes a function that is represented by a multi-dimensional integral. We can separate every integral into time-integrals and loop-momentum-integrals. If we perform the time-integrals first, the resulting loop \textit{integrands} are rational functions of the vertex and edge variables. Let the number of vertices be $n$. We express the rational functions as $R_{\sigma_1 ... \sigma_n}$ where $\sigma_{i}=\pm$ depending on whether the $i^{\text{th}}$ vertex is on the $+$ or $-$ branch.

Not all graphs give rise to distinct rational functions. The correlators obey a symmetry that relates a given graph to the complex conjugate of its reflection about the late time surface. Furthermore, for conformally coupled scalars in flat space the rational functions are real. This implies in the example of the bubble that $R_{--}=(R_{++})^*=R_{++}$, and $R_{-+}=(R_{+-})^*=R_{+-}$. 

The rational functions for the correlator come with a rigid structure that is fully captured by time-ordered diagrams. Every $R_{\sigma_1...\sigma_n}$ can be expressed as a sum of rational functions of time-ordered diagrams. For the bubble this amounts to summing the functions corresponding to $r=0$ and $r=2$:
\\
\begin{minipage}{.195\textwidth}
    \vspace{1em}
    \includegraphics{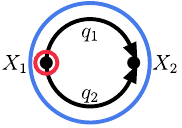}
\end{minipage}
\hspace{-.7em}
\begin{minipage}{.285\textwidth}
    \begin{equation*}
        :\frac{1}{4q_1q_2}\textcolor{blue2}{\frac{1}{X_1+X_2}}\textcolor{red2}{\frac{1}{X_1+q_1+q_2}}
    \end{equation*}
\end{minipage}
\ \
\begin{minipage}{.195\textwidth}
    \vspace{1em}
    \includegraphics{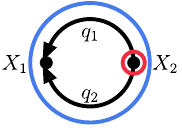}
\end{minipage}
\hspace{-.7em}
\begin{minipage}{.285\textwidth}
    \begin{equation*}
        :\frac{1}{4q_1q_2}\textcolor{blue2}{\frac{1}{X_1+X_2}}\textcolor{red2}{\frac{1}{X_2+q_1+q_2}}
    \end{equation*}
\end{minipage}
\begin{center}
\vspace{-1.5em}
\begin{minipage}{.18\textwidth}
    \centering
    \vspace{1em}
    \includegraphics{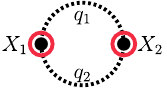}
\end{minipage}\hspace{-.3em}
\begin{minipage}{.32\textwidth}
    \begin{equation*}
        : \frac{1}{4q_1q_2}\textcolor{red2}{\frac{1}{X_1+q_1+q_2}\frac{1}{X_2+q_1+q_2}}
    \end{equation*}
\end{minipage}
\end{center}
We conclude that the loop integrand for the bubble correlator is
\begin{equation}\label{eq:Bubble_Rat}
    R_C^{\circ}=\frac{1}{4q_1 q_2} \left[ \frac{1}{X_1+X_2}\left( \frac{1}{X_1+q_1+q_2} + \frac{1}{X_2 + q_1 + q_2} \right) + \frac{1}{X_1+q_1+q_2} \frac{1}{X_2 + q_1 + q_2} \right]\ .
\end{equation}
It will be useful to define a normalised rational function, $\tilde R_C^{\circ}\equiv4q_1q_2 R_C^{\circ}$.
For the loop integral, it is useful to simplify the piece coming from the disconnected sector further in terms of partial fractions:
\begin{equation}
    \frac{1}{X_1+q_1+q_2} \frac{1}{X_2 + q_1 + q_2} = \frac{1}{X_1-X_2}\left( -\frac{1}{X_1+q_1+q_2} +\frac{1}{X_2 + q_1 + q_2} \right)
\end{equation}

Importantly, apart from the prefactor $1/(4 q_1 q_2)$, {\it the correlator's integrand depends only on a single combination of loop energies, $q_1+q_2 \equiv u$}. On the contrary, the $r=1$ sector that we have to include for the bubble wavefunction coefficient contributes functions that also depend on $q_1$ or $q_2$ in additional to $u$.

\subsubsection{The Triangle Correlator}

\begin{figure}[h!]
    \centering
    \includegraphics{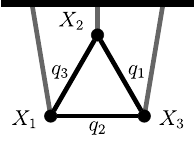}
    \caption{Assignment of variables to the edges and vertices for the $+++$ triangle graph.}
    \label{fig:Triangle_InIn_+++_1}
\end{figure}
As the loop integrand for the triangle becomes more involved, it pays off to make use of the correlator's permutation symmetry. Suppose we assign the variables $X_1,X_2,X_3$ and $q_1,q_2,q_3$ to respectively the vertices and edges as done in Fig.~\ref{fig:Triangle_InIn_+++_1}. The loop \textit{integrand} is invariant under the dihedral permutation group $D_3$ consisting of two cyclic rotations and three reflections acting on the vertices and edges of the graph. The variables are labelled in such a way that they all transform `covariantly'. As an example, consider the permutation $1\rightarrow 3\rightarrow 1, 2\rightarrow 2$. The vertex associated to $X_2$ and the edge associated to $q_2$ remain fixed whereas $X_1\leftrightarrow X_3$ and $q_1 \leftrightarrow q_3$.

In total, there are four sectors of time-ordered diagrams. The correlator for the triangle contains again only two sectors: the $r=0$ and $r=2$ sectors. The $+++$ (and $---$ by reflection symmetry) contribution maps to the fully time-ordered sector. We can endow the three edges with six acyclic time-orderings of which we highlight one below:
\begin{center}
\begin{minipage}{.25\textwidth}
    \centering
    \vspace{2em}
    \includegraphics{Figures/Triangle_Graphs/Triangle_InIn_+++.pdf}
\end{minipage}
\begin{minipage}{.1\textwidth}
\begin{equation*}
    \ni
\end{equation*}
\end{minipage}
\begin{minipage}{.25\textwidth}
    \centering
    \vspace{2em}
    \includegraphics{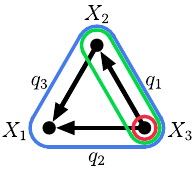}
\end{minipage}
\end{center}
The other five terms are obtained by acting on the rational function with the elements of $D_3$.

The $+-+$ contribution is a sum of two time-ordered diagrams which are related by the $1\leftrightarrow 3$ permutation  
\begin{center}
\begin{minipage}{.25\textwidth}
    \centering
    \includegraphics{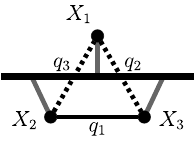}
\end{minipage}
\begin{minipage}{.1\textwidth}
\begin{equation*}
    =
\end{equation*}
\end{minipage}
\begin{minipage}{.25\textwidth}
    \centering
    \includegraphics{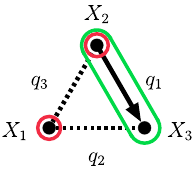}
\end{minipage}
\begin{minipage}{.1\textwidth}
\begin{equation*}
    +
\end{equation*}
\end{minipage}
\begin{minipage}{.25\textwidth}
    \centering
    \includegraphics{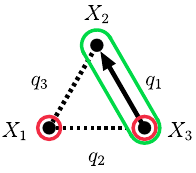}
\end{minipage}
\end{center}
Notice that the $+-+$ and $++-$ terms are obtained from $-++$ by the two cyclic permutations of $D_3$. 

We would like to completely mod out the $D_3$ group when writing down the loop's integrand. In this way, we only have to analyse and compute two terms. If we then want to retrieve the full function, we have to symmetrise this individual `covariant' building block by summing over all permutations. We choose to work with the following two diagrams that are mapped to the two rational functions displayed on the right hand-side:
\begin{center}
\begin{minipage}{.25\textwidth}
    \centering
    \includegraphics{Figures/Triangle_Graphs/Triangle_Tubing_321.pdf}
\end{minipage}
\begin{minipage}{.05\textwidth}
\begin{equation*}
    \rightarrow
\end{equation*}
\end{minipage}
\begin{minipage}{.64\textwidth}
    \begin{align}
        \frac{1}{8q_1q_2q_3}\textcolor{blue2}{\frac{1}{X_1+X_2+X_3}}\textcolor{green2}{\frac{1}{X_2+X_3+q_2+q_3}}\textcolor{red2}{\frac{1}{X_3+q_1+q_2}} 
    \label{eq:R_Triangle_321}
    \end{align}
\end{minipage}
\begin{minipage}{.25\textwidth}
    \centering
    \includegraphics{Figures/Triangle_Graphs/Triangle_Tubing_0032.pdf}
\end{minipage}
\begin{minipage}{.05\textwidth}
\begin{equation*}
    \rightarrow
\end{equation*}
\end{minipage}
\begin{minipage}{.64\textwidth}
    \begin{align}
        \frac{1}{8q_1q_2q_3}\textcolor{red2}{\frac{1}{X_1+q_2+q_3}}\textcolor{green2}{\frac{1}{X_2+X_3+q_2+q_3}}\textcolor{red2}{\frac{1}{X_3+q_1+q_2}} 
    \end{align}
\end{minipage}
\end{center}

Summing the two terms:
\begin{equation}
    R^{\triangle}_C=\frac{1}{8 q_1 q_2 q_3} \left[ \textcolor{blue2}{\frac{1}{X_1+X_2+X_3}}+\textcolor{red2}{\frac{1}{X_1+q_2+q_3}}  \right] \textcolor{red2}{\frac{1}{X_3+q_2+q_3}}\textcolor{green2}{\frac{1}{X_2+X_3+q_2+q_3}}+ \text{perms.}
    \label{eq:Triangle_Corr_Rat}
\end{equation}
Let us also define the normalised function, $\tilde R^{\triangle}_C\equiv8q_1q_2q_3R^{\triangle}_C$.
By inspecting an individual covariant term of $\tilde R^\triangle_C$, one straightforwardly infers that there are only two linear combinations of loop variables that appear: $q_1+q_2\equiv u$ and $q_2+q_3 \equiv v$. In App.~\ref{app:wfcoeffs}, we write down a few examples of rational functions for the other sectors. Constructing an analogous covariant building block reveals that the $r=1$ and $r=3$ sectors depend on an additional linear combination of loop variables.

\subsection{The Measure}

In general, the integrand of a $1$-loop diagram in cosmology will depend on all the loop energies, $q_i \equiv | \vec q_i|$ for $i \in 1,...,E$. In the previous section we have highlighted the structure of the integrands which combined with the measure, lead to the following schematic form:
\begin{equation}
    \mathcal{I}_{1-\text{loop}}^{(V)} = \int\frac{\d^d q_1}{(2\pi)^d} \frac{1}{2^Vq_1\cdots q_V} \left(\frac{1}{L_1 \cdots L_V} + \cdots \right),
\end{equation}
where the $L_i$ are linear factors that depend on the loop energies, $q_i$, and external energies, $X_i$. 

The Baikov representation~\cite{Frellesvig:2017aai,Baikov:1996iu} is one of the most convenient ways of parametrising the loop integrals since it naturally only depends on the loop energies. In general, one obtains the cosmological Baikov form of the measure by changing variables from the components of $\vec q_1$ with respect to some orthonormal coordinate system to the magnitudes of the loop momenta, $q_i \equiv |\vec q_i|$. We call these, following the customary abuse of terminology, loop energies. 

For 1-loop diagrams, there are as many loop energies as vertices, $V$. The change of variables is therefore from the first $V$ components of $\vec q_1$ to the $V$ loop energies\footnote{For $V>3$, this change of variables is, strictly speaking, valid only in generic dimension $d$. For $V=4$, however, this restriction can be bypassed by working in generic $d$, following the procedure explained below, and setting $d=3$ only at the very end.}. Since the space dimension is $d$, we will let the last $d-V+1$ components of the loop momentum parametrise the volume of a sphere,
\begin{equation}
    \d^d q_1 = \d q_1^{(1)}\wedge \d q_1^{(2)} \wedge \cdots \wedge \left( {q_1^{(V)}}^{d-V} \d q_1^V \wedge \d\Omega_{d-V+1}\right).
\end{equation}
The determinant Jacobian for the change of variables from $\vec q_1 = (q_1^{(1)},...,q_1^{(V)})$ to $\vec Q = (q_1, q_2, ..., q_V)$ is easiest obtained from the inverse,
\begin{equation}
    \d q_1^{(1)} \wedge \ ... \ \wedge \d q_1^{(V)} = \left| \frac{\partial q_1^i}{\partial Q^j} \right| \d q_1 \wedge \ ... \ \wedge \d q_V= \left| \frac{\partial Q^i}{\partial q_1^j} \right|^{-1} \d q_1 \wedge \ ... \ \wedge \d q_V,
\end{equation}
Every $\vec q_i$ can be linearly related to $\vec q_1$ with a coefficient $\pm$ and thus the matrix corresponding to the inverse of the Jacobian takes the following form,
\begin{equation}
    \left(\frac{\partial Q^i}{\partial q_1^j}\right)=
    \begin{pmatrix}
        \vec q_1/q_1 \\
        \vec q_2/q_2 \\
        \vdots \\
        \vec q_V/q_V
    \end{pmatrix}.
\end{equation}
The determinant of this matrix is proportional to the volume of the simplex formed by the vectors of the loop momenta~\cite{Sommerville1958}:
\begin{equation}
    \left|\frac{\partial Q^i}{\partial q_1^j}\right|=\frac{1}{q_1\cdots q_V} \left|\begin{matrix}
        \vec q_1 \\
        \vec q_2 \\
        \vdots \\
        \vec q_V
    \end{matrix}\right|
    =\frac{V!}{q_1\cdots q_V} \mathcal V(\vec q_1,...,\vec q_V).
\end{equation}
It turns out that we can write the radius $q_V$ in terms of this volume as well. Starting with the matrix expression, we can subtract the first row from the second, the second from the third and so forth without changing the determinant,
\begin{equation}
    \left|\begin{matrix}
        \vec q_1 \\
        \vec q_2 \\
        \vdots \\
        \vec q_V
    \end{matrix}\right| = \left|\begin{matrix}
        \vec q_1 \\
        \vec q_2-\vec q_1 \\
        \vdots \\
        \vec q_V - \vec q_{V-1}
    \end{matrix}\right|= \left|\begin{matrix}
        \vec q_1 \\
        \vec k_1 \\
        \vdots \\
        \vec k_{V-1}
    \end{matrix}\right|
\end{equation}
We can then choose the $V$-th basis vector of the orthonormal coordinate system to be perpendicular to the external momenta, $\vec k_1 , ... , \vec k_{V-1}$. This way the determinant reduces to 
\begin{equation}
    V!\times\mathcal{V}_V=\left|\begin{matrix}
        \vec q_1 \\
        \vec k_1 \\
        \vdots \\
        \vec k_{V-1}
    \end{matrix}\right| = q_1^{(V)} \left|\begin{matrix}
        \vec k_1 \\
        \vec k_2 \\
        \vdots \\
        \vec k_{V-1}
    \end{matrix}\right|
\end{equation}
Consequently,
\begin{equation}
    q_1^{(V)} = V\times\frac{\mathcal{V}(\vec q_1,...,\vec q_V)}{\mathcal V(\vec k_1,...,\vec k_{V-1})}
\end{equation}
Collecting the results, the measure can be expressed solely in terms of the edge-lengths of the $V$-dimensional simplex formed by the loop momenta,
\begin{align}
    \d^d q_1 
    &= \d q_1^{(1)}\wedge \d q_1^{(2)} \wedge \cdots \wedge \left( {q_1^{(V)}}^{d-V} \d q_1^V \wedge \d\Omega_{d-V+1}\right) \nonumber \\
    &
    =
    \frac{ V^{d-V}}{V!}
    \frac{\mathcal{V}(\vec q_1,...,\vec q_V)^{d-V-1}}{\mathcal{V}(\vec k_1,...,\vec k_{V-1})^{d-V}}q_1 \cdots q_V \ \d q_1\wedge \ \cdots \ \wedge \d q_V \wedge \d\Omega_{d-V+1}
\end{align}
By multiplying the matrix of momenta vectors by its transpose from the right, we can express its determinant in terms of the corresponding Gram-determinant
\begin{equation}
    \left| \begin{matrix}
        \vec q_1 \\
        \vec q_2 \\
        \vdots\\
        \vec q_V
    \end{matrix}\right| = 
    \sqrt{\left|\begin{pmatrix}
        \vec q_1 \\
        \vec q_2 \\
        \vdots\\
        \vec q_V
    \end{pmatrix} \cdot \begin{pmatrix}
        \vec q_1 \\
        \vec q_2 \\
        \vdots\\
        \vec q_V
    \end{pmatrix}^T \right|} = \sqrt{|q_i \cdot q_j|}.
\end{equation}
Repeating the same step for the $\vec k_i$ determinant, we obtain the following compact expression~\cite{Benincasa:2024lxe},
\begin{equation}
    \d^d q_1 = \d V(\mathbb S_{d-V}) \times \d q_1 \cdots \d q_E \times q_1 \cdots q_E \times \frac{|\vec q_i \cdot \vec q_j|^{\frac{d-V-1}{2}}}{|\vec k_i\cdot \vec k_j|^{\frac{d-V}{2}}}.
\end{equation}
Every dot product, $\vec q_i \cdot \vec q_j$, in the Gram determinant can be written as a sum of $q_i^2$'s and $k_i^2$'s by using three-momentum conservation at every vertex,
\begin{equation}
    2\vec q_i \cdot \vec q_j = \Big|\sum_{n=i}^{j-1}\vec k_n\Big|^2-q_i^2 - q_j^2
\end{equation}
assuming the internal lines associated to $q_i$ and $q_{i+1}$ are connected to the vertex $\nu_i$. This leads to expressing the volume of a simplex in terms of the Cayley-Menger determinant. 

An important property of this measure is that it is always proportional to the product of loop energies, $q_1 \cdots q_V$. Working with the time-ordered decompositions of the correlators (and wavefunction coefficients), this product will exactly cancel against the same factor in the denominator of every term in the sum of rational functions. Therefore, the final `partial fraction decomposition' is in a certain sense minimal. To be more precise, in the space of loop energies, $(q_1,q_2,...,q_V)\in \mathbb R^V$, the number of linearly independent factors in the denominator is smallest. More explicitly, the schematic form of the 1-loop integrals in terms of the Baikov representation reads
\begin{equation}
    \mathcal{I}_{1-\text{loop}}^{(V)} = \frac{ V^{d-V}}{V!}\int \frac{\mathcal{V}(\vec q_1,...,\vec q_V)^{d-V-1}}{\mathcal{V}(\vec k_1,...,\vec k_{V-1})^{d-V}} \left(\frac{1}{L_1 \cdots L_V} + \cdots \right)\d q_1 \d q_2 \cdots \d q_V,
\end{equation}

Notice that the cosmological Baikov measure favours the time-ordered decompositions of correlators (and wavefunction coefficients) over any other representation.

Another striking property of the cosmological Baikov parametrisation is the integration region. The original integration domain is simply given by $\mathbb R^d$ while implicitly three-momentum is conserved at the vertices. The momentum conservation equations build a simplex from the $\vec q_i$ embedded in $\mathbb R^d$. In other words, we are integrating over all configurations of the vectors, $\vec q_i$, that form a simplex\footnote{Modulo possible singular hypersurfaces.} by letting a single vector vary over all $\mathbb R^d$. In terms of the edge-lengths, $q_1,...,q_V$, the simplex of loop momenta only exists if the loop energies satisfy a positivity condition~\cite{Benincasa:2024lxe}:
\begin{equation}
    \frac{\mathcal V(\vec q_1,...,\vec q_V)}{\mathcal V(\vec k_1,...,\vec k_{V-1})} \geq  0.
\end{equation}
The configurations of external kinematics for which the bound is saturated for all $q_1,...,q_V$ instead of strictly satisfied are associated to singularities which will be discussed in the next section. 
\\

\noindent 
\textbf{Bubble:}
\begin{figure}[h!]
    \centering
    \includegraphics{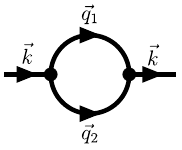}
    \caption{The bubble Feynman graph with the arrows indicating the directions of the three-momenta.}
    \label{fig:Bubble_Momenta}
\end{figure}
In general, the volume of the loop-momentum simplex in terms of the edge-lengths, $\mathcal V(\vec q_1,...,\vec q_V)$, is an irreducible polynomial. The only special case is the bubble for which the simplex is a triangle and the polynomial factorises into linear factors,
\begin{equation}
    \mathcal V ( \vec q_1,\vec q_2) = \frac{1}{4}\sqrt{(q_1+q_2+k)(q_1+q_2-k)(q_1-q_2+k)(-q_1+q_2+k)}.
\end{equation}
This expression is known as Heron's formula. It is easy to recognise how the three triangle inequalities starting with $q_1+q_2-k \geq 0$ imply positivity of the triangle's area. Notice that permuting the vertices and edges correspondingly, $ q_1 \leftrightarrow q_2$, does not affect Heron's formula. This invariance is also reflected in the integration region as a reflection symmetry.

The correlator's rational function multiplied by the factor $q_1 \cdots q_V$ depends only on the specific combination $q_1+q_2$ (see \eqref{eq:Bubble_Rat}). After a simple linear change of variables from $(q_1,q_2) \rightarrow (u,v)= (q_1+q_2,q_1-q_2)$, the integral factorises:
\begin{equation}
    \mathcal I_{\text{bubble}} = \frac{2^{3-d}}{k^{d-2}}\int_{-k}^{k} (k^2-v^2)^{\frac{d-3}{2}}\d v\int_k^\infty (u^2-k^2)^{\frac{d-3}{2}} \left( \frac{1}{(X_1+X_2) (X_1+u)} + \cdots \right) \d u \ .
    \label{eq:Bubble_Schematic}
\end{equation}
One could regard `integrating out' the $v$ variable as a projection of the original integration region onto a constant $v$ section. Although in this case, the projection is trivial since the original domain is rectangular, this picture will guide us for the more complicated examples.
\begin{figure}[htb!]
    \centering
    \includegraphics[width=.22\textwidth]{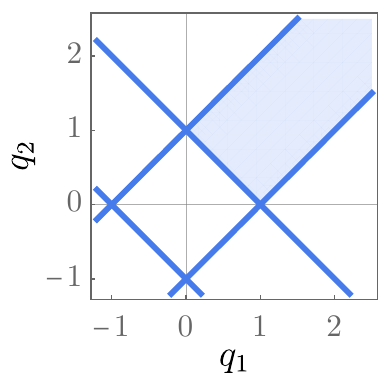}
    \caption{The integration domain for the bubble in terms of the side-lengths $q_1$ and $q_2$ with $k=1$. A reflection about the diagonal does not affect the integration region.}
    \label{fig:Bubble_Region}
\end{figure}

\begin{figure}[htb!]
    \centering
    \includegraphics{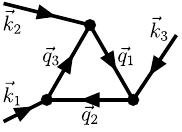}
    \caption{The triangle Feynman graph with the arrows indicating the direction of the three-momenta. }
    \label{fig:Triangle_Momenta}
\end{figure}

\noindent
\textbf{Triangle:}
With the triangle we make an encounter with the more general structure of cosmological loop integrals. Let $q_i$ be the loop energy of the edge opposite to $\nu_i$ as displayed in Fig.~\ref{fig:Triangle_Momenta}. 
This assignment is slightly different from the canonical one we used before. However, the permutational symmetry of the vertices captured by the $D_3$ group, acts more naturally in this way.

The external momenta and the loop momenta form a tetrahedron. Its volume in terms of the edge-lengths is an irreducible sextic polynomial:
\begin{equation}
    \mathcal V(\vec q_1,\vec q_2,\vec q_3)^2 = 
    \frac{1}{3!}{\left|  
    \begin{matrix}
        q_1^2 & \frac{q_1^2+q_2^2-k_3^2}{2} & \frac{q_1^2+q_3^2-k_2^2}{2} \\
        \frac{q_1^2+q_2^2-k_3^2}{2}& q_2^2 & \frac{q_2^2+q_3^2-k_1^2}{2} \\
         \frac{q_1^2+q_3^2-k_2^2}{2}& \frac{q_2^2+q_3^2-k_1^2}{2} & q_3^2 
    \end{matrix}
    \right|}.
\end{equation}
Importantly, the tetrahedron's volume is invariant under the $D_3$ permutation group. 

\begin{figure}[!htb]
    \centering
    \begin{minipage}{.4\textwidth}
        \centering
        \includegraphics[width=0.9\linewidth]{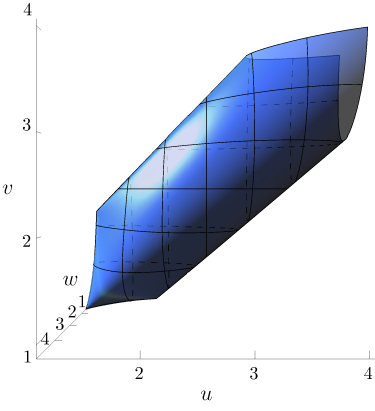}
    \end{minipage}\hspace{.1\textwidth}
    \begin{minipage}{.4\textwidth}
        \centering
        \includegraphics[width=0.9\linewidth]{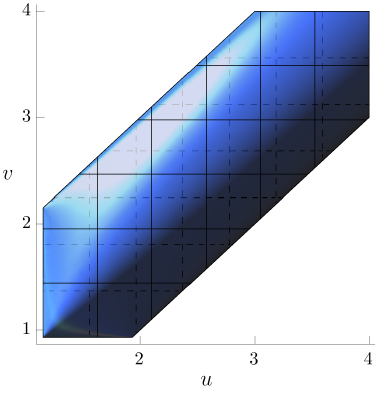}
    \end{minipage}%
    \caption{The variety defined by the null-volume of a tetrahedron in terms of the $u,v$ and $w$ variables plotted for the values $k_1=1.17,k_2=1.37$ and $k_3=1.05$. The left panel shows a generic viewpoint; the right panel the projection onto a constant $w$ section. }
    \label{fig:Volume_Variety}
\end{figure}

In analogy to the bubble, we perform a linear change of variables: $(q_1,q_2,q_3) \rightarrow (u,v,w)=(q_1+q_2,q_2+q_3,q_1+q_3)$. These variables transform respectively as $(q_3,q_1,q_2)$ under $D_3$. The zero-volume variety in $(u,v,w)\in \mathbb R^3$ defined by $\mathcal V(\vec q_1,\vec q_2,\vec q_3)=0$ is plotted in Fig.~\ref{fig:Volume_Variety}. 
Plugging $R_C^\triangle$ from \eqref{eq:Triangle_Corr_Rat} into the schematic expression of the loop integral,
\begin{equation}
    \mathcal{I}_{\text{triangle}} = \frac{3^{d-3}}{3!\times 8}\int \frac{\mathcal V(\vec q_1,\vec q_2,\vec q_3)^{d-4}}{\mathcal V(\vec k_1,\vec k_2)^{d-3}} \ \tilde R^{\triangle}_C \ \d q_1 \d q_2 \d q_3 \ .
    \label{eq:Triangle_Schematic}
\end{equation}
Again, as for the general case, the product $q_1 q_2 q_3$ cancels in numerator and denominator. As we have mentioned before, a single covariant building block only depends on two linear combinations of the loop energies. For example, the two terms in \eqref{eq:Triangle_Corr_Rat} only depend on $u$ and $v$ but not on $w$:
\begin{equation}
    \tilde R_C^\triangle = \frac{1}{(X_3+u)(X_2+X_3+v)(X_1+X_2+X_3)}+\frac{1}{(X_1+v)(X_3+u)(X_2+X_3+v)} + \text{perms.}
    \label{eq:Triangle_Rat_uv}
\end{equation}

The integral linearly distributes over all $12$ time-ordered terms and thus the integral also splits up into $12$ corresponding parts. Since the measure and integration region do not transform under the action of $D_3$, the $12$ integrals can all be related to only $2$ by permutations. 

Furthermore, one can split the second term of \eqref{eq:Triangle_Rat_uv} into two simpler terms using partial fractions,
\begin{equation}
    \frac{1}{X_3+u}\frac{1}{(X_1+v)(X_2+X_3+v)} = \frac{1}{X_3+u}\left( \frac{1}{X_1 + v}-\frac{1}{X_2+X_3+v} \right) \frac{1}{-X_1+X_2+X_3}.
    \label{eq:Triangle_Integrand_PF}
\end{equation}
We can thus obtain the full integral by considering just a single rational function in the integrand,
\begin{equation}
    \frac{1}{\pm X_1+X_2+X_3}\frac{1}{X_3 + u}\frac{1}{X_v+v},
\end{equation}
where $X_w = X_1$ or $X_2+X_3$ and the minus sign is chosen correspondingly.

\subsection{The Reduced Loop Integrals}\label{sec:ReducedIntegrals}

In general, the rational functions corresponding to the correlator depend on one fewer loop energy than there are integration variables. Performing the integrals explicitly will be simplified by first `integrating out' the redundant loop energy. Moreover, the measure of a reduced integral turns out to resemble the leading Landau singularity for the corresponding amplitude\footnote{Technically, Landau singularities of the second kind.}. We will first illustrate the `integrating out'-procedure with the bubble and triangle and then show how the polynomials we find in the measures can be reformulated in terms of formal four-vectors. The number of space dimensions is fixed throughout this section to be $d=3+2\epsilon$. 

\paragraph{Bubble.}
For the special case of the bubble, the redundant variable is trivially integrated out since \eqref{eq:Bubble_Schematic} is already factorised. The redundant $v$ integral is straightforward:
\begin{equation}
    \int_{-k}^{k} (k^2-v^2)^{\epsilon}\d v =  k^{1+2\epsilon}\frac{\sqrt{\pi } \Gamma (\epsilon +1)}{\Gamma \left(\epsilon +\frac{3}{2}\right)}.
\end{equation}
By virtue of the same factorisation, the reduced integral is directly read off:
\begin{equation}
    \mathcal I_{\text{bubble}}^{\text{(reduced)}}=\int_k^\infty (u^2-k^2)^{\epsilon} 
    \ R_{C,\text{bubble}} \
    \d u \ ,
\end{equation}
where we have ignored all numerical factors in the definition of the reduced integral. 

The polynomial in the measure can be written in terms of bookkeeping four-vectors. Let $Q_1^\mu=(q_1,\vec q_1)$ and $Q_2^\mu=(q_2,\vec q_2)$, then using the definition of $u$ and momentum conservation at one of the two vertices, 
\begin{equation}
    u^2-k^2=(q_1+q_2)^2-|\vec q_1+\vec q_2|^2 = \eta_{\mu\nu}(Q_1+Q_2)^\mu(Q_1+Q_2)^\nu.
\end{equation}

\paragraph{Triangle.}
Suppose that we take as our generating rational function
\begin{equation}
    \frac{1}{X_3+u}\frac{1}{X_v+v},
\end{equation}
then the reduced integral is obtained from integrating out $w$. Changing variables in $\mathcal{V}(\vec q_1,\vec q_2,\vec q_3)$ from $(q_1,q_2,q_3)$ to $(u,v,w)$ we observe that the highest order of $w$ that appears is quadratic. We may always factorise the polynomial in terms of one variable (by continuing to complex space $(u,v,w)\in \mathbb C^3$) in the following way:
\begin{equation}
    8\times 3!\times\mathcal V^2 = \Omega(u,v) \times \Big(w-w_1(u,v)\Big) \Big(w_2(u,v)-w\Big).
    \label{eq:Volume_fact}
\end{equation}
The factor $8\times 3!$ serves solely as an aesthetic normalisation.

Focusing only on the $w$ integral, we can see that integration region is bounded by the two solutions of the volume polynomial $w_1$ and $w_2$ defined by \eqref{eq:Volume_fact}. As a result, we can readily evaluate the integral,
\begin{align}
    (8\times 3!)^{\epsilon-1/2}\int 
    \mathcal V(u,v,w)^{-1+2\epsilon}
    \d w &= 
    \Omega^{\epsilon-1/2}
    \int_{w_1}^{w_2} \Big((w-w_1)(w_2-w)\Big)^{\epsilon-1/2} \d w \\
    &= B(\epsilon+\tfrac12,\epsilon+\tfrac12) \ \Omega^{\epsilon-1/2} \ (w_2-w_1)^{2\epsilon}.
\end{align}

The full integration region defined by $\mathcal{V}(\vec q_1,\vec q_2,\vec q_3)\geq 0$ projects onto a $(u,v)$-surface by integrating out $w$. Algebraically, this surface is defined by the positivity of the discriminant, $\Omega(u,v) \times (w_2-w_1) \geq 0$, since it tells us for a pair of $(u,v)$ whether there are one or two real solutions for $w$. In other words, whether we are inside the `tube' depicted in Fig.~\ref{fig:Volume_Variety}. The reduced integration region is thus defined as
\begin{equation}
    \Sigma = \Big\{ (u,v) \in \mathbb R^2 \ \Big| \ \Omega(w_2-w_1)\propto(k_2+u-v) (k_2-u+v) \geq 0,\
    u\geq k_3, \ v \geq k_1 \Big\} \ ,
\end{equation}
where the latter two conditions are enforced by three-momentum conservation respectively at the vertices $\nu_3$ and $\nu_1$.
\begin{figure}[htb!]
    \centering
    \includegraphics[width=0.35\linewidth]{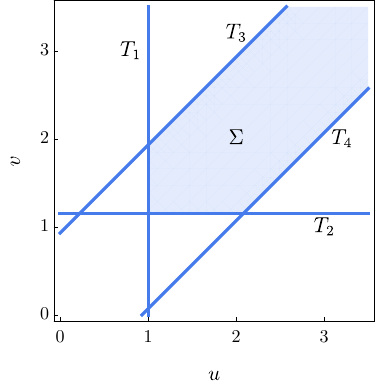}
    \caption{The reduced integration region for the triangle. }
    \label{fig:Triangle_Region}
\end{figure}

From a geometrical point of view this region is obtained by a projection onto a constant $w$ section as shown in the right panel of Fig.~\ref{fig:Volume_Variety}. Moreover, the inequalities that define $\Sigma$ can also be directly concluded from three-momentum conservation at the vertices. On top of requiring that the magnitudes of all three-vectors should be positive, the four inequalities bounding the integration domain are the four remaining triangle inequalities that we can write down using only $u$ and $v$. For example, $\pm u \mp v+k_2=\pm q_1 \mp q_3+k_2 \geq 0$ correspond to the two remaining triangle inequalities for vertex $\nu_2$. The other two, $w\pm k_2\geq 0$ are now automatically `integrated in' by integrating out $w$. 

The triangle integral is UV and IR finite and therefore we can set $\epsilon=0$. If we ignore all numerical factors, the triangle integral in $d=3$ reduces to
\begin{equation}
    \mathcal I_{\text{triangle}}^{\text{(reduced)}} = \int_\Sigma  \frac{1}{\sqrt{\Omega(u,v)}} \ \frac{1}{(X_3+u)(X_v+v)} \ \d u \, \d v \ .
    \label{eq:Triangle_Reduced_1}
\end{equation}
where the function $\Omega(u,v)$ is an irreducible quadratic polynomial\footnote{It is quartic and homogeneous in $u,v,k_1,k_2,k_3$.},
\begin{equation}
    \Omega(u,v)=k_1^4+k_2^4+k_3^4-2 k_1^2k_2^2-2k_1^2k_3^2-2k_2^2k_3^2-2uv(k_1^2-k_2^2+k_3^2)+4k_1^2u^2+4 k_3^2 v^2.
    \label{eq:Omega}
\end{equation}

\begin{figure}[htb!]
    \centering
    \includegraphics{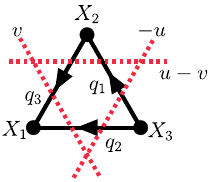}
    \caption{A triangle Feynman diagram with the arrows specifying the time-ordering. We assigned $\pm q_i$ according to the orientation of the time-ordering with respect to the triangle made of cuts (\textcolor{red2}{red} dashed lines).}
    \label{fig:Triangle_Slicing}
\end{figure}

Let us construct the non-Lorentz four-vectors from the loop energies and momenta: $Q_i^\mu=(q_i,\vec q_i)$. Consider now the connected time-ordered diagram corresponding to the rational function in \eqref{eq:Triangle_Reduced_1} with $X_v=X_2+X_3$ as in Fig.~\ref{fig:Triangle_Slicing}.
For any time-ordering diagram we can assign an energy to a cut as the signed sum of loop energies that intersect the cut. The sign is determined by the orientation of the time ordering relative to the cut triangle: if time flows into the triangle, we assign a minus sign, whereas if time flows out, we assign a plus sign. In the example at hand, there are three cuts respectively with $-u, u-v$ and $v$ as associated energies. 

Similarly, we can also specify a three-momentum to a cut as the sum of three-momenta that run across it. The signs for the three-momenta are only to be consistent with three-momentum conservation and do not vary for different time-orderings. For the current case, the three-momenta of the cuts are simply $\vec k_3,\vec k_2$ and $\vec k_1$ in the same fixed order. 

The four-momenta associated to the cuts can then trivially be constructed from the energy in the zeroth component and the three-momentum in the last $d$ components. Thus, to complete the example, we write $L_1^\mu=(-u,\vec k_3), \ L_2^\mu = (u-v,\vec k_2)$ and $L_3^\mu = (v,\vec k_1)$. Notice that these four-momenta satisfy conservation by construction: $L_1^\mu+L_2^\mu+L_3^\mu=0$.

The polynomial in the measure is the Gram determinant of two cut-four-momenta:
\begin{equation}
    \Omega(u,v) = -4 \left|\begin{matrix}
        L_1^2 & L_1 \cdot L_2 \\
        L_1 \cdot L_2 & L_2^2 
    \end{matrix}
    \right|,
    \label{eq:Triangle_Reduced_Measure_4Mom}
\end{equation}
where the dot product is now understood to be with respect to the $\mathbb R^{1,3}$ metric. 

\newpage

\section{Singularities}
\label{sec:Sing}

In this section, we discuss two complementary approaches to identifying the possible singularities of correlator loop integrals. Arguably, the most physical method is based on inspecting for which external kinematics a loop diagram can be interpreted as the scattering of classical particles~\cite{Coleman:1965xm,Mizera:2023tfe}. Correlators generalise scattering amplitudes while probing the same underlying physical processes, and as a result their singularity structures closely resemble one another. That said, there are a few essential differences that require additional care and may even shed new light on certain aspects of the singularity structure of amplitudes, such as second-type singularities. Finally, as the triangle diagram offers a new type of cosmological correlator that has not yet been fully understood, its singularity structure might shed some light into the general singularities of one-loop correlators.

While this physical approach provides valuable intuition, it tends to slightly ``overshoot'' by predicting more singularities than are actually present in the integral. Moreover, the resulting equations quickly become intractable. An alternative approach, which is both precise and readily generalisable, deals directly with the mathematical structure of the integrals themselves. Following the general logic of~\cite{Eden:1966dnq}, we show how to classify all possible ``pinch'' and ``end-point'' singularities directly from the polynomials appearing in the integrals. These singularities admit a clear interpretation on either the physical sheet or on other sheets of the analytic continuation. To validate this approach, we compare the sets of singularities obtained for triangle integrals with those produced by the \texttt{PLD.jl} code~\cite{Fevola:2023fzn}.

Let us also note that the present Landau analysis of time-ordered correlators overlaps with the analysis of wavefunction coefficients carried out in~\cite{Salcedo:2022aal}. The singularities identified in~\cite{Salcedo:2022aal} form a subset of the singularity set derived here.

Finally, we explain how to compute discontinuities associated with an important class of kinematic configurations in cosmology: the partial-energy singularities. To illustrate the procedure, we derive the discontinuities for both partial-energy singularities of a time-ordered diagram of the triangle. We show that, for a specific subset of partial-energy singularities, the naive tree-level expectation continues to hold: even at one loop, a generic correlator in this regime factorises into a lower-point correlator multiplied by an amplitude.

\subsection{Cosmological Landau Equations}

Tree-level amplitudes typically diverge at points in kinematic space where the particles being exchanged go on-shell and propagate classically. For single-exchange diagrams, such points are known as \textit{resonances}. For loop diagrams, amplitudes likewise peak at configurations that admit a classical interpretation. First of all, this requires that the particles propagating in the loop go on-shell. For a one-loop diagram with $V$ vertices and loop four-momenta $Q_{ij}^\mu$ on every internal line connecting the vertices $\nu_i$ and $\nu_j$, this condition reads
\begin{equation}
    \eta_{\mu\nu} Q_{ij}^\mu Q_{ij}^\nu = m_{ij}^2
    \qquad \forall\, ij \in E^* \subseteq E,
    \label{eq:Landau_Onshell}
\end{equation}
where $m_{ij}$ is the mass of the particle propagating along the internal line (edge) $ij$, and $E^*$ denotes a subset of the set of all edges, $E$, of the diagram.

In addition, classicality requires that particles propagate along straight trajectories between interaction points, much like billiard balls~\cite{Coleman:1965xm,Mizera:2023tfe}. Concretely, if each vertex of the Feynman diagram is associated with a definite spacetime position $x_i^\mu$, then the four-momentum on an internal line must be parallel to the vector connecting its endpoints,
\begin{equation}
    \alpha_{ij} Q_{ij}^\mu = (x_i - x_j)^\mu .
\end{equation}
Summing all such difference vectors around the loop yields zero, and therefore
\begin{equation}
    \sum_{ij\in E^*} \alpha_{ij} Q_{ij}^\mu = 0,
    \label{eq:Landau_Circuit}
\end{equation}
which imposes $d+1$ additional constraints on top of the on-shell conditions.

If the number of equations exceeds the number of free loop variables, these classicality conditions impose constraints on the external kinematics. The points in kinematic space where these constraints are satisfied are known as \textit{Landau singularities}. When $E^* = E$ and thus all $\alpha_{ij}$ are non-zero, the corresponding solution is referred to as a \textit{leading singularity}. There also exists a so-called \textit{second-type} singularity, which also arises when $\alpha_{ij}\neq 0$ but is located at infinity in loop-momentum space. If instead some of the $\alpha_{ij}$ vanish, and thus $E^*\subset E$, the singularity is termed \textit{subleading}.

Cosmological loop integrals exhibit a different structure. Although we work in flat space for simplicity, the presence of a fixed time-slice at which correlators are evaluated requires us to integrate time explicitly. As a consequence, there are no four-vectors associated with internal propagators, and particles appearing in cosmological diagrams are automatically \textit{on-shell}. The conditions \eqref{eq:Landau_Onshell} therefore become superfluous in the cosmological setting. Classicality still requires particles to propagate between definite spatial positions, which implies
\begin{equation}
    \sum_{ij\in E^*} (\vec x_i - \vec x_j)
    =
    \sum_{ij\in E^*} \alpha_{ij} \vec q_{ij}
    =
    0,
    \label{eq:CosmoLandau_Vectors}
\end{equation}
together with the requirement that there are no time delays,
\begin{equation}
    \sum_{ij\in E^*} |\vec x_i - \vec x_j|
    =
    \sum_{ij\in E^*} \pm \alpha_{ij} q_{ij}
    =
    0,
    \label{eq:CosmoLandau_Energies}
\end{equation}
where we used $|\vec q_{ij}| = q_{ij}$ for massless scalars, and where the sign of $q_{ij}$ is fixed by the time ordering.

At this stage, there is no fundamental difference between the cosmological Landau equations and their conventional counterparts. However, in contrast to the usual approach, we do not impose energy conservation at each vertex in cosmology. Such relations instead appear as additional constraints that can be imposed on top of \eqref{eq:CosmoLandau_Vectors} and \eqref{eq:CosmoLandau_Energies}. Which partial-energy conservation equations can be imposed is determined by the set of admissible tubings for the corresponding time-ordered graph.

A further difference between the conventional Landau equations and the cosmological version presented here is the explicit $(d+1)$ split between space and time. In contrast to scattering amplitudes, where all time orderings are accounted for simultaneously by allowing the sign of the zeroth component of the four-momentum $Q_i^\mu$ to vary, we instead fix $q_i>0$ in cosmology. As a result, each time-ordered graph comes with its own distinct set of cosmological Landau equations.

\begin{figure}[htb!]
    \centering
    \includegraphics{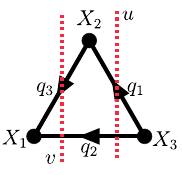}
    \caption{A time-ordered triangle diagram with intermediate energies $u$ and $v$.}
    \label{fig:Triangle_Slicing_2}
\end{figure}

\paragraph{Triangle.}
We illustrate the procedure for solving the cosmological Landau equations by considering the connected time-ordered triangle diagram shown in Fig.~\ref{fig:Triangle_Slicing_2}.

The vector equations \eqref{eq:CosmoLandau_Vectors} are straightforward and read
\begin{equation}
    \alpha_1 \vec q_1 + \alpha_2 \vec q_2 + \alpha_3 \vec q_3 = 0.
    \label{eq:Triangle_Landau_Vectors}
\end{equation}
To determine the relative signs in the energy equation, we observe that the particle emitted from $\nu_3$ must reach vertex $\nu_1$ at the same time as particle $q_1$ propagates from $\nu_3$ to $\nu_2$, followed by $q_3$ from $\nu_2$ to $\nu_1$. This leads to
\begin{equation}
    \Delta t_{32}+\Delta t_{21}-\Delta t_{31}
    =
    \alpha_1 q_1 + \alpha_3 q_3 - \alpha_2 q_2 .
    \label{eq:Triangle_Landau_Energies}
\end{equation}
Finally, since the classical scattering involves all particles circulating in the loop, we require all time slices up to (but not including) the final vertex $\nu_1$ to conserve energy,
\begin{equation}
    u+X_3 = 0,
    \qquad
    v+X_2+X_3 = 0.
    \label{eq:Triangle_Landau_ECons}
\end{equation}

To solve these equations, we first rescale $\alpha_2\rightarrow\alpha_1\alpha_2$ and $\alpha_3 \rightarrow \alpha_1\alpha_3$, and then take scalar products of \eqref{eq:Triangle_Landau_Vectors} with two independent external momenta:
\begin{align}
    \vec k_1 \cdot \vec q_1
    + \alpha_2 \, \vec k_1 \cdot \vec q_2
    + \alpha_3 \, \vec k_1 \cdot \vec q_3
    &= 0 , \\
    \vec k_2 \cdot \vec q_1
    + \alpha_2 \, \vec k_2 \cdot \vec q_2
    + \alpha_3 \, \vec k_2 \cdot \vec q_3
    &= 0 .
\end{align}
These two equations determine $\alpha_2$ and $\alpha_3$. Making use of the freedom to rescale $\alpha_1$, $\alpha_2$, and $\alpha_3$ once more, we can write
\begin{align}
    \alpha_1
    &=
    \left(v^2-k_1^2\right)
    \left(-k_1^2+k_2^2+k_3^2-2 u(u-v)\right), \\
    \alpha_2
    &=
    \left((u-v)^2-k_2^2\right)
    \left(k_1^2-k_2^2+k_3^2-2 u v\right), \\
    \alpha_3
    &=
    \left(u^2-k_3^2\right)
    \left(k_1^2+k_2^2-k_3^2+2 v (u-v)\right).
    \label{eq:Triangle_Landau_AlphaSols}
\end{align}

Having fixed $\alpha_1$, $\alpha_2$, and $\alpha_3$, the energy equation \eqref{eq:Triangle_Landau_Energies} imposes an additional constraint on the energies $q_1$, $q_2$, and $q_3$, or after the change of variables of Sec.~\ref{sec:setup}, on the energies $u$, $v$, and $w$. We use this equation to solve for $w$ and substitute the result into the component of \eqref{eq:Triangle_Landau_Vectors} orthogonal to both $\vec k_1$ and $\vec k_2$,
\begin{equation}
    \alpha_1 q_1^{(T)} + \alpha_2 q_2^{(T)} + \alpha_3 q_3^{(T)} = 0,
\end{equation}
which yields, without imposing any assumptions on the $\alpha_i$,
\begin{equation}
    \left(u^2-k_3^2\right)
    \left(v^2-k_1^2\right)
    \left((u-v)^2-k_2^2\right)
    \sqrt{\Omega(u,v)}
    =
    0.
    \label{eq:Triangle_Landau_Master}
\end{equation}
As anticipated earlier, the Landau equations automatically reproduce the measure of the cosmological Baikov representation, $\Omega(u,v)$.

Solving the Landau equations now amounts to finding all solutions of \eqref{eq:Triangle_Landau_Master}. Since $\Omega(u,v)$ is a quadratic polynomial, \eqref{eq:Triangle_Landau_Master} admits eight solutions for generic momenta $k_1$, $k_2$, and $k_3$ consistent with three-momentum conservation. Six of these solutions correspond to setting one of the $\alpha_i$ to zero. Of course, multiple conditions can be satisfied simultaneously, for instance $u=k_3$ and $v=k_1$.

However, solving \eqref{eq:Triangle_Landau_Master} alone does not impose any constraints on the external kinematics $X_2$ and $X_3$. The only relations available that connect $u$ and $v$ to the vertex energies are the two partial-energy conditions,
\begin{equation}
    u+X_3=0,
    \qquad
    v+X_2+X_3=0,
\end{equation}
which correspond to the red and green tubings in the diagram next to \eqref{eq:R_Triangle_321}.

In summary, the singular loci of the triangle in terms of $X_2$ and $X_3$ are obtained by solving \eqref{eq:Triangle_Landau_Master} together with $u+X_3=0$ and/or $v+X_2+X_3=0$. As we will show in the next section, not \textit{all} solutions to this combined system correspond to genuine singularities of the integral. For example, the condition $u+k_3=0$ can be combined with several other solutions of \eqref{eq:Triangle_Landau_Master}, but only in conjunction with $u+X_3=0$ does it lead to an actual singular point of the integral, namely $X_3-k_3=0$.

We will see that, for the singularity $X_3-k_3=0$, the condition $u+k_3=0$ is satisfied only accidentally. The fundamental origin of this singularity is instead that the hyperplane $u+X_3=0$ becomes tangent to $\Omega(u,v)=0$, and similarly for $v+k_1=0$.

Let us begin by considering the case in which all three coefficients $\alpha_i$ are non-zero. In this situation, the only condition that can be imposed on the loop variables $u$ and $v$ is
\[
\Omega(u,v)=0.
\]
Imposing in addition the energy-conservation conditions \eqref{eq:Triangle_Landau_ECons}, a single constraint on the external kinematics remains,
\[
\Omega(-X_3,-X_2-X_3)=0.
\]
Defining the four-vectors $P_3^\mu=(X_3,\vec k_3)$ and $P_1^\mu=(-X_2-X_3,\vec k_1)$, this locus can be identified with the vanishing of a Gram determinant,
\begin{equation}
    \left|
    \begin{matrix}
        P_1^2 & P_1 \cdot P_3 \\
        P_1 \cdot P_3 & P_3^2
    \end{matrix}
    \right|
    = 0 .
\end{equation}
The vanishing of this Gram determinant is a well-known condition for a \textit{second-type} Landau singularity~\cite{Fairlie2ndType}.

There is, however, an important caveat. For scattering amplitudes, the zeroth component of $P_1$ can be rewritten as $-X_2-X_3=X_1$ by virtue of total energy conservation. For correlators, we are instead left with $-X_2-X_3$, and the variable $X_1$ does not enter the constraint at all. 

More importantly, we find that it is not possible for all $\alpha_i$ to be strictly positive simultaneously, for a simple geometric reason.
The triangle in real space formed by the triplet $(\vec x_1,\vec x_2,\vec x_3)$ saturates a triangle inequality as a consequence of \eqref{eq:Triangle_Landau_Energies}, which collapses the internal vectors $\vec q_1$, $\vec q_2$, and $\vec q_3$ onto a single line. Parallel internal vectors in turn force the external momenta $\vec k_1$, $\vec k_2$, and $\vec k_3$ to form a degenerate (collapsed) triangle. We therefore conclude that the second-type singularity of the correlator is, in direct analogy with the triangle amplitude, not located on the physical sheet for generic external three-momenta.

A subleading singularity solves \eqref{eq:Triangle_Landau_Master} together with one of the conditions $\alpha_i=0$ given in \eqref{eq:Triangle_Landau_AlphaSols}. In \eqref{eq:CosmoLandau_Vectors}, setting one of the $\alpha_i$ to zero while assuming the associated $\vec q_i$ remains finite implies that the corresponding difference of position vectors vanishes. We can therefore interpret $\alpha_i=0$ as a pinching of the corresponding edge in the time-ordered graph.

Suppose we do not involve the $q_1$ edge in the Landau equations. The three-momentum and energy equations then reduce to
\begin{align}
    \alpha_2 \vec q_2 + \alpha_3 \vec q_3 &= 0 , \\
    \alpha_2 q_2 - \alpha_3 q_3 &= 0 .
\end{align}
Which partial-energy conservation equation must be imposed can be inferred most transparently by pinching the edge corresponding to the vanishing $\alpha_i$.
\begin{figure}[htb!]
\centering
\begin{subfigure}[t]{.3\textwidth}
    \centering
    \includegraphics{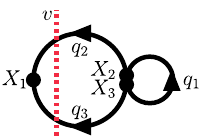}
    \caption{The pinched time-ordered graph obtained by identifying the vertices $\nu_2$ and $\nu_3$.}
    \label{fig:Triangle_Pinch_1}
\end{subfigure}
\hspace{.02\textwidth}
\begin{subfigure}[t]{.3\textwidth}
    \centering
    \includegraphics{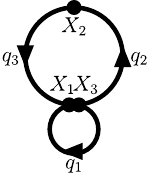}
    \caption{A pinched time-ordered graph that results in a cycle.}
    \label{fig:Triangle_Pinch_2}
\end{subfigure}
\hspace{.02\textwidth}
\begin{subfigure}[t]{.3\textwidth}
    \centering
    \includegraphics{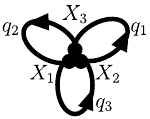}
    \caption{The fully pinched graph.}
    \label{fig:Triangle_Pinch_3}
\end{subfigure}
\caption{Three possible pinches of the triangle diagram.}
\label{fig:Triangle_Pinches}
\end{figure}

In this case, $\alpha_1=0$ and we collapse the internal line $q_1$ as in Fig.~\ref{fig:Triangle_Pinch_1}. This implies the partial-energy condition
\begin{equation}
    v+X_2+X_3 = 0.
\end{equation}
We follow a strategy analogous to the one used previously. By contracting the vector equation with $\vec k_1$ and $\vec k_2$, we obtain two scalar equations. Solving these together with the energy equation $\alpha_2 q_2=\alpha_3 q_3$ fixes two loop energies and determines the pair $(\alpha_2,\alpha_3)$ up to an overall rescaling. The latter admits a particularly simple solution, $\alpha_2=\alpha_3=1$. Finally, the component of the three-vector Landau equation orthogonal to both $\vec k_1$ and $\vec k_2$ yields
\begin{equation}
    \sqrt{v^2-k_1^2} = 0.
\end{equation}
Imposing the positivity condition $v \geq k_1$, the only solution to the singular locus in kinematic space is
\begin{equation}
    X_2+X_3+k_1=0.
\end{equation}
All physical conditions are satisfied: $\alpha_i\geq 0$, $u>k_3$, and $v>k_1$.

In a similar vein, we can solve the $\alpha_3=0$ subleading equations,
\begin{align}
    \alpha_1 \vec q_1+\alpha_2 \vec q_2 &= 0, \\
    \alpha_1 q_1 - \alpha_2 q_2 &= 0, \\
    u + X_3 &= 0,
\end{align}
to obtain the second subleading locus,
\begin{equation}
    X_3 + k_3 = 0.
\end{equation}
This subleading singularity also lies on the physical sheet. The corresponding collapsed diagram is analogous to Fig.~\ref{fig:Triangle_Pinch_1}.

The final single-pinch configuration corresponds to $\alpha_2=0$:
\begin{align}
    \alpha_1 \vec q_1 + \alpha_3 \vec q_3 &= 0, \\
    \alpha_1 q_1 + \alpha_3 q_3 &= 0.
\end{align}
In this case, the singular loci in kinematic space would require imposing both
\begin{equation}
    u+X_3=0,
    \qquad
    v+X_2+X_3=0,
\end{equation}
which leads to
\begin{equation}
    X_2 \pm k_2 = 0.
\end{equation}
However, the two Landau equations involve only positive quantities and appear as sums. For physical configurations, they therefore cannot be satisfied simultaneously.

In terms of pinched diagrams, these subleading Landau equations correspond to the creation of a cycle: the vertex $\nu_2$ would have to occur both before and after $\nu_1$ and $\nu_3$ in time, as illustrated in Fig.~\ref{fig:Triangle_Pinch_2}. We therefore conclude that this subleading singularity does not lie on the physical sheet.

Our procedure above consisted of setting $\alpha_i=0$ for some $i$ from the outset and then solving the Landau equations anew. While this is conceptually clear, it is not the most efficient strategy. A more economical approach is to recycle the general solutions \eqref{eq:Triangle_Landau_AlphaSols} together with \eqref{eq:Triangle_Landau_Master}.

For instance, from \eqref{eq:Triangle_Landau_AlphaSols} one immediately sees that the condition $v^2-k_1^2=0$ solves $\alpha_1=0$ and \eqref{eq:Triangle_Landau_Master} simultaneously, without the need to re-derive the Landau equations for the pinched graph.

It is also important to stress that setting a subset of the $\alpha_i$ to zero should not be viewed as an independent assumption. Ultimately, the only condition we require is the vanishing of \eqref{eq:Triangle_Landau_Master}. The appearance of solutions with one or more $\alpha_i=0$ should therefore be understood as a consequence of solving \eqref{eq:Triangle_Landau_Master}, rather than as a starting point of the analysis. From this perspective, the set of vanishing $\alpha_i$ merely serves as a convenient way to classify the different solutions of \eqref{eq:Triangle_Landau_Master}.

At subsubleading order, the relevant configurations arise when two of the coefficients $\alpha_i$ vanish simultaneously. The two non-trivial possibilities are $\alpha_1=\alpha_2=0$ and $\alpha_2=\alpha_3=0$. Combining these conditions with the two partial-energy constraints generates a total of four candidate singular loci,
\begin{align}
    \alpha_1=\alpha_2=0
    &\quad \Longrightarrow \quad
    \left\{
    \begin{matrix}
        X_3+k_1+k_2=0 , \\
        X_3+k_1-k_2=0 ,
    \end{matrix}
    \right.
    \\
    \alpha_2=\alpha_3=0
    &\quad \Longrightarrow \quad
    \left\{
    \begin{matrix}
        X_2+X_3+k_2+k_3=0 , \\
        X_2+X_3-k_2+k_3=0 .
    \end{matrix}
    \right.
    \label{eq:Triangle_Double_Pinch}
\end{align}

Both sets of $\alpha$-constraints imply that all three vertices of the triangle coincide in coordinate space, as illustrated in Fig.~\ref{fig:Triangle_Pinch_3}. Although the pinched diagrams corresponding to $\alpha_1=\alpha_2=0$ and $\alpha_2=\alpha_3=0$ are topologically identical, they represent distinct physical situations. In the first case, only the particle associated with $\vec q_3$ is required to propagate classically, which is possible only if $\vec q_3=0$. Similarly, imposing $\alpha_2=\alpha_3=0$ forces $\vec q_1=0$.

\begin{figure}[h!]
    \centering
    \includegraphics{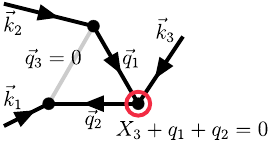}
    \caption{The $\alpha_1=\alpha_2=0$ double-pinched graph. To make the momentum-conservation relations transparent, we do not collapse the vertices in this representation, even though they coincide in coordinate space.}
    \label{fig:Triangle_DoublePinch}
\end{figure}

Setting one of the internal three-momenta $\vec q_i$ to zero effectively reduces the loop diagram to a tree-level configuration. For instance, $\vec q_3=0$ implies $\vec q_1=\vec k_2$ and $\vec q_2=-\vec k_1$. Combining this with the partial-energy constraint $X_3+q_1+q_2=0$ yields the singular locus $X_3+k_1+k_2=0$. The corresponding diagram is shown in Fig.~\ref{fig:Triangle_DoublePinch}. By the same physical reasoning, only the ``all-plus'' solutions in \eqref{eq:Triangle_Double_Pinch} lie on the physical sheet.

The final class of solutions corresponds to configurations in which all three coefficients vanish at the same time, $\alpha_i=0$ for all $i$. Singularities of this type can only be obtained by first solving the system of Landau equations for generic $\alpha_i$ consistent with viewing the set of vanishing $\alpha_i$ as a classification of solutions. The associated pinched diagram has the same topology as the double-pinched case and is depicted in Fig.~\ref{fig:Triangle_Pinch_3}. Solving \eqref{eq:Triangle_Landau_Master} under these conditions leads to six distinct candidate singular loci, which can be organised as follows:
\begin{equation}
\begin{split}
    2 k_1 X_3+k_1^2-k_2^2+k_3^2=0\\
    \pm 2 k_2 X_3+k_1^2-k_2^2-k_3^2=0\\
    2 k_3(X_2+X_3)+k_1^2-k_2^2+k_3^2=0\\
    \pm 2 k_2 (X_2+X_3)+k_1^2+k_2^2-k_3^2=0
\end{split}
\end{equation}

\subsection{General Singularity Analysis for Correlator Loops}

Analytically continuing the loop variables from $\mathbb{R}^{V-1}$ to $\mathbb{C}^{V-1}$ allows us to probe kinematic configurations that are not physically accessible. We now regard the original integration region as a contour embedded in $\mathbb{C}^{V-1}$. As long as no singularities are crossed, this contour may be continuously deformed without changing the value of the integral, by virtue of Cauchy’s theorem. Such contour deformations are essential both for identifying singular loci and for computing discontinuities. Before turning to the cosmological integrals of interest, we briefly sketch the general logic underlying this analysis.

Singular loci arise when contour deformations fail to avoid singular hypersurfaces that move onto the integration contour. These situations fall into two broad classes: so-called \emph{pinch} singularities and \emph{end-point} singularities. Pinch singularities occur when two (or more) singular hypersurfaces collide and trap the contour between them. As a simple illustration, consider the one-dimensional integral
\begin{equation}
    \mathcal I = \int_0^1 \frac{1}{(z-z_1)(z-z_2)}\,\d z ,
\end{equation}
with $z\in\mathbb{C}$. A pinch singularity arises when the points $z_1$ and $z_2$ approach each other from opposite sides of the contour, preventing any deformation that would keep the integration path free of singularities.

End-point singularities, on the other hand, occur when a fixed boundary of the integration domain collides with a singular hypersurface. In the same simple example, an end-point singularity is encountered when $z_1 \to 0$, such that the pole hits the lower integration limit. 

In higher-dimensional complex spaces, identifying singularities by inspection quickly becomes infeasible. An additional complication arises from the fact that, even when singular hypersurfaces collide, the contour might still be able to ``escape'' the pinch by deforming into transverse complex directions. To ensure that the contour is genuinely trapped, one must require that the allowed variations around the would-be singular point are linearly dependent.

Concretely, suppose that the singular hypersurfaces in $\mathbb{C}^{V-1}$ are defined by $f_i(u_1,\dots,u_{V-1})=0$ with $i\in I$. A necessary condition for a singularity of the integral is that
\begin{equation}
    f_i(u_1,\dots,u_{V-1}) = 0, \qquad i\in I^* \subseteq I,
    \label{eq:Landau_Mathy1}
\end{equation}
together with the existence of coefficients $x_i$ such that
\begin{equation}
    \frac{\partial}{\partial u_j}\!\left(\sum_i x_i f_i(u_1,\dots,u_{V-1})\right)=0
    \qquad \forall\, j .
    \label{eq:Landau_Mathy2}
\end{equation}
Solutions to these equations define candidate singular loci of the integral~\cite{Eden:1966dnq}. In the special case where only a single function $f_i$ is involved, these conditions reduce to the requirement that the solution is a critical point of $f_i$.

\subsubsection{Bubble}

For the correlator bubble, it suffices to inspect two integrals instead of three: one of the two connected contributions and the disconnected piece. Moreover, we will see that the analysis of the non-time-ordered integral can be reduced to that of the connected pieces by means of partial fractions. The singularities of the bubble have been studied extensively in the context of hypergeometric functions (see Sec.~\ref{sec:ExplicitResults}). Although the presence of a non-zero dimensional-regularisation parameter renders the integrand multivalued, the bubble still serves as a useful training ground for the analysis of the triangle.

Let us first study a particular permutation of the connected contribution, denoted by $\mathcal I_\bubA$. The other connected piece, together with its corresponding singularity analysis, is obtained by exchanging $X_1 \leftrightarrow X_2$. The rational prefactor of $\mathcal I_\bubA$ contains a total-energy pole that does not depend on the loop momentum. As a consequence, $\mathcal I_\bubA$ inherits this simple pole. Factoring it out allows us to focus on the singularities that are intrinsic to the integral itself:
\begin{equation}
    (X_1+X_2)\, \mathcal I_\bubA
    =
    \int_\Gamma T_1^\epsilon \, T_2^\epsilon \,
    \d\log(L_1) ,
\end{equation}
where $\Gamma$ denotes a so-called \textit{twisted cycle}, which facilitates the analytic continuation of the integration region $u\in(k,\infty)$. The hyperplanes appearing in the integrand are
\begin{equation}
    T_1 = u+k, \qquad T_2=u-k, \qquad L_1=X_1+u .
\end{equation}

By forming pairwise intersections of these hyperplanes and solving for their common zeros, we obtain the possible singular loci in kinematic space. Since we are working in one complex dimension, the linear-dependence conditions are automatically satisfied. The solutions to these equations form a subspace of the combined space of integration and external variables parametrised by $(X_1,X_2,k,u)$. By projecting out $u$ from this subset, we obtain the singular loci in the space of external variables\footnote{Strictly speaking, the singular loci we write down correspond to the irreducible components of the Zariski closure of the projected subset in the space of external variables.}
\begin{center}
\begin{tabular}{|l|l|}
    \hline
    Intersection & Singular locus \\ \hline
    $T_1\cap T_2$ & $k=0$ \\
    $T_1 \cap L_1$ & $X_1-k=0$ \\
    $T_2 \cap L_1$ & $X_1+k=0$ \\ \hline
\end{tabular}
\end{center}

Since $T_2=0$ corresponds to a boundary point of the integration contour, its incidence with $L_1=0$ results in a genuine divergence of the integral. The collision of $T_1=0$ and $T_2=0$ instead produces a singularity only for $\epsilon<1/2$, where the integral transitions from being ultraviolet-divergent to infrared-divergent. Finally, the locus $X_1-k=0$ does not manifest itself as a branch point in the explicit expression for $\mathcal I_\bubA$.

We can treat the disconnected contribution either separately or as a slight modification of the analysis above. Let us first derive its singularities without reference to the connected case. The disconnected integral is
\begin{equation}
    \mathcal I_\bubC
    =
    \int_k^\infty (u^2-k^2)^{\epsilon} 
    \frac{1}{(X_1+u)(X_2+u)} 
    \d u \ .
\end{equation}
Introducing the additional hyperplane $L_2=X_2+u$, the singularity equations are summarised by
\begin{center}
\begin{tabular}{|l|l|}
    \hline
    Intersection & Zariski closure \\ \hline
    $T_1\cap T_2$ & $k=0$ \\
    $T_1 \cap L_1$ & $X_1-k=0$ \\
    $T_2 \cap L_1$ & $X_1+k=0$ \\
    $T_1 \cap L_2$ & $X_2-k=0$ \\
    $T_2 \cap L_2$ & $X_2+k=0$ \\
    $L_1 \cap L_2$ & $X_1-X_2=0$ \\ \hline
\end{tabular}
\end{center}

As a qualitatively new singular locus, we find $X_1-X_2=0$. In contrast to the other points in kinematic space, sitting on $X_1=X_2$ can lead to a genuine pinching of the contour. To see this, let $L_1=0$ lie above the contour by requiring $\Re(-X_1)>k$ and $\Im(X_1)>0$. Similarly, let $\Re(-X_2)>k$ but choose $\Im(X_2)<0$, so that $L_2=0$ lies below the contour. If we now equate the real parts, $\Re(-X_1)=\Re(-X_2)$, and send both imaginary parts to zero, the contour becomes trapped between $L_1=0$ and $L_2=0$. However, since this requires $\Im(X_2)<0$, the corresponding singularity does not occur on the physical sheet.

Alternatively, one may use partial fractions,
\begin{equation*}
    \frac{1}{X_1+u}\frac{1}{X_2+u}
    =
    \frac{1}{X_1-X_2}
    \left(
        \frac{1}{X_2+u}-\frac{1}{X_1+u}
    \right),
\end{equation*}
and directly recycle the analysis of the time-ordered contribution. In this representation, the subtle singularity that arises when $L_1=0$ and $L_2=0$ approach the contour from opposite sides manifests itself as a pole in $X_1-X_2$, with a non-vanishing residue whenever the two integrals are evaluated on different branches.

This strategy of recycling simpler analyses will be particularly useful when dealing with loop integrals that are more complicated than the bubble.

\subsubsection{Triangle}\label{sec:TriangleAllLandau}

To make the discussion concrete, we choose to study a single, fully time-ordered piece:
\begin{equation}
    \mathcal I = \int_\Sigma  \frac{1}{\sqrt{\Omega(u,v)}} \ \frac{1}{(X_3+u)(X_2+X_3+v)} \ \d u \, \d v \ .
    \label{eq:Triangle_Reduced_2}
\end{equation}
The analysis carries straightforwardly over to other time-ordered diagrams as was emphasised in Sec.~\ref{sec:setup}. 
The singular and bounding hypersurfaces are given by
\begin{align}
    \Sigma:\left\{\begin{matrix}
        T_1 = u-k_3 \\
        T_2 = v-k_1 \\
        T_3 = u-v+k_2 \\
        T_4 = -u+v+k_2
    \end{matrix}\right.
    \qquad 
    \text{Integrand}: \left\{\begin{matrix}
        \Omega(u,v) \\
        L_1 = X_3+u \\
        L_2 = X_2+X_3+v
    \end{matrix}\right.
    \label{eq:Triangle_HyperplanesAndMore}
\end{align}

To analyse the singularity structure of \eqref{eq:Triangle_Reduced_2}, it is convenient to extend the integration space to $\mathbb{CP}^2$. The original integration domain $\Sigma$ has a boundary on the union of the hyperplanes defined by $T_1,T_2,T_3$, and $T_4$. However, its definition is rigid: \emph{a priori} there is no notion of contour deformations within $\Sigma$ itself. This limitation can be overcome by viewing $\Sigma$ as a particular representative of a relative homology class, which we denote by $[\Gamma]$.

Concretely, let us define the singular locus of the integrand as
\begin{equation*}
    S = \left\{ (u,v)\in \mathbb{CP}^2 \ \big| \ L_1=L_2=\Omega=0 \right\},
\end{equation*}
and the union of all boundary hypersurfaces as
\begin{equation*}
    A = \left\{ (u,v)\in \mathbb{CP}^2 \ \big| \ T_1=T_2=T_3=T_4=0 \right\}.
\end{equation*}
The integration domain may then be regarded as a representative of a class in the relative homology group
\begin{equation*}
    H_2(\mathbb{CP}^2\setminus S, A).
\end{equation*}
Within this framework, contour deformations correspond to choosing different representatives of the same relative homology class. In particular, the original domain $\Sigma$ corresponds to one such representative.

This construction admits a simple geometric interpretation. The representative $\Gamma$ may be visualised as a two-dimensional membrane in $\mathbb{CP}^2$ whose boundary lies on four planes. These planes intersect pairwise at isolated points. While the one-dimensional boundary of $\Gamma$ may be freely deformed along each plane, it is constrained to pass through the intersection points of pairs of planes. As we will see, this geometric picture plays an essential role in understanding the origin of singularities.

Both $S$ and $A$ are complex codimension-one submanifolds of $\mathbb{CP}^2$. The triangle integral can therefore be interpreted as a pairing between a holomorphic top-form—since $\Omega$, $L_1$, and $L_2$ depend only on $u$ and $v$, and not on their complex conjugates—representing a class $\omega\in H^2(\mathbb{CP}^2\setminus S, A)$, and a homology class $[\Gamma]\in H_2(\mathbb{CP}^2\setminus S, A)$. In close analogy with the one-dimensional case, such pairings are invariant under contour deformations, i.e.\ under changes of the representative $\Gamma$ within the same homology class.

\begin{wrapfigure}{r}{6cm}
\centering
\includegraphics{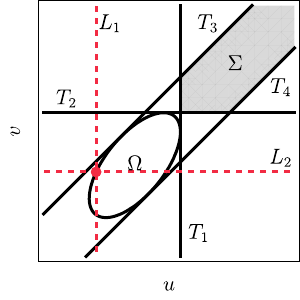}
\caption{The red filled circle indicates a possible location of the intersection point of $L_1=0$ and $L_2=0$ (red dashed lines) on $\Omega=0$. We have fixed $k_1=1.15$, $k_2=0.93$, $k_3=1.00$, and $X_2=-0.70$.}
\label{fig:Sings_2type}
\end{wrapfigure}
Within this setup, the Landau equations \eqref{eq:Landau_Mathy1} and \eqref{eq:Landau_Mathy2} provide a systematic way of identifying potential singularities by diagnosing when the integration cycle becomes pinched. As a first illustration, consider the system
\begin{align}
    &L_1 = 0, \quad L_2 = 0, \quad \Omega(u,v)=0, \\
    &\frac{\partial}{\partial u}\!\left(L_1 + x_1 L_2 + x_2 \Omega \right)=0, \quad 
    \frac{\partial}{\partial v}\!\left(L_1 + x_1 L_2 + x_2 \Omega \right)=0.\nonumber
\end{align}
Its solution corresponds to the familiar second-type singularity,
\begin{equation}
    \Omega(-X_3,-X_2-X_3)=0.
\end{equation}
However, the intersection point of $L_1$, $L_2$, and $\Omega$ does not, in general, lie on the boundary $A$, nor inside the region bounded by $T_i\geq0$ as illustrated in Fig.~\ref{fig:Sings_2type}. One therefore does not expect this hypersurface in $(X_2,X_3)$-space to give rise to a singularity on the physical sheet.

Below, we scrutinise every singularity implied by \eqref{eq:Landau_Mathy1} and \eqref{eq:Landau_Mathy2} accompanied by plots that illustrate the corresponding positions of $L_1=0$ and $L_2=0$ with respect to $\Omega=0$ and $T_i=0$. In the plots, we have fixed $k_1=1.15$, $k_2=0.93$, and $k_3=1.00$, and omitted the labels $L_1$ and $L_2$. The analytic results are summarised in Table~\ref{tab:Triangle_Singularities} where we classify the singularities according to how the corresponding hypersurfaces intersect the boundary of the integration cycle $\Gamma$,
\begin{equation*}
    A_+ = \left\{ (u,v) \in \mathbb{CP}^2 \ \big| \ 
    \bigcup_i \left\{ T_i=0 \ \text{and} \ T_{j\neq i}\geq0 \right\} \right\}.
\end{equation*}

\emph{A posteriori}, we observe that the codimension of the intersection of hypersurfaces, as a subvariety of $A_+$, appears to distinguish between two types of singular behaviour. In cases where the intersection has codimension zero, the integral develops logarithmic divergences, whereas for codimension one, the first derivative of the imaginary part is discontinuous, while the real part remains finite and continuous. At present, we do not have a complete understanding of the origin of this distinction.

The singularities implied by \eqref{eq:Landau_Mathy1} and \eqref{eq:Landau_Mathy2} can then be organised into the following classes:

\begin{itemize}
    \item $X_3+k_3=0$ and $X_2+X_3+k_1=0$: 
\end{itemize} 
\begin{minipage}{6cm}
    \centering
    \includegraphics{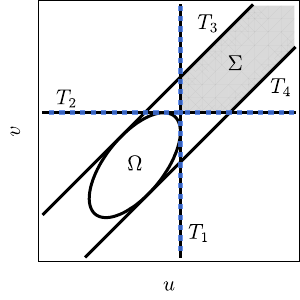}
\end{minipage}
\begin{minipage}{\linewidth-6cm}
\vspace{-.5cm}
These two kinematic loci solve the Landau equations by aligning $L_1$ with $T_1$ and $L_2$ with $T_2$, respectively. 
These configurations are indicated by the blue dashed lines in the plot, and a concrete example of $L_1=T_1$ is shown in Fig.~\ref{fig:Triangle_Hyperplane_Spec}. In complex four-dimensional space, one may think of the planes $L_1$ and $T_1$ (or $L_2$ and $T_2$) as overlapping, thereby obstructing any contour deformation. Since the singular surface in this case defines a codimension-zero submanifold of $A_+$, we identify $X_3+k_3=0$ and $X_2+X_3+k_1=0$ as physical singularities.
\end{minipage}

\begin{itemize}
    \item $X_3+k_1-k_2=0$ and $X_2+X_3-k_2+k_3=0$:
\end{itemize}
\begin{minipage}{6cm}
    \centering
    \includegraphics{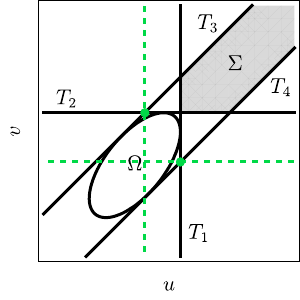}
\end{minipage}
\begin{minipage}{\linewidth-6cm}
\vspace{-.6cm}
The triplets $(L_1,T_2,T_3)$ and $(L_2,T_1,T_4)$ intersect at isolated points in $\mathbb{CP}^2$, respectively, as illustrated by the green dots in the plot on the left. The corresponding positions of $L_1=0$ and $L_2=0$ are indicated by green dashed lines. However, these intersection points lie neither on $A_+$ nor on $\Sigma$, and we therefore do not expect the corresponding loci to appear on the physical sheet. In the plot on the left, these cases correspond to the green markers located on $T_1=0$ and $T_2=0$ that do not touch $\Sigma$.
\end{minipage}

\newpage
\begin{itemize}
    \item $X_3+k_1+k_2=0$ and $X_2+X_3+k_2+k_3=0$:
\end{itemize}
\begin{minipage}{6cm}
\vspace{-1cm}
    \centering
    \includegraphics{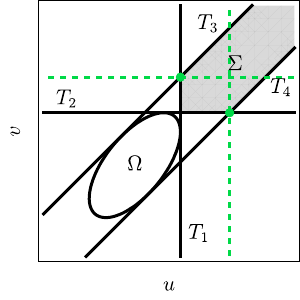}
\end{minipage}
\begin{minipage}{\linewidth-6cm}
By contrast, the corresponding intersections $(L_1,T_2,T_4)$ and $(L_2,T_1,T_3)$ lie on $A_+$. We therefore expect these loci to manifest as branch points of the integral. When all parameters are taken to be real and the contour is chosen as $\Gamma=\Sigma\subset\mathbb R^2$, the hypersurfaces $L_1$ and $L_2$ intersect $\Sigma$ transversely, rendering the integral ill defined. In the present setup, however, the contour may be deformed around $L_1$ or $L_2$ everywhere except at their intersections with the boundary $A_+$. At these points, $\Gamma$ is forced to pass through $T_1\cap T_3$ and $T_2\cap T_4$ and cannot be deformed. These intersection loci have complex codimension one as subvarieties of $A_+$, as indicated by the corresponding markers in the plot on the left.
\end{minipage}

\begin{itemize}
    \item $X_2+k_2=0$ and $X_2-k_2=0$:
\end{itemize}
\begin{minipage}{6cm}
\vspace{-2.6cm}
    \centering
    \includegraphics{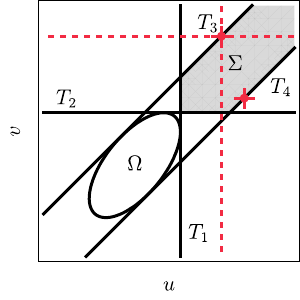}
\end{minipage}
\begin{minipage}{\linewidth-6cm}
The integral may diverge if the contour becomes trapped on $A_+$ by a pinching between $L_1$ and $L_2$. The intersections $(L_1,L_2,T_3)$ and $(L_1,L_2,T_4)$ correspond to the kinematic loci $X_2+k_2=0$ and $X_2-k_2=0$, respectively. For such a pinching to occur, $L_1$ and $L_2$ must approach $T_3$ or $T_4$ from opposite sides in complex space. This can be implemented by assigning opposite $i\varepsilon$ prescriptions, $L_1\to L_1+i\varepsilon$ and $L_2\to L_2-i\varepsilon$. Geometrically, one may picture the planes $L_1$ and $L_2$ piercing the planes $T_3$ and $T_4$ at points, with the boundary of $\Gamma$ tracing a one-dimensional curve on these planes. The boundary can then become trapped by a pinching of the two intersection points, in direct analogy with the bubble singularity at $X_1-X_2=0$. Since this mechanism requires at least one vertex energy to have $\varepsilon<0$, we conclude that the singularities $X_2\pm k_2=0$ do not lie on the physical sheet.
\end{minipage}

\begin{itemize}
    \item $X_3^2-k_3^2=0$ and $(X_2+X_3)^2-k_1^2=0$:
\end{itemize}
\begin{minipage}{6cm}
    \centering
    \includegraphics{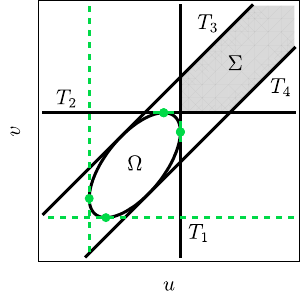}
\end{minipage}
\begin{minipage}{\linewidth-6cm}
\vspace{-.55cm}
Finally, there are two additional potential singularities arising from the pinching of $\Omega$ with either $L_1$ or $L_2$. The intersection of $L_1$ with $\Omega$ defines the locus $X_3^2-k_3^2=0$, while the intersection of $L_2$ with $\Omega$ implies $(X_2+X_3)^2-k_1^2=0$. For physical values of the $k_i$, the corresponding intersection points lie outside $A_+$. This is illustrated in the plot on the left by the vertically dashed green markers on the $\Omega$ curve. We therefore conclude that these pinches do not occur on the physical sheet.
\end{minipage}

Throughout this discussion, the external three-momenta $k_i$ are held at a generic point inside the momentum-conserving region. Allowing the $k_i$ to vary can degenerate the integration region; for instance, setting $k_2=0$ collapses $\Sigma$ to a line. More generally, on the boundary of the momentum-conserving region, where one of the relations $\pm k_1\pm k_2\pm k_3=0$ holds, the polynomial $\Omega(u,v)$ becomes singular and the variety $\Omega(u,v)=0$ degenerates into a single hyperplane.

The results of the singularity analysis are summarised in Table~\ref{tab:Triangle_Singularities}. Only singularities supported on $A_+$ that do not require a pinching of $L_1$ and $L_2$ are expected to appear on the physical sheet. The horizontal line separates physical-sheet singularities from the rest. We further expect the first two loci to lead to divergences, while the third and fourth appear only as regular branch points. 
\begin{table}[h!]
\centering
\begin{tabular}{|l|l|l|}
    \hline
    Hypersurfaces Integral & Kinematic Space & Type \\ \hline
    $L_1,T_1$ & $X_3+k_3=0$ & $\text{codim}=0$ on $A_+$ \\
    $L_2,T_2$ & $X_2+X_3+k_1=0$ & $\text{codim}=0$ on $A_+$ \\
    $L_1,T_2,T_4$ & $X_3+k_1+k_2=0$ & $\text{codim}=1$ on $A_+$ \\
    $L_2,T_1,T_3$ & $X_2+X_3+k_2+k_3=0$ & $\text{codim}=1$ on $A_+$\\ \hline
    $L_1,L_2,T_3$ & $X_2+k_2=0$ & $\text{codim}=1$ on $A_+$ \\
    $L_1,L_2,T_4$ & $X_2-k_2=0$ & $\text{codim}=1$ on $A_+$ \\
    $L_1,T_2,T_3$ & $X_3+k_1-k_2=0$ & $\text{codim}=1$ outside $A_+$ \\
    $L_2,T_1,T_4$ & $X_2+X_3-k_2+k_3=0$ & $\text{codim}=1$ outside $A_+$ \\
    $\Omega, L_1$ & $X_3^2-k_3^2=0$ & $\text{codim}=1$ outside $A_+$ \\
    $\Omega, L_2$ & $(X_2+X_3)^2-k_1^2=0$ & $\text{codim}=1$ outside $A_+$ \\
    $\Omega,L_1,L_2$ & $\Omega(-X_3,-X_2-X_3)=0$ & $\text{codim}=1$ outside $A_+$ \\ \hline
\end{tabular}
\caption{The intersections of singular and bounding hypersurfaces and their corresponding kinematic configurations that lead to singularities of the integral. We also highlight the codimension of the subvariety on $A_+$ defined by the intersection. }
\label{tab:Triangle_Singularities}
\end{table}

\subsubsection{Validation with \texttt{PLD.jl}}

We would like to validate the list of candidate singularities derived above using the \texttt{PLD.jl} algorithm~\cite{Fevola:2023fzn,Fevola:2023kaw}. 
In essence, this algorithm provides a consistent and computationally tractable way of determining the analogue of a \emph{principal $A$-determinant} for specialised parameter choices. 
Principal $A$-determinants characterise the singular locus of GKZ systems: $D$-modules whose solution spaces are described by \emph{generalised Euler integrals}. 
These encompass a broad class of integrals of the form
\begin{equation}
    I_\Gamma (z_{1},\ldots,z_{m}) 
    =
    \int_\Gamma 
    f_1^{s_1} \cdots f_l^{s_l} \,
    \alpha_1^{\nu_1} \cdots \alpha_{n}^{\nu_n} \,
    \frac{\d \alpha_1}{\alpha_1} \cdots \frac{\d \alpha_n}{\alpha_n},
    \label{eq:EulerInt}
\end{equation}
where $s=(s_1,\ldots,s_l)\in\mathbb C^l$ and $\nu=(\nu_1,\ldots,\nu_n)\in\mathbb C^n$. 
We use a shorthand notation $z_i=(z_{i,1},\ldots,z_{i,s})$, so that $(z_1,\ldots,z_m)\in\mathbb C^{m\times s}$. 
The integration variables $\alpha=(\alpha_1,\ldots,\alpha_n)$ take values on the \emph{very affine variety}
\begin{equation}
    X
    =
    \left\{
    \alpha \in (\mathbb C^\ast)^n
    \;\middle|\;
    \alpha_1 \cdots \alpha_n \, f_1 \cdots f_l \neq 0
    \right\},
\end{equation}
and the integral is taken over a so-called \emph{twisted cycle} $\Gamma$ that keeps track of the multivaluedness of the integrand~\cite{Matsubara-Heo:2023ylc}. 
The functions $f=(f_1,\ldots,f_l)$ are Laurent polynomials with coefficients depending on the parameters $z$,
\begin{equation}
    f_i = \sum_{u\in A_i} z_{i,u}\,\alpha^u,
\end{equation}
where each $A_i$ is a finite set of integer vectors and we use the shorthand $\alpha^u=\alpha_1^{u_1}\cdots\alpha_n^{u_n}$.

The principal $A$-determinant is a polynomial in the coefficients $z_{i,u}$ whose zero locus captures all singular points of the integral in parameter space. 
Equivalently, it detects loci where the topology of the variety $X$ changes, i.e. when its signed Euler characteristic drops.
For a generic generalised Euler integral of the form~\eqref{eq:EulerInt}, every monomial in each $f_i$ carries an independent parameter $z_{i,u}$. 
In physical applications, however, the parameters are almost never generic: they satisfy algebraic relations among themselves, thereby restricting the system to a subvariety of $\mathbb C^{m\times s}$. 
On such a subvariety, the Euler characteristic of $X$ is typically already reduced relative to the fully generic case. 
Singularities of the \emph{specialised} integral then arise when the Euler characteristic drops further as the parameters are varied. 
This locus is captured by the \emph{Euler discriminant}.

\begin{figure}[h!]
\centering
\begin{subfigure}[b]{.45\textwidth}
    \centering
    \includegraphics[width=.8\textwidth]{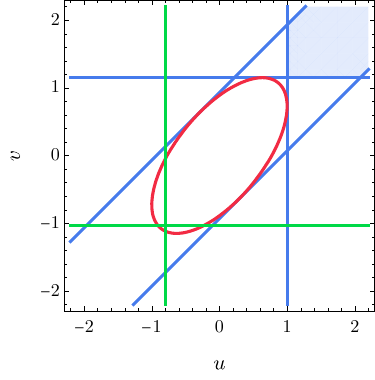}
    \caption{The arrangement for generic parameters. Using \texttt{HomotopyContinuation.jl}, we find $\chi=15$.}
    \label{fig:Triangle_Hyperplane_Gen}
\end{subfigure}
\hfill
\begin{subfigure}[b]{.45\textwidth}
    \centering
    \includegraphics[width=.8\textwidth]{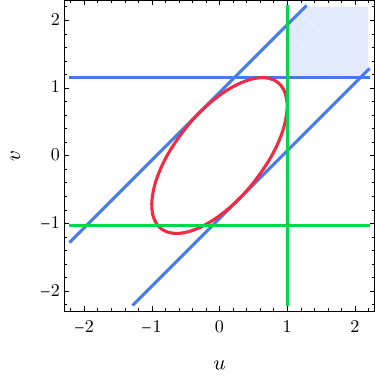}
    \caption{The arrangement with $X_3+k_3=0$, where the Euler characteristic drops to $\chi=10$.}
    \label{fig:Triangle_Hyperplane_Spec}
\end{subfigure}
\caption{Hyperplane arrangements of the singular and bounding hyperplanes augmented by the degree-two hypersurface defined by $\Omega=0$.}
\label{fig:Triangle_Hyperplane_Arrangements}
\end{figure}

It is often instructive to visualise this phenomenon by plotting the relevant curves in $(u,v)\in\mathbb R^2$ and identifying configurations where the number of bounded regions drops from its maximal value, as illustrated in Fig.~\ref{fig:Triangle_Hyperplane_Arrangements}. 
We emphasise that counting bounded regions in a hyperplane arrangement augmented by a quadratic polynomial does \emph{not} in general compute the Euler characteristic of the corresponding variety~\cite{Reinke2024}. 
Nevertheless, in the two configurations shown in Figs.~\ref{fig:Triangle_Hyperplane_Gen} and~\ref{fig:Triangle_Hyperplane_Spec}, this counting turns out to be correct, as verified by an explicit computation of critical points.

To apply the Euler-discriminant machinery to the triangle integral, we first rewrite it in a form compatible with~\eqref{eq:EulerInt}. 
This naturally arises once the dimensional-regularisation parameter is taken to be non-zero, $\epsilon\neq0$. 
Up to overall proportionality factors, the reduced integral then reads
\begin{equation}
    \mathcal I
    \propto
    \int
    \frac{(T_1 T_2 T_3 T_4)^{\epsilon}}{\Omega^{1/2+\epsilon}}
    \frac{\d u'}{u'}
    \frac{\d v'}{v'},
\end{equation}
where $u'=u-X_3$ and $v'=v-X_v$, and the functions $T_1,T_2,T_3,T_4$ and $\Omega$ are defined as in~\eqref{eq:Triangle_HyperplanesAndMore} after the shifts $u=u'+X_3$ and $v=v'+X_v$.

By Schwinger-parameterising $\Omega$ and each $T_i$ individually, rescaling all Schwinger parameters but one, and subsequently performing the exponential integral over the overall scale, we obtain an integral over the Schwinger parameters $y_i$ together with the original variables $u'$ and $v'$. 
The resulting polynomial appearing in the integrand takes the Cayley form
\begin{equation}
    F(u',v')
    =
    \Omega(u',v')
    +
    \sum_i y_i\, T_i(u',v').
\end{equation}
We then set
\begin{verbatim}
pars = (X_1,X_2,X_3,k_1,k_2,k_3),
vars = (u',v',y_1,y_2,y_3,y_4),
\end{verbatim}
and run the command
\begin{verbatim}
getSpecializedPAD(F,pars,vars)
\end{verbatim}
on the Cayley configuration.
The output recovers all singularities that we found above but reveals the following additional class of candidate singularities:
\begin{center}
\begin{tabular}{|l|l|}
    \hline
    Potential singularity & Landau equations${}^\ast$ \\
    \hline
    $2 k_1 X_3+k_1^2-k_2^2+k_3^2=0$ & $\Omega,\,L_1,\,T_2$ \\
    $2 k_2 X_3+k_1^2-k_2^2-k_3^2=0$ & $\Omega,\,L_1,\,T_3$ \\
    $2 k_2 X_3-k_1^2+k_2^2+k_3^2=0$ & $\Omega,\,L_1,\,T_4$ \\
    $2 k_2 (X_2+X_3)+k_1^2+k_2^2-k_3^2=0$ & $\Omega,\,L_2,\,T_3$ \\
    $2 k_2 (X_2+X_3)-k_1^2-k_2^2+k_3^2=0$ & $\Omega,\,L_2,\,T_4$ \\
    $2 k_3 (X_2+X_3)+k_1^2-k_2^2+k_3^2=0$ & $\Omega,\,L_2,\,T_1$ \\
    \hline
\end{tabular}
\end{center}

We have marked the corresponding Landau equations with an asterisk to emphasise that, when solving the linear-dependence conditions, the coefficient of either $L_1$ or $L_2$ must vanish. 
This explains why we did not find them using the conventional Landau equations.
Equivalently, these loci correspond to configurations where $\Omega$ meets one of the boundary hyperplanes $T_i$ and already induces a pinch without involving $L_1$ or $L_2$. 
Such \emph{permanent pinches}, involving only $\Omega$ and a single $T_i$, do not constrain the external kinematics: their Zariski closure is therefore empty. 
Since these pinches lie outside $A_+$, they do not generate UV or IR divergences requiring regularisation.

\subsection{Partial Energy Singularities and Discontinuities}

The partial energy singularities of a given time-ordered graph can be inferred from its tubings by reattaching the external legs to the corresponding Feynman graphs. We slightly modify the rules for constructing the rational function that serves as the loop integrand. Using the same tubings, and still summing over all vertex energies, we now additionally include the magnitude of the total three-momentum flowing into or out of a given tubing. The graphical rules explained here can be seen as a generalisation of~\cite{Bhowmick:2025mxh} to higher-site graphs.

\begin{figure}[htb!]
    \centering
    \includegraphics{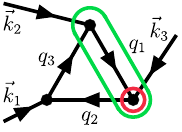}
    \caption{A time-ordered triangle graph with the consistent tubings drawn on it. The external legs are explicitly included, with arrows indicating the direction of the three-momentum $\vec k_i$ flowing into the graph.}
    \label{fig:Triangle_Momenta_Tubing}
\end{figure}

As an example, consider our preferred time-ordering of the triangle diagram and assign the tubings according to the rules explained in Sec.~\ref{sec:Rational}. This yields the configuration shown in Fig.~\ref{fig:Triangle_Momenta_Tubing}.

In contrast to the linear map used to construct the rational function, we now introduce a map that associates to each subgraph $H_j$ the quantity
\begin{equation}
    S(H_j) = \sum_{\nu_i\in H_j} X_i 
    + \left| \sum_{\nu_i\in H_j}\vec k_i \right|.
\end{equation}
Applied to the example in Fig.~\ref{fig:Triangle_Momenta_Tubing}, this prescription yields
\begin{equation}
    S(\textcolor{red2}{H_1}) = X_3 + k_3, \qquad 
    S(\textcolor{green2}{H_2}) = X_2 + X_3 + \lvert \vec k_2 + \vec k_3 \rvert 
    = X_2 + X_3 + k_1 .
\end{equation}

We are interested in studying the monodromies of integrals associated with these partial energy singularities. The cosmological Baikov representation is particularly well suited for this purpose, as it naturally accommodates multi-dimensional contour deformations and the application of Leray’s multivariate residue calculus. We refer to~\cite{Hannesdottir:2022xki,Britto:2024mna,Abreu:2017ptx} for comprehensive reviews and applications in a physics context. In the present treatment, however, we will bypass many formal aspects and instead employ a more heuristic version of this well-developed framework.

The bubble diagram with a hard cutoff $\Lambda$\footnote{For the sake of the example, we avoid addressing the multi-valuedness of the measure.} provides a convenient setting to illustrate the basic ideas underlying the computation of monodromies. Consider again the connected time-ordered integral
\begin{equation}
    \mathcal I_{\bubA}
    = \frac{1}{X_1 + X_2}
    \int_k^\Lambda
    \frac{\d u}{X_1 + u} \ .
\end{equation}
We identify $X_1 + k = 0$ as a partial energy singularity. To analytically continue $X_1$ away from its physical values, we introduce an $i\varepsilon$ prescription and take $X_1 \to X_1 + i\varepsilon$, in accordance with the prescription that renders the time integrals convergent. 

\begin{figure}[h!]
\centering
\begin{subfigure}[t]{.45\textwidth}
    \centering
    \includegraphics{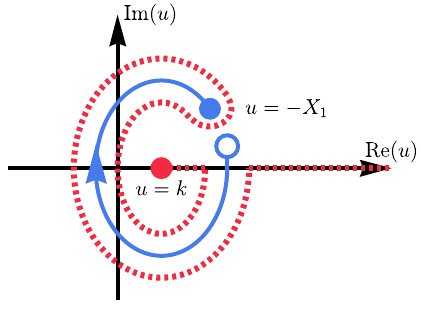}
    \caption{Deforming the original integration path $(k,\infty)$ to compute the monodromy of $X_1$ about $-k$.}
    \label{fig:Bubble_Contour_Def}
\end{subfigure}
\hspace{.05\textwidth}
\begin{subfigure}[t]{.45\textwidth}
    \centering
    \includegraphics{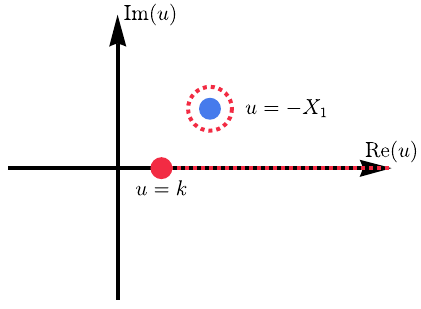}
    \caption{The tube in the left panel can be collapsed, yielding the original contour together with a small circle around the pole at $u = -X_1$.}
    \label{fig:Bubble_Contour_Res}
\end{subfigure}
\caption{An illustration of the contour deformation due to revolving $X_1$ about $k$.}
\label{fig:Bubble_Contour_}
\end{figure}

Now continue $X_1$ to negative values such that the singular hyperplane $u = -X_1$ moves onto the integration contour. The endpoint $u = k$, with $k>0$, remains fixed. Place the singularity at $u = -X_1$ with $-X_1 > k$ and rotate it clockwise about the point $u = k$. Deforming the contour accordingly leads to the configuration shown in Fig.~\ref{fig:Bubble_Contour_Def}. By Cauchy’s theorem, this deformation is equivalent to the original integral plus the contribution from the residue at $u = -X_1$, as illustrated in Fig.~\ref{fig:Bubble_Contour_Res}. 

As a result, the monodromy of $\mathcal I_{\bubA}$ about the partial energy singularity $X_1 + k = 0$ is given by a single residue,
\begin{equation}
    (\mathcal M_{X_1 + k} - \mathbb I)\, \mathcal I_{\bubA}
    = \frac{1}{X_1 + X_2} \, 2\pi i \,
    \operatorname{Res}_{u = -X_1} \frac{1}{X_1 + u}
    = \frac{2\pi i}{X_1 + X_2}.
\end{equation}
We will interpret this result within a factorisation framework in the following sections.

\paragraph{Triangle.}
The reduced integral for the triangle diagram,
\begin{equation}
    \mathcal I = \int_\Sigma \frac{1}{\sqrt{\Omega(u,v)}} \,
    \frac{1}{(X_3+u)(X_v+v)} \, \d u \, \d v \ ,
    \label{eq:Triangle_Reduced_3}
\end{equation}
exhibits two distinct partial energy channels, located at $X_3+k_3=0$ and $X_v+k_1=0$.
As before, the integrand contains a multivalued function. However, for the purpose of studying partial energy singularities, we may safely assume that the associated branch cuts can be avoided, since the hypersurface $\Omega(u,v)=0$ lies sufficiently far away from the integration contour.

Let $\gamma$ denote a relative cycle in the homology group $H_1(S,A)$. The differential form $\omega$ on $\mathbb {CP}^2\setminus S$ is required to be closed and to vanish upon restriction to $S$. Leray’s multivariate residue formula for a simple pole then reads
\begin{equation}
    \int_{\delta \gamma} \omega = 2\pi i \int_\gamma \operatorname{res}(\omega),
\end{equation}
where $\delta$ denotes Leray’s coboundary map. Intuitively, this map replaces the cycle on $S$ by a thin tubular neighbourhood fibred by circles. If the variety $S$ is defined by an algebraic equation $s(u,v)=0$, then for a point $y\in S$ the residue is given by
\begin{equation}
    \operatorname{res}(\omega)
    = \left. \frac{s_y \, \omega}{\d s_y} \right|_{S}.
\end{equation}

In our case, the integration region $\Gamma$ defines a relative cycle in $H_2(\mathbb {CP}^2\setminus S,A)$, while $\gamma$ is the curve connecting the two intersection points of, for example, the hyperplane $X_3+u=0$ with $\Gamma$ (namely $L_1\cap T_2$ and $L_1\cap T_3$). The integrand in \eqref{eq:Triangle_Reduced_3} is naturally interpreted as a holomorphic two-form $\omega$, with the factor $\d s_y/s_y$ already extracted. Holomorphicity ensures that the form is closed, and implies that it vanishes trivially on the boundary of $\Gamma$.

Using the same reasoning as in the one-dimensional case, the monodromy is obtained by deforming the contour such that the difference between the original and deformed integration regions forms a tube wrapping the singular hypersurface, $\Gamma' - \Gamma = \delta \gamma$.

An alternative approach is to regard the integral over $\Gamma$ as an iterated integral. One may then apply the one-dimensional contour deformation to the inner integral and subsequently perform the outer integration. 

For simplicity, we focus on the discontinuity associated with the monodromy around a branch point, which already captures the essential information about the corresponding singularity. 

Integrals of the form \eqref{eq:Triangle_Reduced_3} diverge near a singular hypersurface $G(X_3,X_v)=0$ as
\begin{equation}
    \mathcal I \sim \log(G) \times \text{finite}.
\end{equation}
Approaching $G=0$ and encircling it by analytically continuing $X_3$ and/or $X_v$, one finds
\begin{equation}
    (\mathcal M_{X_3+k_3=0}-\mathbb I)\, \mathcal I
    \sim 2\pi i \times \text{finite}\big|_{G=0}.
\end{equation}
Computing the monodromy first and subsequently taking the limit $G\to0$ thus directly extracts the finite coefficient multiplying the logarithmic divergence.

\begin{wrapfigure}{r}{7cm}
    \centering
    \includegraphics{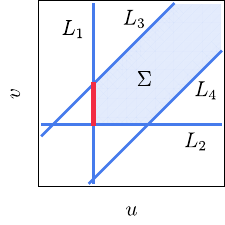}
    \caption{Taking the residue at $u=-X_3$ and imposing the limit $X_3\to -k_3$ restricts the original integration region ($\Sigma$, light blue) to a boundary (red line segment).}
    \label{fig:Triangle_Region_Mon}
\end{wrapfigure}

We illustrate this procedure by computing the coefficient of the partial energy singularity at $X_3+k_3=0$:
\begin{align}
    \lim_{X_3+k_3\rightarrow0}
    \Big[(\mathcal M_{X_3+k_3=0}-\mathbb I)\, \mathcal I\Big]
    &= \lim_{X_3+k_3\rightarrow0}
    \int_{\mathcal P(\Sigma)}
    \operatorname{Res}_{X_3+u}
    \left\{
    \frac{1}{\sqrt{\Omega(u,v)}} \,
    \frac{1}{(X_3+u)(X_v+v)}
    \right\} \, \d v \nonumber\\
    &= \int_{k_1}^{k_2+k_3}
    \frac{2\pi i}{\big(-k_3 (k_3-2v)-k_1^2+k_2^2\big)\,(X_v+v)} \, \d v .
\end{align}
Here, $\mathcal P(\Sigma)$ denotes the restriction of the original integration region induced by taking the residue and subsequently enforcing the limit. Suppressing the imaginary directions of $\mathbb{CP}^2$, this restriction can be visualised as in Fig.~\ref{fig:Triangle_Region_Mon}.

Since the resulting cut integral no longer involves a square root of an irreducible polynomial, it can be evaluated in terms of logarithms and dilogarithms, yielding
\begin{equation}
    \lim_{X_3+k_3\rightarrow0}
    (\mathcal M_{X_3+k_3}-\mathbb I)\, \mathcal I
    =
    \frac{2\pi i
    \left(
    \log (k_1+X_v)
    -\log (k_2+k_3+X_v)
    -\log \!\left(\frac{k_1+k_2-k_3}{k_1+k_2+k_3}\right)
    \right)}
    {2k_3 X_v +k_1^2-k_2^2+k_3^2},
    \label{eq:Triangle_PartialEnergy_Disc1}
\end{equation}
provided three-momentum conservation holds, implying the positivity of the triangle inequalities involving $k_1$, $k_2$, and $k_3$.

Encircling the branch point at $X_3+k_3=0$ moves the integral onto a different sheet, where additional singularities appear. In particular, one finds new singular loci at $X_v+k_2+k_3=0$, $k_1+k_2\pm k_3=0$, and $2k_3X_v+k_3^2-k_2^2+k_1^2=0$. These singularities were already anticipated by the Landau analysis of Sec.~\ref{sec:TriangleAllLandau} for $X_v=X_2+X_3$, although the comparison is not entirely direct, since we are probing a limit of a different branch of $\mathcal I$.

Proceeding analogously, the monodromy of $X_v$ about $k_1$ yields
\begin{equation}
    \lim_{X_v+k_1\rightarrow0}(\mathcal M_{X_v+k_1}-\mathbb I)\, \mathcal I
    =
    \frac{2\pi i
    \left(
    \log (k_3+X_3)
    -\log (k_1+k_2+X_3)
    +\log \!\left(\frac{k_1+k_2+k_3}{-k_1+k_2+k_3}\right)
    \right)}
    {2 k_1 X_3+k_1^2-k_2^2+k_3^2},
    \label{eq:Triangle_PartialEnergy_Disc2}
\end{equation}
with corresponding second-sheet singularities at $X_3+k_1+k_2=0$, $-k_1+k_2+k_3=0$, and $2 k_1 X_3+k_1^2-k_2^2+k_3^2=0$.

To obtain the discontinuities of the full triangle correlator in an efficient manner, we proceed as follows. First, we enumerate all integrands that develop a pole at, for example, $u=-X_3$. Diagrammatically, this amounts to selecting all time-ordered graphs admitting the corresponding tubing; in this case, the tubing associated with the single vertex $\nu_3$. After taking the residue, one is left with an integral over either $v$ or $w$. If the remaining variable is $v$, one substitutes the appropriate expression for $X_v$ (for our preferred time-ordering, $X_v=X_2+X_3$) into \eqref{eq:Triangle_PartialEnergy_Disc1}. If instead the integration variable is $w$, one identifies the permutation that maps the corresponding rational integrand to that of \eqref{eq:Triangle_PartialEnergy_Disc1}, and applies the same permutation to the final result. Although somewhat tedious, this procedure is entirely straightforward and reduces to summing over permutations.

\subsubsection{Factorisation Theorem}

The leading behaviour of a correlator in a partial energy limit involving $V-1$ vertices obeys a particularly simple relation. We first illustrate this relation for the triangle diagram and discuss the underlying physical picture, after which we formulate the general statement.

The partial energy of interest corresponds to the red tubing shown in Fig.~\ref{fig:Triangle_Momenta_Tubing},
\begin{equation}
    X_2+X_3+k_1 .
\end{equation}
By computing all relevant partial energy discontinuities, summing them, and evaluating the result on the locus $X_2+X_3+k_1=0$, we find
\begin{multline}
    \lim_{X_2+X_3+k_1\rightarrow 0}\mathcal I_{\text{triangle}}
    =
    \log(X_2+X_3+k_1)\times
    \left(\frac{1}{X_1+k_1}+\frac{1}{X_1-k_1}\right)
    \times \\[0.2cm]
    \frac{
    \log (k_2-X_2)-\log (-k_1+k_3-X_2)
    +\log (k_2+X_2)-\log (k_1+k_3+X_2)
    }
    {2 k_1 X_2+k_1^2+k_2^2-k_3^2}
    + \text{finite}.
\end{multline}
The term proportional to $1/(X_1-k_1)$ arises from summing all fully connected contributions, whereas the $1/(X_1+k_1)$ term originates entirely from disconnected ones. This separation suggests that the two sectors may be treated independently.

Introducing the four-vectors $P_2=(X_2,\vec k_2)$ and $P_3=(X_3,\vec k_3)$, the expression can be rewritten as
\begin{align}
    \lim_{X_2+X_3+k_1\rightarrow 0}\mathcal I_{\text{triangle}}
    &=
    \log(X_2+X_3+k_1)
    \times
    \left(\frac{1}{X_1+k_1}+\frac{1}{X_1-k_1}\right)
    \times
    \frac{\log(-P_2^2)-\log(-P_3^2)}{P_3^2-P_2^2} \nonumber \\
    &=
    \big(\psi(X_1,k_1)+\psi(X_1,-k_1)\big)
    \times
    \lim_{P_1^2\rightarrow0}
    \mathcal A_\triangle(P_1,P_2,P_3).
\end{align}
This relation is an incarnation of the well-known factorisation property of wavefunction coefficients and correlators at tree level, now appearing at one loop.

At tree level, partial energy limits correspond to sending the associated set of vertices to the infinite past. The diagram splits into two subgraphs separated by an infinite time interval. The part close to the boundary produces a shifted correlator, while the distant part reduces to a flat-space scattering amplitude.

\begin{figure}[h!]
    \centering
    \includegraphics{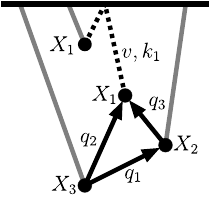}
    \caption{Schematic representation of a time-ordered triangle graph in the partial energy factorisation limit.}
    \label{fig:Triangle_PartialEnergy}
\end{figure}

Although the same structure appears here almost verbatim, it is not immediately obvious why the shifted correlator depends on $k_1$, nor why the amplitude must be evaluated in the soft limit $P_1^2\to0$. Both features admit a natural explanation in terms of time separation. In the partial energy limit, the time-ordered triangle graph splits into two parts: the vertex $\nu_1$, which remains close to the boundary, and the vertices $\nu_2$ and $\nu_3$, which are dragged to the infinite past. The two internal lines connecting $\nu_1$ to the pair $(\nu_2,\nu_3)$ are stretched into a thin tube, as depicted in Fig.~\ref{fig:Triangle_PartialEnergy}. This tube may be interpreted as an effective single particle carrying energy $v$ and three-momentum $k_1$.

From the Landau analysis, we know that for this singularity $v=-k_1$, implying $v^2-k_1^2=0$. Physically, if this effective particle propagates for an arbitrarily long time, it must be on shell. The loop correlator therefore factorises into two pieces connected by an on-shell internal state.

Further insight can be gained by manipulating the residue integrand before performing the remaining integrations. Focusing again on the Landau singularity $X_2+X_3+k_1=0$, the residue of the connected rational integrand takes the simple form
\begin{equation}
    \operatorname{Res}_{q_2+q_3+X_2+X_3=0}(R_{\triangle,C})
    =
    -\frac{1}{
    4 q_2 q_3
    (X_1+X_2+X_3)
    (-q_1+q_3+X_2)
    (q_1+q_3+X_2)
    }.
\end{equation}
We introduce the four-vectors
\begin{equation}
    L_1=(X_2+q_3,\vec q_1),\quad
    L_2=(X_1-q_3,\vec q_2),\quad
    L_3=(q_3,\vec q_3),
\end{equation}
and continue to work in the original momentum representation. The three-momentum integral can be uplifted to four dimensions by inserting a delta function,
\begin{equation}
    \int\frac{\d^3 q_3}{(2\pi)^3}
    =
    \int\frac{\d^4 L_3}{(2\pi)^4}\,
    2\pi\,\delta(s-q_3).
\end{equation}
Using the distributional identity
\begin{equation}
    \frac{1}{2y}\delta(x-y)\Theta(x)
    =
    \delta(x^2-y^2)\Theta(x),
\end{equation}
the factor $1/(4q_2q_3)$ can be rewritten in terms of Cutkosky delta functions, yielding
\begin{equation}
    \frac{-2\pi}{X_1+X_2+X_3}
    \int \frac{\d^4 L_3}{(2\pi)^4}
    \delta(L_3^2)\Theta(L_3^0)
    \delta(L_2^2)\Theta(L_2^0)
    \frac{1}{L_1^2}.
\end{equation}
Projecting the partial energy singularity onto external kinematics via $X_2+X_3+k_1=0$, the connected contribution becomes
\begin{equation}
    \frac{-2\pi}{X_1-k_1}
    \,\mathcal{C}ut_{2,3}
    \big\{
    \mathcal A_\triangle(P_1^2,P_2^2,P_3^2)
    \big\},
\end{equation}
where $\mathcal{C}ut_{2,3}$ places internal particles $2$ and $3$ on shell. This reproduces the result obtained from the explicit evaluation of the discontinuity.

\begin{figure}[h!]
    \centering
    \includegraphics{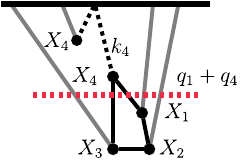}
    \caption{The same intuition applies to the partial energy singularity $X_1+X_2+X_3+k_4=0$ of the box diagram. The energies of the stretched edges sum to $q_1+q_4$.}
    \label{fig:Box_PartialEnergy}
\end{figure}

The relation between partial energy singularities of correlators involving $V-1$ vertices and cut amplitudes generalises straightforwardly to the box diagram. For the configuration in Fig.~\ref{fig:Box_PartialEnergy}, corresponding to dragging the vertices $X_1$, $X_2$, and $X_3$ to the infinite past, the relevant residue is
\begin{equation}
    \underset{q_1+q_4+X_1+X_2+X_3=0}{\operatorname{Res}}
    (R^{\Box}_C)
    =
    \frac{1}{
    4 q_1 q_4
    (X_1+X_2+X_3+X_4)
    \big((q_1+X_1)^2-q_2^2\big)
    \big((q_1+X_1+X_2)^2-q_3^2\big)
    }.
\end{equation}
Defining
\begin{equation}
    L_1=(q_1,\vec q_1),\quad
    L_2=(q_1+X_1,\vec q_2),\quad
    L_3=(q_1+X_1+X_2,\vec q_3),\quad
    L_4=(q_1+X_1+X_2+X_3,\vec q_4),
\end{equation}
and repeating the steps used for the triangle, we obtain
\begin{equation}
    \frac{2\pi}{X_1+X_2+X_3+X_4}
    \int \frac{\d^4 L_1}{(2\pi)^4}
    \delta(L_1^2)\Theta(L_1^0)
    \delta(L_4^2)\Theta(L_4^0)
    \frac{1}{L_2^2 L_3^2}
    =
    \frac{2\pi}{X_1+X_2+X_3+X_4}
    \mathcal{C}ut_{1,4}\{\mathcal A_\Box\}.
\end{equation}

We therefore expect the following relation to hold for a generic one-loop $V$-gon:
\begin{equation}
    \lim_{\sum_{i\neq j} X_i+k_j \rightarrow 0}
    \bigg(\sum_i X_i\bigg)\,
    \mathcal I_{\text{one-loop}}
    =
    (-1)^V
    \,
    \mathcal{C}ut_{e_{j-1,j},e_{j,j+1}}
    \big\{
    \mathcal A_{\text{one-loop}}
    \big\},
\end{equation}
where $j$ labels a single vertex and $e_{i-1,i}$ and $e_{i,i+1}$ denote the edges adjacent to it. We have explicitly verified this conjecture up to and including the pentagon, beyond which the expressions rapidly become intractable.

\newpage
\section{Explicit Results}\label{sec:ExplicitResults}

In this section, we explicitly compute the bubble and triangle loop integrals. 
In both cases, it suffices to obtain a closed-form expression for a single reduced integral; the full result is then constructed by summing over permutations of the external kinematics.

We first observe that the reduced bubble integral in $d$ dimensions can be written as a hypergeometric integral. 
Its relative simplicity allows for many generalisations (see, e.g.,~\cite{Cacciatori:2023tzp,Cacciatori:2024zrv,Westerdijk:2025ywh,Chowdhury:2023arc,Qin:2024gtr,Zhang:2025nzd}). 
Moving from the bubble to the triangle, however, the complexity increases considerably—a pattern that appears to persist as the number of sites grows. 
We present an elegant method that tames this complexity for the triangle, reducing the evaluation of the corresponding integral to the computation of residues of iterated integrals.

Throughout, we express the loop contributions to the correlators in terms of the principal branches of the logarithm and dilogarithm. 
Their branch cuts are chosen according to the standard convention,
\begin{equation}
    \log(z), \quad \mathrm{Li}_2(1-z), \qquad -\pi < \arg(z) \leq \pi .
\end{equation}

\subsection{Warm-up: Bubble}

We begin with the bubble, which we treat as a warm-up due to its relative simplicity. 
The bubble correlator depends on three kinematic variables: the vertex energies $X_1$ and $X_2$, and the exchanged momentum $k$. 
In this case, the rational function depends on a single integration variable, $u$, which allows the original two-dimensional integral to factorise into a numerical prefactor multiplied by a function expressible as a sum of reduced integrals. In particular, we evaluate
\begin{align}
    \mathcal I_{\text{bubble}}^{\text{(red.)}}(X,k)
    &= \int_k^\infty \d u\, (u^2-k^2)^{\epsilon} 
    \frac{1}{X+u} \nonumber \\
    &= -\pi \csc (2\pi\epsilon)\left(X^2-k^2\right)^{\epsilon}
    + \frac{k^{2\epsilon+1}\Gamma\!\left(-\epsilon-\frac{1}{2}\right)\Gamma(\epsilon+1)\,
    {}_2F_1\!\left(\frac{1}{2},1;\epsilon+\frac{3}{2};\frac{k^2}{X^2}\right)}
    {2\sqrt{\pi}\,X},
\end{align}
where $X=X_1$ or $X_2$.

Dimensional regularisation with $d=3+2\epsilon$ renders the integral finite. 
To extract the divergent and finite contributions, we perform a Laurent expansion in $\epsilon$, obtaining
\begin{equation}
    \mathcal I_{\text{bubble}}^{\text{(red.)}}(X,k)
    = -\frac{1}{2\epsilon} - \log(X+k) + \mathcal O(\epsilon).
\end{equation}

The construction of the full bubble correlator from the reduced integral is dictated by the structure of the rational function,
\begin{equation}
    R_{C,\text{bubble}}
    = \frac{1}{X_1+X_2}\!\left(\frac{1}{X_1+u}+\frac{1}{X_2+u}\right)
    - \frac{1}{X_1-X_2}\!\left(\frac{1}{X_1+u}-\frac{1}{X_2+u}\right).
\end{equation}
The total integral is therefore obtained by replacing each $u$-dependent piece with the corresponding reduced integral,
\begin{equation}
    \mathcal I_{\text{bubble}}
    = \frac{\mathcal I_{\text{bubble}}^{\text{(red.)}}(X_1,k)
    + \mathcal I_{\text{bubble}}^{\text{(red.)}}(X_2,k)}{X_1+X_2}
    -
    \frac{\mathcal I_{\text{bubble}}^{\text{(red.)}}(X_1,k)
    - \mathcal I_{\text{bubble}}^{\text{(red.)}}(X_2,k)}{X_1-X_2}.
\end{equation}

Here we have suppressed an overall numerical prefactor and constants sitting on the total-energy pole. 
Substituting the Laurent-expanded reduced integrals, we arrive at the explicit result
\begin{equation}
    \mathcal I_{\text{bubble}}
    = -\frac{\frac{1}{\epsilon}+\log(X_1+k)+\log(X_2+k)}{X_1+X_2}
    + \frac{\log(X_1+k)-\log(X_2+k)}{X_1-X_2}.
\end{equation}

\subsection{Triangle}
As discussed in Section~\ref{sec:setup}, the partially fractioned integrand for the triangle consists of twelve terms, related by permutations of the external parameters: the vertex energies $X_1,X_2,X_3$ and the exchanged momenta $k_1,k_2,k_3$. 
After modding out these symmetries, we are left with two distinct time-orderings: a fully time-ordered and a partially time-ordered contribution. 

By further applying partial fractions, these two rational functions can be related to a single function depending on two integration variables, $u$ and $v$. 
As a result, the entire calculation reduces to the evaluation of a single two-dimensional integral, defined as
\begin{equation}
    \mathcal I_{\text{triangle}}^{\text{(reduced)}} 
    = \int_\Sigma \frac{1}{\sqrt{\Omega(u,v)}} 
    \frac{1}{(X_3+u)(X_v+v)} \, \d u \, \d v \, ,
    \label{eq:Triangle_Reduced_4}
\end{equation}
with all ingredients having been presented in previous sections. 

In this section we give the detailed derivation of a closed-form expression for \eqref{eq:Triangle_Reduced_4}. A companion Mathematica notebook, available at \cite{Notebook}, records the full formula explicitly and gives a brief outline of how it is assembled from the objects introduced below.

\subsubsection{The Measure as a Contour Integral}

As a first step, we rewrite the square root appearing in the measure in terms of an integral representation of the three-dimensional Laplacian Green’s function~\cite{Whittaker1902},
\begin{equation}
    \oint \frac{\d \theta}{x\cos\theta+y\sin\theta+i z}
    = \pm \frac{2\pi i}{\sqrt{x^2+y^2+z^2}} .
\end{equation}
Here the integral is understood to run from $0$ to $2\pi$.

This representation is conveniently evaluated using the residue theorem by analytically continuing the angular variable $\theta$ to the complex plane. The $\theta$ integral can be embedded by regarding $\theta$ as the phase of a complex parameter,
\begin{equation}
    w=\rho\, e^{i\theta},
\end{equation}
which allows us to rewrite the integral as
\begin{equation}
    \oint \frac{\d \theta}{x\cos\theta+y\sin\theta+i z}
    = \oint \frac{2\,\d w}
    {i\left(w^2+1\right)x+\left(w^2-1\right)y-2 w z}.
\end{equation}
In terms of the complex variable $w$, the integrand has two simple poles,
\begin{equation}
    w_\pm=\frac{z\pm\sqrt{x^2+y^2+z^2}}{x+i y}.
\end{equation}
Which residue contributes depends on the position of these poles relative to the unit circle. If $|w_-|<1$, the result carries the minus sign, whereas $|w_+|<1$ corresponds to the plus sign. For $x,y>0$, this condition simplifies to $z<0$ for the $-$ sign and $z>0$ for the $+$ sign.

There is no obstruction to Wick rotating $z$ to obtain the Lorentzian version,
\begin{equation}
    \oint \frac{\d \theta}{x\cos\theta+y\sin\theta+z}
    = \pm \frac{2\pi i}{\sqrt{x^2+y^2-z^2}}.
\end{equation}
When the square root is real, \emph{i.e.} for $x^2+y^2>z^2$, the sign choice is ambiguous. This ambiguity is resolved by giving $z$ a small imaginary part, $z\to z+i\delta$. The absolute values of the poles then become
\begin{align}
    |w_+|^2
    &=\left|\frac{\sqrt{x^2+y^2-z^2}-i z}{x+i y}\right|^2
    =1+\frac{2\delta}{\sqrt{x^2+y^2-z^2}}>1, \\
    |w_-|^2
    &=\left|\frac{-\sqrt{x^2+y^2-z^2}-i z}{x+i y}\right|^2
    =1-\frac{2\delta}{\sqrt{x^2+y^2-z^2}}<1,
    \label{eq:TriangleComp_wPoles}
\end{align}
where the sign of $\delta$ fixes the location of the poles relative to the unit circle. As a convention, we take $\delta>0$. One may still choose which residue to evaluate by reversing the contour orientation: traversing the unit circle anti-clockwise picks up the residue at $w_-$, while going clockwise selects the pole at $w_+$. Although this discussion may seem overly pedantic, the location of the poles $w_\pm$ will play a crucial role in what follows. In either case, the result carries an overall minus sign,
\begin{equation}
    \oint \frac{\d \theta}{x\cos\theta+y\sin\theta+z+i\delta}
    = -\frac{2\pi i}{\sqrt{x^2+y^2-(z+i\delta)^2}},
    \qquad x,y,z\in\mathbb{R},\ \delta>0.
\end{equation}

With this representation in hand, the measure of the reduced triangle loop can be written in terms of the Klein-Gordon Green’s function as
\begin{equation}
    \frac{1}{\sqrt{\Omega(u,v)}}
    = \frac{1}{-2\pi i}
    \oint \frac{\d \theta}{x\cos\theta+y\sin\theta+z},
    \qquad
    \left\{
    \begin{matrix}
        x=\frac{\delta_2}{k_1}v-2k_1 u,\\
        y=-\frac{\sqrt{p}}{k_1}v,\\
        z=\sqrt{p},
    \end{matrix}
    \right.
    \label{eq:TriangleComp_Measure_Rep}
\end{equation}
where we introduced the shorthand combinations
\begin{align}
    \delta_1 &= -k_1^2+k_2^2+k_3^2, \qquad
    \delta_2 = k_1^2-k_2^2+k_3^2, \qquad
    \delta_3 = k_1^2+k_2^2-k_3^2, \\
    p &= 16\,\mathcal V(\vec k_1,\vec k_2)^2
    =(k_1+k_2+k_3)(k_1+k_2-k_3)(k_1-k_2+k_3)(-k_1+k_2+k_3).
\end{align}
For completeness, we record the explicit expressions for the poles,
\begin{equation}
    w_\pm
    = -\frac{i k_1 \sqrt{\Omega(-X_3,-X_v)^2-2 i\delta\sqrt{p}}
    \pm (\sqrt{p}+i\delta)k_1}
    {2k_1^2 X_3-(i\sqrt{p}+\delta_2)X_v}.
    \label{eq:wpm_Poles}
\end{equation}

\subsubsection{The Dilogarithmic Integrals}

We have now learnt how to represent the square root of the irreducible quadratic polynomial in $u$ and $v$ as an additional contour integral. As a result, the integrand becomes linear in both $u$ and $v$:
\begin{equation}
    \mathcal I
    = \frac{1}{-2\pi i}
    \oint \d\theta
    \int_\Sigma
    \frac{1}{Uu+Vv+W}
    \frac{1}{X_3+u}
    \frac{1}{X_v+v}
    \d u\, \d v \, ,
    \label{eq:Triangle_ThetaRep}
\end{equation}
where we have defined
\begin{equation}
    U = -2 k_1 \cos(\theta), \qquad
    V = \frac{\delta_2 \cos(\theta)-\sqrt{p}\sin(\theta)}{k_1}, \qquad
    W = \sqrt{p}.
\end{equation}

This representation opens up a range of techniques for manipulating the integrals and ultimately obtaining explicit expressions in terms of dilogarithms. As a first step, we use partial fractions to decompose the rational function into simpler pieces,
\begin{multline}
    \frac{1}{Uu+Vv+W}\frac{1}{X_3+u}\frac{1}{X_v+v}= \frac{1}{-U X_3 - V X_v + W} \times \\
    \left(
    -\frac{U}{(v+X_v)(uU+vV+W)}
    -\frac{V}{(u+X_3)(uU+vV+W)}
    +\frac{1}{(u+X_3)(v+X_v)}
    \right).
    \label{eq:Triangle_PartialFracs}
\end{multline}
We refer to the three rational functions appearing inside the brackets as the \emph{single-$u$}, \emph{single-$v$}, and \emph{factorised} terms, respectively. We begin with the factorised term, which is by far the simplest, and then discuss the single-$u$ contribution in detail. For the single-$v$ term, we only state the final result, as the procedure closely parallels that of the single-$u$ case.

\paragraph{Factorised term.}
The name of this contribution reflects the fact that the $\theta$ contour integral and the $u,v$ integrals factorise completely.

To evaluate the contour integral, we first reinstate the explicit expressions for $U$, $V$, and $W$:
\begin{equation}
    \frac{1}{-2\pi i}
    \oint \frac{\d \theta}{-U X_3 - V X_v + W}
    =
    \frac{1}{-2\pi i}
    \oint \frac{\d \theta}{x\cos\theta + y\sin\theta + z},
    \qquad
    \left\{
    \begin{matrix}
        x = -\frac{\delta_2}{k_1} X_v + 2k_1 X_3,\\
        y = \frac{\sqrt{p}}{k_1} X_v,\\
        z = \sqrt{p},
    \end{matrix}
    \right.
\end{equation}
where $x$, $y$, and $z$ coincide with the expressions in \eqref{eq:TriangleComp_Measure_Rep} evaluated at $u=-X_3$ and $v=-X_v$. By construction, this contour integral is therefore nothing but the measure evaluated at these values,
\begin{equation}
    \frac{1}{-2\pi i}
    \oint \frac{\d\theta}{-U(\theta) X_3 - V(\theta) X_v + W}
    = \frac{1}{\sqrt{\Omega(-X_3,-X_v)}}.
    \label{eq:Fact_Denominator}
\end{equation}

\begin{figure}[h!]
    \centering
    \includegraphics[width=0.35\linewidth]{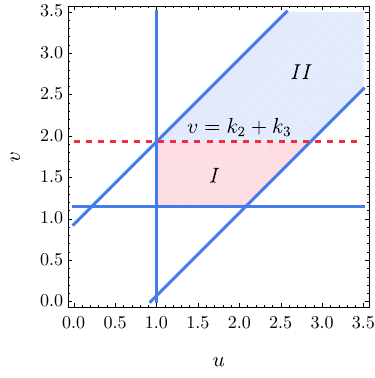}
    \caption{Subdivision of the integration domain into regions I and II.}
    \label{fig:Triangle_Region_Split_V}
\end{figure}

We now perform the $u$ and $v$ integrals explicitly. To this end, we split the integration domain $\Sigma$ into two regions, labelled I and II, as shown in Fig.~\ref{fig:Triangle_Region_Split_V}. Region~I is finite and bounded by the hyperplanes $u=k_3$, $u-v=k_2$, $v=k_1$, and $v=k_2+k_3$. Region~II is unbounded in the $u-v$ direction and can be parametrised by taking $u$ to range from $v-k_2$ to $v+k_2$, with $v$ running from $k_2+k_3$ to $\infty$.

As a result, the integral naturally organises itself into a sum of iterated integrals,
\begin{equation}
    \int_\Sigma \frac{\d u}{X_3+u}\frac{\d v}{X_v+v}
    =
    \int_{k_1}^{k_2+k_3} \frac{\d v}{X_v+v}
    \int_{k_3}^{v+k_2} \frac{\d u}{X_3+u}
    +
    \int_{k_2+k_3}^\infty \frac{\d v}{X_v+v}
    \int_{v-k_2}^{v+k_2} \frac{\d u}{X_3+u}.
\end{equation}

We regard these integrals as functions of the complex variables $X_v$ and $X_3$, while keeping $k_1$, $k_2$, and $k_3$ fixed and real, consistent with three-momentum conservation. Restricting momentarily to $X_3,X_v\in\mathbb{R}$, the integrals require a deformation of the integration contour whenever the external kinematics satisfy $X_3\leq -k_3$ and/or $X_v\leq -k_1$. Equivalently, we introduce an $i\varepsilon$ prescription,
\begin{equation}
    X_3 \to X_3 + i\varepsilon, \qquad
    X_v \to X_v + i\varepsilon,
\end{equation}
so that the hyperplanes $X_3+u=0$ and $X_v+v=0$ never intersect the integration contour. In accordance with the sign required for convergence of the time integrals, we choose $\varepsilon>0$.

It is worth noting that the $i\delta$ and $i\varepsilon$ prescriptions can interfere, and their relative magnitudes must be ordered appropriately. We postpone a detailed discussion of this point to later.

Evaluating the two iterated integrals explicitly and combining the results, we arrive at
\begin{multline}
    F_{\mathrm{F}}(X_3,X_v) \equiv \int_\Sigma \frac{\d u}{X_3+u}\frac{\d v}{X_v+v}
    =
    -\text{Li}_2\!\left(\frac{k_1+k_2+X_3}{k_2-X_v+X_3}\right)
    -\text{Li}_2\!\left(\frac{k_3+X_3}{k_2+k_3+X_v}\right)
    \\
    +\log(X_v-X_3-k_2)\log(k_1+k_2+X_3)
    -\log(k_1+k_2+X_3)\log(k_1+X_v)
    \\
    -\log(k_2+X_v-X_3)\log(k_3+X_3) 
    +\log(k_2+X_v-X_3)\log(k_2+k_3+X_v) 
    \\
    +\log(k_3+X_3)\log(k_1+X_v)
    -\frac{1}{2}\log^2(k_2+k_3+X_v)
    -\frac{1}{2}\log^2(X_v-X_3-k_2)
    .
    \label{eq:Triangle_FactorisedExplicit}
\end{multline}

The hypersurfaces $X_v-X_3\pm k_2=0$ correspond to spurious singularities. While individual logarithms develop branch points there, the full function $F_{\mathrm{F}}$ is continuous across these loci. Consequently, the cancellation of the $i\varepsilon$ prescriptions in differences such as $X_3-X_v$ has no effect on the analytic structure: no branch cuts are associated with letters containing $X_3-X_v$.

Alternatively, the factorised contribution can be characterised through a system of differential equations,
\begin{equation}
    \d\,\mathcal I_{\mathrm{fact.}} = A \cdot \mathcal I_{\mathrm{fact.}},
\end{equation}
where the connection one-form is given by the strictly upper-triangular matrix
\begin{equation}
    A = \d
    \begin{pmatrix}
        0 &
        \log\!\left(\frac{X_3+k_3}{X_3-X_v+k_2}\right) &
        \log\!\left(\frac{X_3-X_v-k_2}{X_3-X_v+k_2}\right) &
        \log\!\left(\frac{X_v+k_1}{X_v-X_3+k_2}\right) &
        \log\!\left(\frac{X_v+k_1}{X_3-X_v+k_2}\right) & 0 \\
        0 & 0 & 0 & 0 & 0 &
        \log\!\left(\frac{X_v+k_2+k_3}{X_v+k_1}\right) \\
        0 & 0 & 0 & 0 & 0 &
        \log\!\left(\frac{1}{X_v+k_2+k_3}\right) \\
        0 & 0 & 0 & 0 & 0 &
        \log\!\left(\frac{1}{X_3+k_3}\right) \\
        0 & 0 & 0 & 0 & 0 &
        \log\!\left(X_3+k_1+k_2\right) \\
        0 & 0 & 0 & 0 & 0 & 0
    \end{pmatrix}.
\end{equation}

The upper-triangular form of $A$ makes it nilpotent (of index three), implying that $\mathcal I_{\mathrm{fact.}}$ is an iterated integral — a fact already apparent from its integral representation. An invariant way to write the solution is
\begin{equation}
    \mathcal I_{\mathrm{fact.}}
    = \mathbb{P}\exp\!\left(\int_\gamma A\right)\cdot\mathcal I_{\mathrm{fact.}}^{(0)},
\end{equation}
where $\gamma:[0,1]\to M$ is a path in the kinematic manifold $M$ parametrised by $(X_3,X_v)$. Since $A$ is exact and satisfies $A\wedge A=0$, the connection is flat. Consequently, $\mathcal I_{\mathrm{fact.}}$ is a homotopy-invariant, multivalued function on $M$ (see e.g.~\cite{Brown2009}).

\paragraph{Single $u$ term.} 
We will follow the same recipe for the $u$ and $v$ integrals of the single-$u$ term, indicated by the boxed expression below:
\begin{multline}
    \frac{1}{-2\pi i}\oint \d \theta \,
    \frac{-1}{-U X_3-V X_v+W}F_u(\theta)
    \\=
    \frac{1}{-2\pi i}\oint \d \theta \,
    \frac{-1}{-U X_3-V X_v+W}
    \boxed{
    \int_\Sigma
    \frac{1}{(v+X_v)\left(u+\frac{V}{U}v+\frac{W}{U}\right)}
    \d u \, \d v
    } .
\end{multline}
Before the explicit calculation, let us roughly analyse where the possible singularities are.

First, we split the integration region according to Fig.~\ref{fig:Triangle_Region_Split_V}:
\begin{multline}
    F_u(\theta)
    =
    \int_\Sigma
    \frac{\d v \, \d u}{(v+X_v)\left(u+\frac{V}{U}v+\frac{W}{U}\right)}
    =
    \int_{k_1}^{k_2+k_3} \frac{\d v}{v+X_v}\int_{k_3}^{v+k_2} \frac{\d u}{u+\frac{V}{U}v+\frac{W}{U}}
    \\+
    \int_{k_2+k_3}^\infty \frac{\d v}{v+X_v}\int_{v-k_2}^{v+k_2}\frac{\d u}{u+\frac{V}{U}v+\frac{W}{U}} \, .
    \label{eq:F_SingleU}
\end{multline}
The iterated integrals can be expressed straightforwardly in terms of logarithms and dilogarithms. However, the resulting formulae are lengthy, and there are many ways of combining the logarithms and dilogarithms that lead to different analytic structures as functions of $\theta$. Therefore, before performing the integrals explicitly, we derive the locations of the branch points and branch cuts of this integral. These will serve as an organising principle and as a consistency check.

First, let us analyse the branch points of the analytic continuation of $F_u(\theta)$. We obtain this function by setting $\theta$ to be the phase of a complex number $w=\rho\, e^{i\theta}$. We regard the remaining parameters, $X_3$, $X_v$, and the $k_i$, as constants. We use a bar to indicate the analytic continuation from the unit circle to the complex plane, $\bar f(w)=f(-i\log w)$. If we allow the contour $\Sigma$ to be deformed, the integral is uniquely represented by a multivalued function on $\mathbb C$.

Differential equations provide a useful way to make contact with this multivalued function. As for the factorised term, we first write a system of first-order differential equations,
\begin{equation}
    \d \,\mathcal I_U = B \cdot \mathcal I_U,
\end{equation}
where the exact connection one-form reads 
\begin{equation}
    B= \d \begin{pmatrix}
        B_1^{\intercal} & 0 \\
        0_{6\times6} & B_2
    \end{pmatrix}
\end{equation}
with
\begin{equation}
    B_1 = \begin{pmatrix}
        0 \\ 
        \log \left(\frac{k_3 U-V X_v+W}{k_2 U-(U+V) X_v+W}\right) \\
        \log \left(-\frac{V \left(k_1+X_v\right)}{k_3 U-V X_v+W}\right) \\
        \log \left(\frac{(U+V) \left(k_1+X_v\right)}{k_2 U+(U+V) X_v-W}\right) \\
        \log \left(\frac{-k_2 U-(U+V) X_v+W}{k_2 U-(U+V) X_v+W}\right) \\
        \log \left(-\frac{(U+V) \left(k_1+X_v\right)}{k_2 U-(U+V) X_v+W}\right)
    \end{pmatrix},
    \qquad \qquad 
    B_2 = \begin{pmatrix}
        \log \left(\frac{k_2+k_3+X_v}{k_1+X_v}\right) \\
        \log \left(\frac{k_3 (U+V)+k_2 V+W}{k_3 U+k_1 V+W}\right) \\
        \log \left(\frac{k_3 (U+V)+k_2 V+W}{U+V}\right) \\
        \log \left(k_2+k_3+X_v\right) \\
        \log \left(\frac{k_1 (U+V)+k_2 U+W}{U+V}\right) \\
        0
    \end{pmatrix}.
\end{equation}
Notice that $U$, $V$, and $W$ always appear linearly in both numerator and denominator. This reflects the fact that only the ratios $V/U$ and $W/U$ appear as variables in the integral. In deriving the differential equations, we treated these ratios as independent variables, even though they are in reality functions of the momenta $k_1,k_2,k_3$ and the complex integration variable $w$. The connection $B$, when expressed in terms of the single variable $w$, can therefore be viewed as the pull-back from a larger space involving $V/U$ and $W/U$ to the space of interest $X\ni(w,X_v,k_1,k_2,k_3)$.
In contrast to the factorised term, this matrix is not flat on the full manifold $X$. We do not have a good understanding of why this is the case.

Currently, we are only interested in the analytic properties of the iterated integral as a function of $w$. We will treat the external variables as constant parameters. We let $\gamma$ describe a path $[0,1]\to \mathbb C\setminus\{w_*^i\}\ni w$, where $w_*^i$ are the eight branch points discussed below. The multivalued function can then be written as
\begin{equation}
    \mathcal{I}_U(w)=\mathbb P \ \exp\left( \int_\gamma B \right).
\end{equation}
Choosing a branch amounts to a fixed choice for $\gamma$. A convenient path is the one that gives rise to the original integral over the region $\Sigma$.

Before diving into the specific branch, let us first find the locations of the branch points in $\mathbb C$. First embed the integration region $\Sigma$ (that we allow to be deformed now) in $\mathbb{CP}^2$ by regarding $u,v$ as homogeneous coordinates in the chart $z=1$ so that $[u,v;1]\in \mathbb{CP}^2$. Writing hyperplanes as $L_i=\delta_{ab}\,P_i^a X^b$ with $X:=(u,v,1)$ and $a,b\in(1,2,3)$, we can write the intersection point between two hyperplanes as
\begin{equation}
    Q_{ij}^c=\epsilon^{c}{}_{ab}\,P_i^aP_j^b,
\end{equation}
where $\epsilon^c{}_{ab}$ is the Levi--Civita tensor. Moreover, taking three hyperplanes, $P_i,P_j$ and $P_k$, they intersect at the same point for vanishing $\det(P_i,P_j,P_k)$. The hyperplanes appearing in the integral are shown below:
\begin{center}
\begin{tabular}{|l|l|}
    \hline
    Line & Coefficients \\ \hline
    $P_1$ & $(1,0,-k_3)$ \\ \hline
    $P_2$ & $(0,1,-k_1)$ \\ \hline
    $P_3$ & $(1,-1,k_2)$ \\ \hline
    $P_4$ & $(-1,1,k_2)$ \\ \hline
    $P_5$ & $(U,V,W)$ \\ \hline
    $P_6$ & $(0,1,X_v)$ \\ \hline
\end{tabular}
\end{center}
Notice that in $\mathbb{CP}^2$, the integration region is bounded: $P_3$ and $P_4$ are parallel and meet at a single point at infinity. Consequently, $\Sigma$ is a quadrilateral embedded in $\mathbb{CP}^2$.

Every genuine singularity of the integral corresponds to a point in parameter space for which three hyperplanes meet at a boundary point of $\Sigma$. Therefore, finding the singular points amounts to scanning through all triplets we can pick from all the hyperplanes and checking whether they intersect somewhere on $\Sigma$ with real $u$ and $v$.

Let us start by looking at singularities which do not involve $P_5$. We take the $k_i$ to generically satisfy the momentum conserving triangle inequalities. The only parameter we can tune is $X_v$. By virtue of the $i\varepsilon$ prescription, $X_v$ is slightly imaginary and there exists no intersection point of $P_6$ with any other line on the real integration contour. However, assuming $\varepsilon=0$ gives a good indication of where $\bar F_u$ and eventually the full correlator peaks as a function of $X_v$. The only two candidates are where $X_v+k_1=0$ such that $P_6$ aligns with $P_2$ and $X_v+k_2+k_3=0$ such that $P_1$, $P_3$ and $P_6$ intersect.

We are not interested in intersections for finite $u$ or $v$ that involve both $P_5$ and $P_6$, since the $i\varepsilon$ prescription forbids $P_6$ from intersecting any other line on the real contour. In contrast to before, the singularities do not give any useful information about the full correlator.

Therefore, it suffices to intersect $P_5$ with the set of lines that bound the integration region. The integration region is bounded by four lines which intersect at the four points $Q_1^c=\epsilon^c_{ab}P^a_1 P_2^b$, $Q_2^c=\epsilon^c_{ab}P^a_1 P_3^b$, $Q_3^c=\epsilon^c_{ab}P^a_2 P_4^b$ and $Q_4^c=\epsilon^c_{ab}P^a_3 P_4^b$. The loci in $w=\rho \, e^{i\theta}$ for which the points $Q_i$ lie on $P_5$ are given by:
\begin{center}
\begin{tabular}{|l|l|}
    \hline 
    Incidence & Locus $w$ \\ \hline \hline
    $P_5\cdot Q_1=0$ & $k_3 U+k_1 V+W=0$ \\ \hline
    $P_5 \cdot Q_2=0$ & $k_2 V+k_3 (U+V)+W=0$ \\ \hline
    $P_5 \cdot Q_3=0$ & $k_2 U+k_1 (U+V)+W=0$ \\ \hline
    $P_5\cdot Q_4=0$ & $2k_2(U+V)=0$ \\ \hline
\end{tabular}
\end{center}
Every equation defines a quadratic equation in $w$ and therefore has two solutions.

\begin{wrapfigure}{r}{6.5cm}
    \centering
    \includegraphics{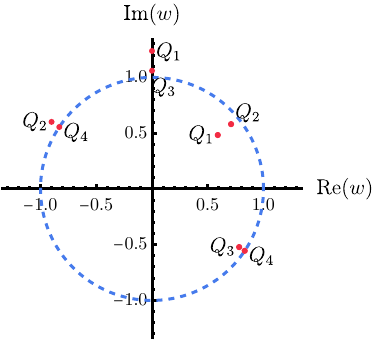}
    \caption{The locations of the branch points as predicted from the intersections $P_5\cdot Q_i$. We have set $k_1=1.15,k_2=0.93,k_3=1.00$ and $\delta=0.2$.}
    \label{fig:Triangle_CPlane_BP}
\end{wrapfigure}
We can solve these equations and find the exact locations of the putative branch points on the complex $w$-plane. The momenta $k_1,k_2$ and $k_3$ are the parameters that appear in these four equations and they are real. Moreover, without the small imaginary deformation $W \rightarrow W+i\delta$, all of the solutions lie on $\mathbb S^1$. This highlights another convenient consequence of the $i\delta$ prescription: except for the two branch points arising from $U+V=0$, every other branch point is moved away from the unit circle, resolving any potential branch-cut issues. The smooth displacement of the singularities as we increase the size of $\delta$ allows for perturbative solutions of the singular loci. We assume that the displacement in the radius is small by complexifying in the following way:
\begin{equation*}
\theta=-i\log(e^{\delta \ r}e^{i\theta})= \theta - i \, \delta \, r + \mathcal O(\delta^2).
\end{equation*}
Most importantly, the radial deformations are sign definite by virtue of the triangle inequalities that the momenta $k_i$ obey.
\begin{center}
\begin{tabular}{|l|l|}
    \hline 
    Incidence & Loci $w$ \\ \hline \hline
    \multirow{2}{*}{$P_5\cdot Q_1=0$} & $\left(1+\frac{\delta}{(-k_1+k_2+k_3)(k_1+k_2-k_3)}\right)i$ \\ \cline{2-2}
    & $\left(1-\frac{\delta}{(-k_1+k_2+k_3)(k_1+k_2-k_3)}\right)\times\left(\frac{\sqrt{p}}{2 k_1 k_3}+i\frac{\delta_2}{2k_1k_3}\right)$ \\ \hline
    \multirow{2}{*}{$P_5\cdot Q_2=0$} & $\left( 1+\frac{\delta}{(-k_1+k_2+k_3)(k_1+k_2+k_3)} \right)\times\left( -\frac{\sqrt{p}}{2k_1k_2}+i\frac{\delta_3}{2k_1 k_2} \right)$ \\ \cline{2-2}
    & $\left( 1-\frac{\delta}{(-k_1+k_2+k_3)(k_1+k_2+k_3)} \right)\times\left(\frac{\sqrt{p}}{2k_1k_2}+i\frac{\delta_2}{2k_1 k_3} \right)$ \\ \hline
    \multirow{2}{*}{$P_5\cdot Q_3=0$} & $\left( 1 + \frac{\delta}{(k_1+k_2-k_3)(k_1+k_2+k_3)} \right)\times i$ \\ \cline{2-2}
    & $\left( 1 - \frac{\delta}{(k_1+k_2-k_3)(k_1+k_2+k_3)} \right)\times\left( \frac{\sqrt{p}}{2 k_1 k_2} -i\frac{\delta_3}{2k_1k_2}\right)$ \\ \hline
    \multirow{2}{*}{$P_5\cdot Q_4=0$} & $1\times\left( -\frac{\sqrt{p}}{2k_1k_2}+i\frac{\delta_3}{2k_1k_2} \right)$ \\ \cline{2-2}
    & $1\times\left( \frac{\sqrt{p}}{2k_1k_2}-i\frac{\delta_3}{2k_1k_2} \right)$ \\ \hline
\end{tabular}
\label{tab:Branch_Points}
\end{center}

Let us now come back to the specific branch of $\bar F_u(w)$. We will fix $\Sigma$ to lie in the real section of $\mathbb{CP}^2$, i.e. where $(u,v)\in\mathbb R^2$. In the end, we want to construct an explicit form for this branch by expressing it in terms of principal branches of logarithms and dilogarithms. There are alternatives to these two single-valued functions, such as Goncharov polylogarithms~\cite{goncharov2000,Brown2009,Duhr:2014woa}. Depending on how we combine the $\log$ and $\mathrm{Li}_2$ functions, $\bar F_u(w)$ will be discontinuous at different locations on the complex plane. 
However, the fixed integration contour $\Sigma$ only allows for a single branch-cut structure.

How do we determine which of these two corresponds to the particular branch we have chosen? From the perspective of the integral \eqref{eq:F_SingleU}, the branch cuts on $\mathbb C \ni w$ are the values of $w$ for which the hyperplane $Uu+Vv+W=L_5$ pierces through the $\Sigma$ surface. As we have pointed out before, every intersection point $P_5\cdot Q_i=0$ corresponds to a pair of points $w\in\mathbb C$. In a similar vein, the region $\Sigma \in \mathbb{CP}^2$ gets mapped to two topologically equivalent patches in $\mathbb C$.
\begin{figure}[h!]
\centering
\begin{minipage}{.45\textwidth}
    \centering
    \includegraphics[width=0.95\linewidth]{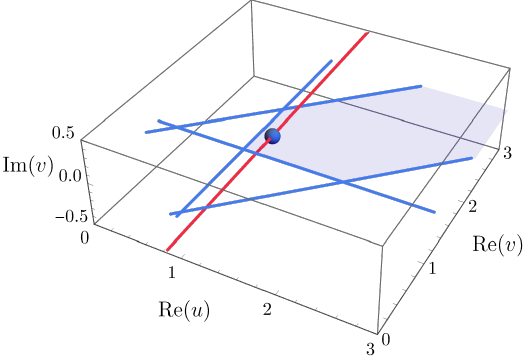}
    \caption{The hyperplane arrangement in the projected space with coordinates $(\mathrm{Re}(u),\mathrm{Re}(v),\mathrm{Im}(v))$. We set the momenta to $k_1=1.15$, $k_2=0.93$, $k_3=1.00$, $\delta=0.10$ and the complex-plane coordinate to $w=(0.94)e^{i\times 0.51}$. The red line corresponds to the hyperplane $L_5$, whereas the blue lines bound the shaded integration region $\Sigma$. The blue dot indicates where $L_5$ pierces $\Sigma$.}
    \label{fig:Triangle_Region_Branch}
\end{minipage}
\hspace{0.05\textwidth}
\begin{minipage}{.45\textwidth}
    \centering
    \includegraphics[width=0.95\linewidth]{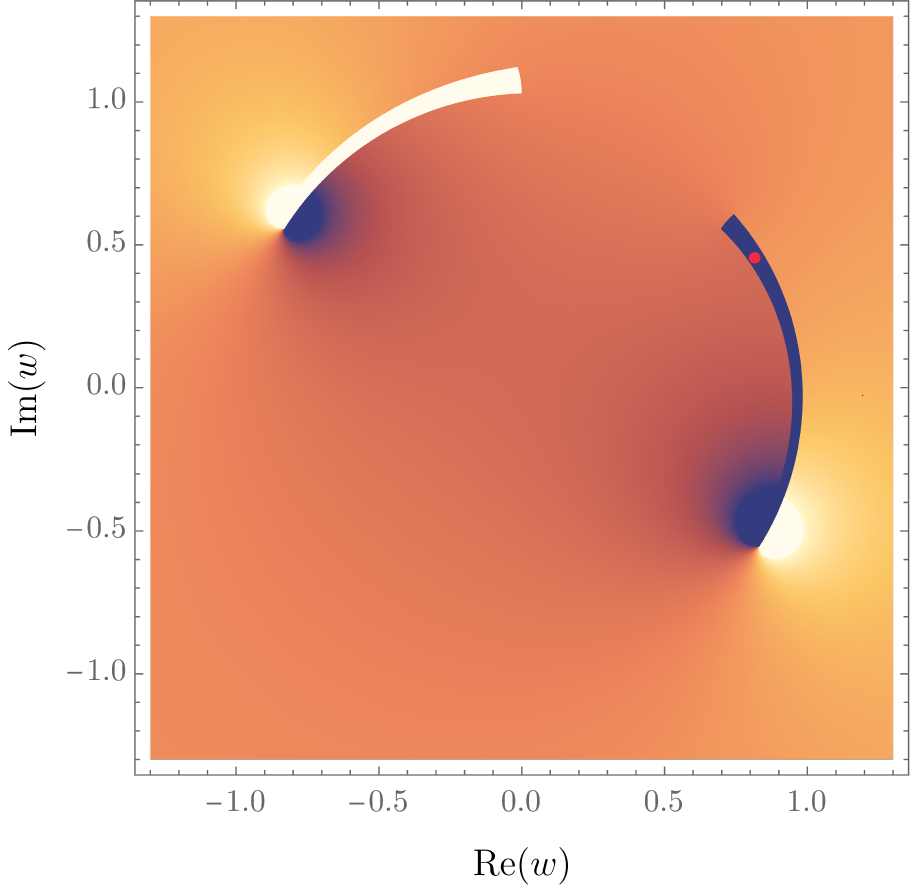}
    \caption{The function $\bar F_u(w)$ plotted on the complex plane for $k_1=1.15$, $k_2=0.93$, $k_3=1.00$, $X_v=5.60$, $\delta=0.10$ and $\varepsilon=1.00 \times 10^{-3}$. The red dot indicates the same location as on the left, $w=(0.94)e^{i\times 0.51}$, and is found on the branch cut.}
    \label{fig:Triangle_CPlane_FOnBC}
\end{minipage}
\end{figure}
In Figures~\ref{fig:Triangle_Region_Branch} and \ref{fig:Triangle_CPlane_FOnBC} we show an example of a point on $\Sigma$ and its image on the complex plane. The plots confirm our expectation that intersections of $L_5$ with $\Sigma$ get mapped to branch cuts of $\bar F_u(w)$ on $\mathbb C$. The plot of $\bar F_u(w)$ shows a few spurious discontinuities. These disappear as we decrease $\delta$. Cleaner plots can be found in the next section.

Let us point out a subtlety concerning the two parameters $\varepsilon$ and $\delta$, controlling respectively the small imaginary parts of $X_v$ and $W$. Their purpose in life is to position the arguments of the logarithms and dilogarithms just above their respective branch cuts. However, with two small parameters, we will generally see interference in a rational function containing both $X_v$ and $W$: one small imaginary contribution might cancel the other. A potentially problematic example is given by
\begin{equation}
    \mathrm{Li}_2(\zeta_1)=\mathrm{Li}_2\left(-\frac{V \left(k_2+k_3+X_v\right)}{k_3 U-V X_v+W}\right).
\end{equation}
Suppose that we relate the two parameters as $\varepsilon=\lambda \,\delta$. The imaginary part of $\zeta_1$,
\begin{equation}
    \Im\left\{ \zeta_1 \right\}
    =
    -\frac{V \left(k_3 (\lambda  U+\lambda  V-1)+k_2 (\lambda  V-1)-X_v+\lambda  W\right)}{\left(k_3 U-V X_v+W\right)^{2}}
\end{equation}
has two zeroes. If $V=0$, then $\zeta_1=0$ and therefore $\zeta_1$ does not cross the real axis through a branch cut. If instead
\begin{equation}
    k_3 (\lambda  U+\lambda  V-1)+k_2 (\lambda  V-1)-X_v+\lambda  W=0
    \qquad \implies \qquad
    \zeta_1=\frac{1}{1-\frac{1}{V\lambda}},
\end{equation}
and $V\lambda \sim 1$, then $\zeta_1$ goes through the branch cut. The dimensionful quantity $V$ attains a maximum value of $2k_3$. To avoid the branch cut discontinuity, we must require $2\lambda k_3<1$. Consequently, $\lambda$ must be sufficiently small compared to $k_3$. More practically, we assume the following relative ordering:
\begin{equation}
    \varepsilon/\delta \ll 1.
\end{equation}

\noindent
\paragraph{Single $v$ term.}
The boxed integral below,
\begin{multline}
    \frac{1}{-2\pi i}\oint \frac{-\d \theta}{- U X_3-V X_v+W} \, F_v(\theta) \\
    =
    \frac{1}{-2\pi i}\oint \frac{-\d \theta}{- U X_3-V X_v+W}
    \boxed{
    \int_\Sigma
    \frac{1}{(u+X_3)\left(\frac{U}{V}u+v+\frac{W}{V}\right)}
    \, \d u \, \d v
    },
    \label{eq:SingleV}
\end{multline}
is most efficiently evaluated using a different subdivision of the integration region~$\Sigma$.
\begin{figure}[h!]
    \centering
    \includegraphics[width=0.35\linewidth]{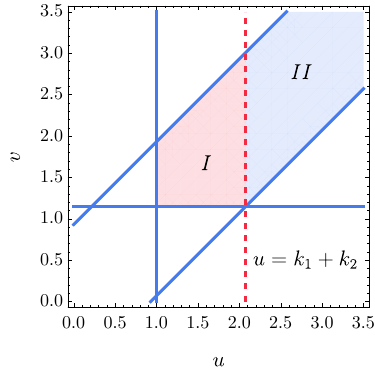}
    \caption{An alternative split of $\Sigma$ into regions I and II.}
    \label{fig:Triangle_Region_Split_U}
\end{figure}

The subdivision shown in Fig.~\ref{fig:Triangle_Region_Split_U} allows us to write the integral over $\Sigma$ directly as a sum of two iterated integrals:
\begin{multline}
    \int_\Sigma
    \frac{\d u \, \d v}{(u+X_3)\left(\frac{U}{V}u+v+\frac{W}{V}\right)}
    =
    \int_{k_3}^{k_1+k_2} \frac{\d u}{u+X_3}
    \int_{k_1}^{u+k_2}
    \frac{\d v}{\frac{U}{V}u+v+\frac{W}{V}} \\
    +
    \int_{k_1+k_2}^{\infty} \frac{\d u}{u+X_3}
    \int_{u-k_2}^{u+k_2}
    \frac{\d v}{\frac{U}{V}u+v+\frac{W}{V}} .
\end{multline}

The structure of these iterated integrals closely mirrors that of the single-$u$ term. In fact, the hyperplane arrangement governing the analytic structure on the complex $w$-plane is the same. As a consequence, we can directly carry over the analysis of the single-$u$ term to the present case. As before, it is crucial to ensure that the functions representing the $u,v$ integrals in \eqref{eq:SingleV} have the correct branch-cut structure.

The same types of functions appear in the explicit analytic expressions for regions~I and~II as in the single-$u$ case. Moreover, even the counting follows the same pattern: four dilogarithms for region~I and two for region~II, which combine into four dilogarithms for the full integration region.

\subsubsection{Final Result from Residues}

Starting from \eqref{eq:Triangle_ThetaRep} and \eqref{eq:Triangle_PartialFracs}, we analytically continue
$\theta \in [0,2\pi)$ to a complex variable $w=\rho e^{i\theta}$,
\begin{equation}
    \frac{1}{-2\pi i}
    \oint
    \frac{-\d \theta}{- U(\theta) X_3 - V(\theta) X_v + W(\theta)}
    \left(F_u(\theta)+F_v(\theta)\right)
    =
    \frac{1}{\pi (x-iy)}
    \int_C
    \frac{\bar F_u(w)+\bar F_v(w)}{(w-w_-)(w-w_+)},
\end{equation}
where $C$ denotes the unit-circle contour in the complex $w$-plane. The two poles $w_-$ and $w_+$ are defined as in \eqref{eq:TriangleComp_wPoles}, with $x$, $y$, and $z$ given by \eqref{eq:TriangleComp_Measure_Rep}. We have shown in the previous sections how to construct the appropriate branches of $\bar F_u(w)$ and $\bar F_v(w)$. Their branch cuts take the shape of two curvilinear quadrilaterals: one lying inside the unit circle and one outside.

\begin{figure}[h!]
\centering
\begin{minipage}{.48\textwidth}
    \centering
    \includegraphics[width=\linewidth]{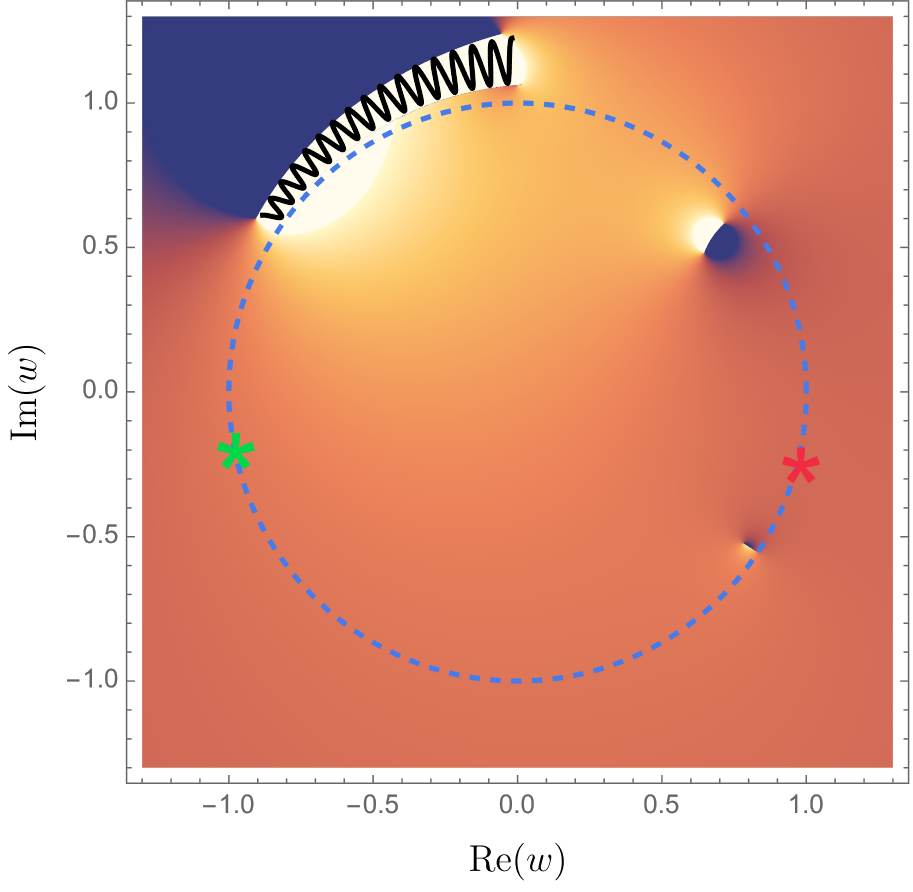}
    \caption{The function $\bar F_O(w)$ plotted for $k_1=1.150,k_2=0.930$, $k_3=1.000$, $X_v=5.600+0.001 i$ and $\delta=0.200$. The green and red stars indicate the positions of respectively the pole inside and outside the unit-circle. We have drawn a black wiggly line through the $\mathcal{O}(1)$ branch cut. In addition, there are two small $\mathcal{O}(\delta)$ branch cuts lying inside the unit circle. }
    \label{fig:Triangle_CPlane_FOutside}
\end{minipage}
\hspace{0.02\textwidth}
\begin{minipage}{.48\textwidth}
    \centering
    \includegraphics[width=\linewidth]{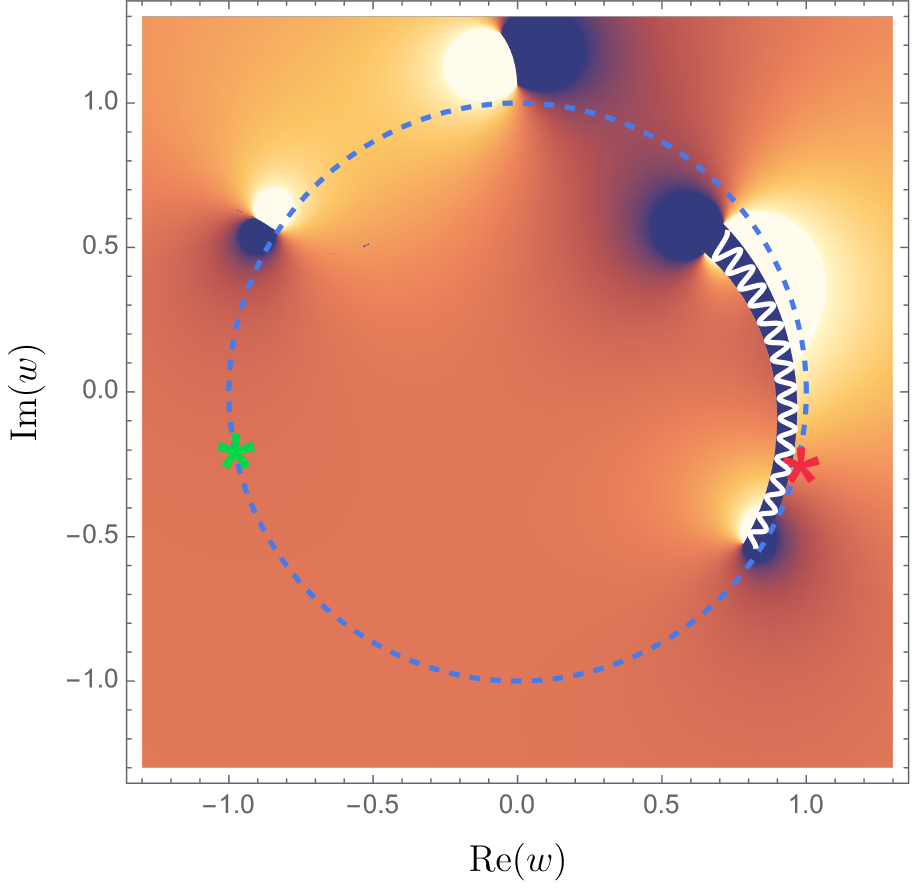}
    \caption{The function $\bar F_I(w)$ plotted for $k_1=1.150,k_2=0.930$, $k_3=1.000$, $X_v=5.600+0.001 i$ and $\delta=0.200$. The green and red stars indicate the positions of respectively the pole inside and outside the unit-circle. The white wiggly line highlights the $\mathcal{O}(1)$ branch cut while the two small $\mathcal{O}(\delta)$ branch cuts lying inside the unit circle are left unmarked. }
    \label{fig:Triangle_CPlane_FInside}
\end{minipage}
\end{figure}

Somewhat surprisingly, the sum of $\log^2$ and $\mathrm{Li}_2$ functions appearing in $\bar F_u(w)+\bar F_v(w)$ can be reorganised into two separate contributions,
\begin{equation}
    \bar F_u(w)+\bar F_v(w) = \bar F_O(w) + \bar F_I(w),
\end{equation}
where $\bar F_O(w)$ has $\mathcal{O}(1)$-sized branch cuts outside the unit circle and $\mathcal{O}(\delta)$-sized branch cuts inside, while the opposite holds for $\bar F_I(w)$. We have illustrated their respective analytic structures on the complex plane in Figures~\ref{fig:Triangle_CPlane_FOutside} and~\ref{fig:Triangle_CPlane_FInside}.

Since the branch points at the same angular position lie a distance of $\mathcal{O}(\delta)$ apart, the small branch cuts in the plots have length of order $\delta$. A contour deformation that places these small branch cuts either inside or outside the unit circle contributes functions of transcendental weight at order $\delta$. These functions may diverge logarithmically in certain kinematic regions. In such regions, $\delta$ must be taken sufficiently close to zero for the result to make sense.

We are integrating over a closed contour, and the inside/outside split makes it possible to apply the residue theorem. Concretely, we evaluate the residue of the pole inside the unit circle for the function with a large branch cut outside the unit circle, and vice versa for the pole outside:
\begin{equation}
    \frac{1}{\pi(x-iy)}\int_C \frac{\bar F_O(w)+\bar F_I(w)}{(w-w_-)(w-w_+)}
    =
    \frac{\bar F_O(w_-)+\bar F_I(w_+)}{\sqrt{\Omega(-X_3,-X_v)}}+\mathcal{O}(\delta),
\end{equation}
where we find exactly the same denominator as for the factorised term, cf.~\eqref{eq:Fact_Denominator}.

Putting all the pieces together, the integral in \eqref{eq:Triangle_ThetaRep} admits the following closed-form expression:
\begin{equation}
    \mathcal I(X_3,X_v)
    =
    \frac{F_F(X_3,X_v)-\bar F_O(w_-)-\bar F_I(w_+)}{\sqrt{\Omega(-X_3,-X_v)}}+\mathcal{O}(\delta).
    \label{eq:Triangle_GenInt_Final}
\end{equation}
Here $F_F$ is defined in \eqref{eq:Triangle_FactorisedExplicit}, and $\bar F_O(w)$ and $\bar F_I(w)$ are given in App.~\ref{app:exptriangle}.
The explicit formulae for the poles $w_\pm$ are given in \eqref{eq:wpm_Poles}.

\begin{figure}[h!]
\begin{subfigure}{.49\textwidth}
    \centering
    \includegraphics[width=0.95\linewidth]{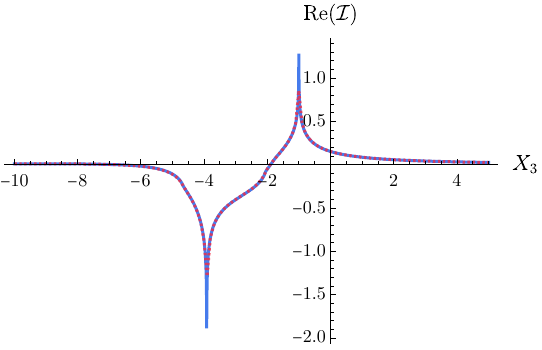}
\end{subfigure}
\begin{subfigure}{.49\textwidth}
    \centering
    \includegraphics[width=0.95\linewidth]{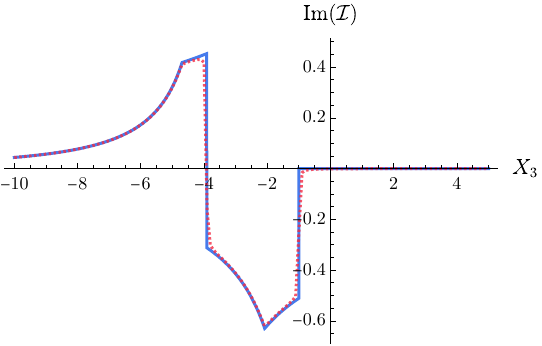}
\end{subfigure}
\caption{The real and imaginary parts of $\mathcal I$ with $X_v=X_2+X_3$ and $\varepsilon\rightarrow0$. We fixed $X_2=2.76,k_1=1.15,k_2=0.93$ and $k_3=1.00$. The dashed red lines are the numerical results with the same values for the parameters and $\varepsilon=0.01$. }
\label{fig:Triangle_GenInt_Final}    
\end{figure}
We plotted the real and imaginary parts of the final result \eqref{eq:Triangle_GenInt_Final} above in Fig.~\ref{fig:Triangle_GenInt_Final}. The choice $X_v=X_2+X_3$ implies that we are considering \eqref{eq:R_Triangle_321}. 

First of all, for physical values $X_3 \geq k_3$, the function is real-valued. The negative peak in the real part corresponds to the kinematic point $X_2+X_3+k_1=0$, while the positive peak occurs at $X_3+k_3=0$. These are precisely the points we identified as singular by inspecting the integral in Sec.~\ref{sec:Sing}.

In addition, there are two locations where the first derivative of the imaginary part appears discontinuous and, correspondingly, the second derivative of the real part develops an extremum. These occur at $X_3+k_1+k_2=0$ and $X_2+X_3+k_2+k_3=0$. In the singularity analysis, these configurations in kinematic space appeared as less divergent singularities of the integral. Away from these loci, the function remains analytic.

Taken together, these plots confirm that analysing the singularities of time-ordered integrals provides an accurate description of the analytic structure of cosmological correlators.

By using the partial-fraction decomposition in \eqref{eq:Triangle_Integrand_PF}, we can write the total covariant building block as
\begin{equation}
    I_{123}(X_1,X_2,X_3)
    =
    \frac{\mathcal{I}(X_3,X_2+X_3)}{X_1+X_2+X_3}
    +
    \frac{\mathcal{I}(X_3,X_1)-\mathcal{I}(X_3,X_2+X_3)}{-X_1+X_2+X_3}.
    \label{eq:Triangle_BuildingBlock_Final}
\end{equation}
The full triangle correlator is then obtained by summing over all $D_3$ permutations,
\begin{equation}
    I_\triangle
    =
    I_{123}+I_{132}+I_{312}+I_{321}+I_{213}+I_{231}.
    \label{eq:Triangle_Perms_Final}
\end{equation}

Interestingly, the combinations $\Omega(X_3,X_2+X_3)$ and $\Omega(X_3,X_1)$ that appear in the covariant building block are invariant under the $2\leftrightarrow 3$ and $1\leftrightarrow 3$ reflections of $D_3$, respectively. Consequently, only three distinct $\Omega$ polynomials appear in $I_\triangle$.

\begin{figure}[h!]
    \centering
    \includegraphics{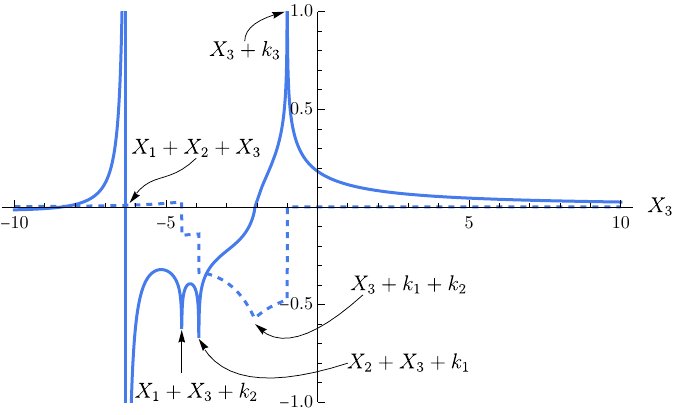}
    \caption{The real (solid line) and imaginary (dashed line) parts of $I_\triangle$. We have fixed $X_1=3.55$, $X_2=2.76$, $k_1=1.15$, $k_2=0.93$ and $k_3=1.00$. The imaginary regulators are taken to be very small: $\delta=1.00\times 10^{-7}$ and $\varepsilon=1.00\times 10^{-9}$.}
    \label{fig:Triangle_Correlator_Result}
\end{figure}

In Fig.~\ref{fig:Triangle_Correlator_Result} we plot the real and imaginary parts of the result as a function of $X_3$. Importantly, the $i\varepsilon$ prescription should be regarded as part of the definition of the function $\mathcal{I}(X_3,X_v)$, with the role of selecting the correct side of the branch cuts. In practice, this can be implemented by replacing every occurrence of $\mathcal{I}(X_3,X_v)$ in \eqref{eq:Triangle_BuildingBlock_Final} by
\[
\mathcal{I}_\varepsilon(X_3,X_v)
=
\mathcal{I}(X_3+i\varepsilon,\,X_v+i\varepsilon),
\]
and restricting $X_3,X_v\in\mathbb R$. This avoids the need to keep track of the $i\varepsilon$ prescription explicitly when performing the permutations.

The plot peaks exactly at the locations where we expect the triangle’s singularities, from left to right:
\[
X_1+X_2+X_3=0,\qquad
X_1+X_3+k_2=0,\qquad
X_2+X_3+k_1=0,\qquad
X_3+k_3=0.
\]
Notice that the minimum of the imaginary part occurs at $X_3+k_1+k_2=0$, where the real part passes through a turning point.

Taking the total-energy limit, $X_1+X_2+X_3 \rightarrow 0$, we can recover the triangle amplitude. On the correlator side, this amounts to taking the residue at $X_1+X_2+X_3=0$ of \eqref{eq:Triangle_Perms_Final}. The resulting expression is valid for any number of external legs.
On the amplitude side, attaching multiple external legs to each vertex implies that the external kinematics are \emph{off-shell}. Defining
$P_i^\mu=(X_i,k_i)$ for $ i=1,2,3$
as the sums of external four-momenta flowing into each vertex, their squares are generically non-zero, $P_i^2\neq 0$. We therefore require the massless triangle loop with off-shell kinematics~\cite{Weinzierl:2022eaz},
\begin{equation}
    I_3
    =
    -\frac{2}{\sqrt{-\Delta_3}}
    \left[
        \text{Cl}_2\!\left(2\arctan\!\left(\frac{\sqrt{-\Delta_3}}{\tilde\delta_1}\right)\right)
        +
        \text{Cl}_2\!\left(2\arctan\!\left(\frac{\sqrt{-\Delta_3}}{\tilde\delta_2}\right)\right)
        +
        \text{Cl}_2\!\left(2\arctan\!\left(\frac{\sqrt{-\Delta_3}}{\tilde\delta_3}\right)\right)
    \right],
    \label{eq:Triangle_Amplitude}
\end{equation}
where $\text{Cl}_2$ denotes the Clausen function. It is related to the classical dilogarithm via
\begin{equation}
    \text{Cl}_2(\theta)
    =
    \frac{1}{2i}
    \left[
        \text{Li}_2\!\left(e^{i\theta}\right)
        -
        \text{Li}_2\!\left(e^{-i\theta}\right)
    \right].
\end{equation}
The quantities $\tilde\delta_i$, constructed from the four-momenta, are defined analogously to the $\delta_i$ built from the three-momenta $k_i$,
\begin{equation}
    \tilde\delta_1 = P_1^2-P_2^2-P_3^2,
    \qquad
    \tilde\delta_2 = -P_1^2+P_2^2-P_3^2,
    \qquad
    \tilde\delta_3 = -P_1^2-P_2^2+P_3^2.
\end{equation}

In the total-energy limit, the three distinct square roots appearing in the denominator of the correlator coincide and are exactly equal to the $\sqrt{-\Delta_3}$ that appears in the amplitude denominator,
\begin{equation}
    \lim_{X_1+X_2+X_3\rightarrow 0}
    \Omega(-X_3,-X_2-X_3)
    =
    -\Delta_3
    =
    (P_1^2)^2+(P_2^2)^2+(P_3^2)^2
    -2P_1^2P_2^2
    -2P_2^2P_3^2
    -2P_3^2P_1^2,
\end{equation}
where $\Delta_3$ is the K\"all\'en function.

\begin{figure}[h!]
    \centering
    \includegraphics{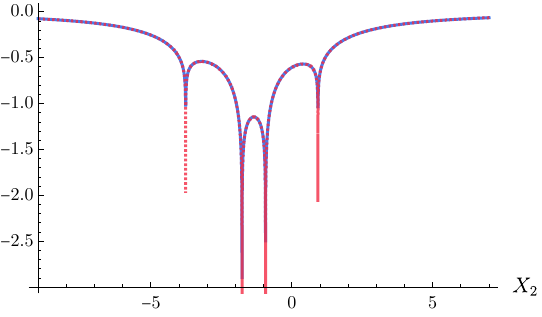}
    \caption{Comparison between the real parts of the off-shell triangle amplitude $-2\times I_3$ (red dashed line) and the total-energy residue of the correlator triangle $\mathrm{Res}_{X_T=0}\,I_\triangle$ (blue solid line). We have set $X_3=-X_1-X_2$, $X_1=2.76$, $k_1=1.15$, $k_2=0.93$ and $k_3=1.00$. The imaginary regulators are again taken to be tiny: $\delta=1.00\times 10^{-7}$ and $\varepsilon=1.00\times 10^{-9}$.}
    \label{fig:Triangle_Amplitude}
\end{figure}

The total-energy residue of the correlator reduces to the amplitude defined in \eqref{eq:Triangle_Amplitude}. Although it is not clear how this correspondence arises analytically, the plots in Fig.~\ref{fig:Triangle_Amplitude} demonstrate perfect agreement.

\newpage
\section{Outlook}
\label{sec:Outlook}

Cosmological correlators encode the interactions among particles produced from vacuum fluctuations, in a manner consistent with locality, causality, and unitarity. Experience with scattering amplitudes teaches us that these cherished principles are reflected in the singularity structure of the observable. Through the study of the bubble and triangle diagrams, we have shown how one-loop cosmological correlators exhibit new types of factorisation singularities, and we have provided an interpretation of their origin. Moreover, to the best of our knowledge, the explicit computation of the triangle in-in correlator employs a set of techniques that have not been widely utilised in cosmological perturbation theory. It is our hope that these techniques will prove useful elsewhere.

With these concrete examples of one-loop cosmological correlators in hand, there remains much to be explored in the near future. Below we list several directions that we find particularly promising:

\begin{itemize}
    \item \textbf{Singularities at one loop}: Our analysis of the singularity structure, which we have applied to these two examples, can be generalised to higher-site and higher-loop diagrams. This would allow us to assess whether the set of singularities identified here is exhaustive, or whether new types of singularities emerge for one-loop diagrams with a larger number of interaction vertices. Moreover, we have not investigated the cuts and their discontinuities away from the locations of branch points; doing so would be very interesting.

    Our flat-space analysis may also be incomplete when applied to other cosmological backgrounds. In particular, it would be interesting to extend our analysis to de Sitter space, where the representation theory is richer and the qualitative behaviour of the correlator may depend on how light the particles are. Many of the tools of Landau analysis can be imported directly into this context.

    Finally, it would be satisfying to formulate a general statement determining the locations (and coefficients) of all singularities of an in-in correlator, which could then be used as the basis for a bootstrap approach.

    \item \textbf{Differential equations:} The organisation of loop diagrams using differential equations is a powerful tool. It not only provides a practical method for direct integration, but also sheds light on the singularity structure of the result. Recently, this approach has been used to uplift flat-space correlators to their counterparts in an FRW universe \cite{Baumann:2025qjx,Baumann:2024mvm,Glew:2025ypb,Arkani-Hamed:2023bsv,Hang:2024xas,Westerdijk:2025ywh,Chen:2024glu,Chen:2023iix},
    where the resulting system of differential equations retains the same structural form across different cosmologies, with only a universal prefactor encoding the time dependence of the scale factor. An analogous uplift for loop correlators would be an ideal target, as it would allow us to investigate how IR divergences—absent in our flat-space results—arise in accelerating cosmologies.

    \item \textbf{A one-loop basis?} Scattering amplitudes in flat space can famously be organised, at one loop, in terms of a small number of master diagrams. In cosmology, we still lack an analogous organising framework. Nevertheless, it seems reasonable to expect that in ``flat-space cosmology'' a finite set of one-loop correlators might form a basis from which all correlators can be constructed. The puzzle is even more pronounced in cosmology, where, for example, in de Sitter space there is not even a basis for tree-level correlators. With the bubble and triangle at hand, we hope that, at least for flat-space cosmology, we are closer to an answer to this problem.

    \item \textbf{Phenomenology:} Our correlators can potentially be combined with weight-shifting operators to provide useful phenomenological input for studies of primordial non-Gaussianity, particularly in connection with Standard Model particles. Since the inflaton is presumably neutral under the Standard Model gauge group, it can interact with the Higgs field, quarks, and leptons only through loop diagrams. Although the squeezed limit is relatively well understood, full shape functions require detailed computations that are not yet available. Understanding the proper interplay between tools from conformal kinematics and the triangle diagram computed in this paper will be essential for constructing a broad set of phenomenological templates for searches for primordial non-Gaussianity.
\end{itemize}

Inflation is our leading theoretical framework for the origin of structure in the Universe. To make its predictions reliable, it is essential to have loop corrections to tree-level results under control, as well as precise templates for decoding the physics of primordial fluctuations. Progress along this path requires a detailed understanding of loop effects. The recent influx of ideas from collider physics and mathematics into cosmology has provided the necessary tools to tackle this problem. In this paper, we have presented a few explicit examples, and although much remains to be explored, we believe we have set the stage for a systematic characterisation of cosmological correlators at one loop.

\vspace*{\fill}
\noindent{\bf Acknowledgments}
We thank Nima Arkani-Hamed, Giacomo Brunello, Bruno Bucciotti, Craig Clark, 
Arthur Lipstein, Ivo Sachs, Francisco Vazão for many useful discussions.

TW thanks Daniel Baumann and the University of Amsterdam for their hospitality and for the opportunity to present this work.
GLP thanks the Yukawa Institute
for Theoretical Physics at Kyoto University for hospitality during its “Progress of Theoretical
Bootstrap” workshop. He also thanks the Centro de Ciencias de Benasque Pedro Pascual for hospitality.
 
GLP and TW are supported by Scuola Normale and by INFN (IS GSS-Pi). 
The research of GLP and TW is moreover supported by the ERC (NOTIMEFORCOSMO, 101126304). 
Views and opinions expressed are, however, those of the author(s) only and do not necessarily reflect those of the European Union or the European Research Council Executive Agency. 
Neither the European Union nor the granting authority can be held responsible for them. 
GLP is further supported by the Italian Ministry of Universities and Research (MUR) under contract 20223ANFHR (PRIN2022).

\appendix
\section{Wavefunction Coefficients}\label{app:wfcoeffs}

\paragraph{Example 1: Wavefunction Bubble}

The wavefunction coefficient for the bubble diagram consists of three sectors: it shares the $r=0$ and $r=2$ sectors with the correlator, but also contains the $r=1$ sector:

\begin{center}
\begin{minipage}{0.49\textwidth}
\begin{minipage}{.4\textwidth}
    \centering
    \includegraphics[width=0.95\linewidth]{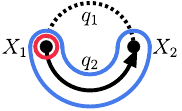}
    \label{fig:Bubble_Tubing_012}
\end{minipage}
\hspace{-1em}
\begin{minipage}{.59\textwidth}
\scalebox{1.2}{$\rightarrow\textcolor{blue}{\frac{1}{X_1+X_2+2q_1}}\textcolor{red}{\frac{1}{X_1+q_1+q_2}}$}
\end{minipage}
\begin{minipage}{.40\textwidth}
    \centering
    \includegraphics[width=0.95\linewidth]{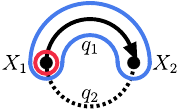}
    \label{fig:Bubble_Tubing_120}
\end{minipage}
\hspace{-1em}
\begin{minipage}{.59\textwidth}
    \scalebox{1.2}{$\rightarrow\textcolor{blue}{\frac{1}{X_1+X_2+2q_2}}\textcolor{red}{\frac{1}{X_1+q_1+q_2}}$}
\end{minipage}
\end{minipage}
\hspace{.002\textwidth}
\begin{minipage}{.49\textwidth}
\begin{minipage}{.4\textwidth}
    \centering
    \includegraphics[width=0.95\linewidth]{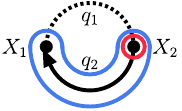}
    \label{fig:Bubble_Tubing_021}
\end{minipage}
\hspace{-1em}
\begin{minipage}{.59\textwidth}
    \scalebox{1.2}{$\rightarrow\textcolor{blue}{\frac{1}{X_1+X_2+2q_1}}\textcolor{red}{\frac{1}{X_2+q_1+q_2}}$}
\end{minipage}
\begin{minipage}{.40\textwidth}
    \centering
    \includegraphics[width=0.95\linewidth]{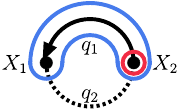}
    \label{fig:Bubble_Tubing_210}
\end{minipage}
\hspace{-1em}
\begin{minipage}{.59\textwidth}
    \scalebox{1.2}{$\rightarrow\textcolor{blue}{\frac{1}{X_1+X_2+2q_2}}\textcolor{red}{\frac{1}{X_2+q_1+q_2}}$}
\end{minipage}
\end{minipage}
\end{center}
To make the difference between the two rational functions more explicit, let us repeat the formula for the correlator bubble:
\begin{equation}
    R_{C,\text{bubble}}=\frac{1}{4q_1 q_2} \left[ \frac{1}{X_1+X_2}\left( \frac{1}{X_1+q_1+q_2} + \frac{1}{X_2 + q_1 + q_2} \right) + \frac{1}{X_1+q_1+q_2} \frac{1}{X_2 + q_1 + q_2} \right]\, .
\end{equation}
To obtain the rational function corresponding to the wavefunction coefficient we have to subtract the $r=1$ sector from the correlator:
\begin{equation}
    R_{\psi,\text{bubble}} = R_{C,\text{bubble}} - \frac{1}{4 q_1 q_2}\left( \frac{1}{X_{12}+2q_1} + \frac{1}{X_{12}+2q_2}\right) \left( \frac{1}{X_1+q_1+q_2}+\frac{1}{X_2+q_1+q_2} \right),
\end{equation}
where we have adopted the notation $X_{i_1...i_n}=X_{i_1}+\cdots +X_{i_n}$.
Let us stress that the rational function for the $r=1$ sector depends on two linearly independent combinations of the integration variables. 

\paragraph{Example 2: Wavefunction Triangle}

Let us illustrate the tubing rules applied to time-ordered diagrams included for the triangle wavefunction coefficient. Below we list eight examples with a special emphasis on the graphs with a single non-time-ordered edge:
\begin{center}
\begin{minipage}{0.49\textwidth}
\begin{minipage}{.4\textwidth}
    \centering
    \includegraphics[width=.95\linewidth]{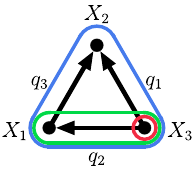}
    \label{fig:Triangle_Tubing_312}
\end{minipage}
\hspace{-2em}
\begin{minipage}{.64\textwidth}
    \scalebox{1.2}{$
        \rightarrow\textcolor{blue2}{\frac{1}{X_{123}}}
        \textcolor{green2}{\frac{1}{X_{13}+q_{13}}}
        \textcolor{red2}{\frac{1}{X_3+q_{12}}}
    $}
\end{minipage}
\begin{minipage}{.40\textwidth}
    \centering
    \includegraphics[width=0.96\linewidth]{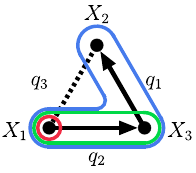}
    \label{fig:Triangle_Tubing_01332}
\end{minipage}
\hspace{-2em}
\begin{minipage}{.64\textwidth}
    \scalebox{1.2}{$
        \rightarrow\textcolor{blue2}{\frac{1}{X_{123}+2q_3}}\textcolor{green2}{\frac{1}{X_{13}+q_{13}}}
        \textcolor{red2}{\frac{1}{X_1+q_{23}}}
    $}
\end{minipage}
\end{minipage}
\hspace{.002\textwidth}
\begin{minipage}{0.49\textwidth}
\begin{minipage}{.4\textwidth}
    \centering
    \includegraphics[width=.95\linewidth]{Figures/Triangle_Graphs/Triangle_Tubing_321.pdf}
    \label{fig:Triangle_Tubing_321}
\end{minipage}
\hspace{-2em}
\begin{minipage}{.64\textwidth}
    \scalebox{1.2}{$
        \rightarrow\textcolor{blue2}{\frac{1}{X_{123}}}
        \textcolor{green2}{\frac{1}{X_{23}+q_{23}}}
        \textcolor{red2}{\frac{1}{X_3+q_{12}}}
    $}
\end{minipage}
\begin{minipage}{.40\textwidth}
    \centering
    \includegraphics[width=0.96\linewidth]{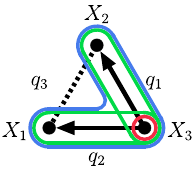}
    \label{fig:Triangle_Tubing_03132}
\end{minipage}
\hspace{-2em}
\begin{minipage}{.64\textwidth}
    \scalebox{1.2}{$
        \rightarrow\textcolor{blue2}{\frac{1}{X_{123}+2q_3}}\textcolor{red2}{\frac{1}{X_3+q_{12}}}
        $}\\
        \scalebox{1.2}{$\quad\times 
        \left(
        \textcolor{green2}{\frac{1}{X_{23}+q_{23}}}+\textcolor{green2}{\frac{1}{X_{31}+q_{31}}}\right)
    $}
\end{minipage}
\end{minipage}
\end{center}
\begin{center}
\begin{minipage}{0.49\textwidth}
\begin{minipage}{.4\textwidth}
    \centering
    \includegraphics[width=.95\linewidth]{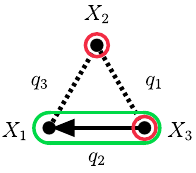}
    \label{fig:Triangle_Tubing_0031}
\end{minipage}
\hspace{-2em}
\begin{minipage}{.64\textwidth}
    \scalebox{1.2}{$
        \rightarrow\textcolor{red2}{\frac{1}{X_{2}+q_{13}}}
        \textcolor{green2}{\frac{1}{X_{13}+q_{13}}}
        \textcolor{red2}{\frac{1}{X_3+q_{12}}}
    $}
\end{minipage}
\begin{minipage}{.40\textwidth}
    \centering
    \includegraphics[width=0.96\linewidth]{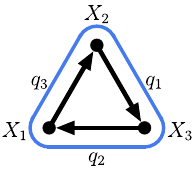}
    \label{fig:Triangle_Tubing_Cycle1}
\end{minipage}
\hspace{-2em}
\begin{minipage}{.64\textwidth}
    \scalebox{1.2}{$
        \rightarrow \qquad \quad \qquad 0 \qquad \qquad \quad
    $}
\end{minipage}
\end{minipage}
\hspace{.002\textwidth}
\begin{minipage}{0.49\textwidth}
\begin{minipage}{.4\textwidth}
    \centering
    \includegraphics[width=.95\linewidth]{Figures/Triangle_Graphs/Triangle_Tubing_0032.pdf}
    \label{fig:Triangle_Tubing_0032}
\end{minipage}
\hspace{-2em}
\begin{minipage}{.64\textwidth}
    \scalebox{1.2}{$
        \rightarrow\textcolor{red2}{\frac{1}{X_{1}+q_{23}}}
        \textcolor{green2}{\frac{1}{X_{23}+q_{23}}}
        \textcolor{red2}{\frac{1}{X_3+q_{12}}}
    $}
\end{minipage}
\begin{minipage}{.40\textwidth}
    \centering
    \includegraphics[width=0.96\linewidth]{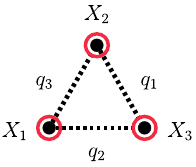}
    \label{fig:Triangle_Tubing_000}
\end{minipage}
\hspace{-2em}
\begin{minipage}{.64\textwidth}
    \scalebox{1.2}{$
        \rightarrow
        \textcolor{red2}{\frac{1}{X_1+q_{23}}\frac{1}{X_2+q_{31}}\frac{1}{X_3+q_{12}}}
    $}
\end{minipage}
\end{minipage}
\end{center}
The graphs with a single non-time-ordered edge and three non-time-ordered edges are not present in the correlator. The correlator only includes the `full cuts' with two non-time-ordered edges. Most importantly, the time-orderings with one or three disconnected edges give rise to distinct rational functions. For example, the first graph with a single non-time-ordered edge is associated to a rational function that depends on the loop variables $2q_3, q_1+q_3$ and $q_2+q_3$. This triplet is linearly independent in $\mathbb R^3 \ni (q_1,q_2,q_3)$ hence precluding a reduced two dimensional integral. The three dimensional integrals appearing uniquely in the wavefunction are expected to be hyperelliptic~\cite{Chowdhury:2023arc,Benincasa:2024ptf}.
\newpage
\section{Explicit Formulae for the Triangle}\label{app:exptriangle}

The analytic continuations $\bar F_u$ and $\bar F_v$ can be computed by performing the iterated integrals in $u$ and $v$. Subsequently, We sum the two functions and reorganise the result into a sum of two new terms, 
\begin{equation}
    \bar F_u(w)+\bar F_v(w) = \bar F_O(w)+\bar F_I(w).
\end{equation}
These two functions, $\bar F_O(w)$ and $\bar F_I(w)$, are characterised by having only $\mathcal{O}(1)$ branch cuts outside the unit circle ($\bar F_O$) or inside the unit circle ($\bar F_I$), while exhibiting only $\mathcal{\delta}$-sized branch cuts in the complementary regions. Their explicit expressions read
\begin{multline}
    F_O = 
    \text{Li}_2\left(\frac{(U+V) \left(k_1+k_2+X_3\right)}{W+(U+V) k_1+U k_2}\right)
    +\frac{1}{2} \log ^2\left(\frac{k_2 U-(U+V) X_v+W}{U+V}\right)
    \\
    -\text{Li}_2\left(-\frac{V \left(k_1+X_v\right)}{W+U k_3-V X_v}\right)+\text{Li}_2\left(-\frac{V \left(k_2+k_3+X_v\right)}{W+U k_3-V X_v}\right)+\text{Li}_2\left(-\frac{(U+V) \left(k_1+X_v\right)}{W+U k_2-(U+V) X_v}\right)
    \\
    +\log \left(k_2+k_3+X_v\right) \log \left(\frac{k_3 (U+V)+k_2 V+W}{k_3 U-V X_v+W}\right)
    +\log \left(k_1+X_v\right) \log \left(\frac{k_1 (U+V)+k_2 U+W}{k_2 U-(U+V) X_v+W}\right)
    \\
    +\log \left(k_3+X_3\right) \log \left(\frac{k_3 U+k_1 V+W}{k_3 (U+V)+k_2 V+W}\right)
    -\log \left(k_1+k_2+X_3\right) \log \left(\frac{-k_1 (U+V)-k_2 U-W}{k_2 V+X_3 (U+V)-W}\right)
    \\
    +\frac{1}{2} \log \left(\frac{k_1 (U+V)+k_2 U+W}{U+V}\right) \log \left(\frac{U+V}{-k_2 V-X_3 (U+V)+W}\right)
    \\
    +\frac{1}{2} \log \left(\frac{k_1 (U+V)+k_2 U+W}{U+V}\right) \log \left(\frac{k_1 (U+V)+k_2 U+W}{-k_2 V-X_3 (U+V)+W}\right)
    \\
    -\log \left(k_1+X_v\right) \log \left(\frac{k_3 U+k_1 V+W}{k_3 U-V X_v+W}\right)
    \label{eq:Triangle_FO}
\end{multline}
\begin{multline}
    F_I = 
    \text{Li}_2\left(\frac{(U+V) \left(k_2+k_3+X_v\right)}{W+V k_2+(U+V) k_3}\right)
    +\text{Li}_2\left(-\frac{U \left(k_1+k_2+X_3\right)}{W+V k_1-U X_3}\right)
    -\text{Li}_2\left(-\frac{U \left(k_3+X_3\right)}{W+V k_1-U X_3}\right)
    \\
    +\text{Li}_2\left(-\frac{(U+V) \left(k_3+X_3\right)}{W+V k_2-(U+V) X_3}\right)
    +\log \left(k_1+X_v\right) \log \left(\frac{k_3 U+k_1 V+W}{k_1 (U+V)+k_2 U+W}\right)
    \\
    -\log \left(k_2+k_3+X_v\right) \log \left(\frac{-k_3 (U+V)-k_2 V-W}{k_2 U+(U+V) X_v-W}\right)
    +\frac{1}{2} \log ^2\left(\frac{k_2 V-X_3 (U+V)+W}{U+V}\right)
    \\
    +\frac{1}{2} \log \left(\frac{k_3 (U+V)+k_2 V+W}{U+V}\right) \log \left(\frac{U+V}{-k_2 U-(U+V) X_v+W}\right)
    \\
    +\frac{1}{2} \log \left(\frac{k_3 (U+V)+k_2 V+W}{U+V}\right) \log \left(\frac{k_3 (U+V)+k_2 V+W}{-k_2 U-(U+V) X_v+W}\right)
    \\
    +\log \left(k_1+k_2+X_3\right) \log \left(\frac{k_1 (U+V)+k_2 U+W}{k_1 V-U X_3+W}\right)
    -\log \left(k_3+X_3\right) \log \left(\frac{k_3 U+k_1 V+W}{k_1 V-U X_3+W}\right)
    \\
    +\log \left(k_3+X_3\right) \log \left(\frac{k_3 (U+V)+k_2 V+W}{k_2 V-X_3 (U+V)+W}\right)
    \label{eq:Triangle_FI}
\end{multline}
\newpage
\section{Feynman Rules}\label{app:FeynRules}

We refer to \cite{BaumannJoyce} for a comprehensive introduction to the computation of in-in correlators using Feynman rules. Two types of propagators appear in these calculations. The first is the Wightman function,
\begin{equation}
    W_k(t,t') = f_k^*(t) f_k(t') = \frac{1}{2k} e^{-ik(t-t')},
\end{equation}
and the second is the Feynman Green’s function,
\begin{equation}
    G_F(k|t,t') = W_k(t,t') \Theta(t-t') + W_k(t',t) \Theta(t'-t)
    = \frac{1}{2k}\!\left(
    e^{-ik(t-t')}\Theta(t-t')
    + e^{-ik(t'-t)}\Theta(t'-t)
    \right).
\end{equation}
Taking the complex conjugate of the Wightman function is equivalent to time reversal:
\begin{align}
    W_k^*(t,t') &= W_k(t',t).
\end{align}
This property manifests itself as a reflection symmetry about the late-time surface.

Using these Feynman rules, the time integrals that compute the bubble loop integrands corresponding to the diagrams shown in Fig.~\ref{fig:Bubble_InIn} can be written as
\begin{align}
    R_{++} &= \frac{(-ig)^2}{\prod_i 2k_i}
    \int_{-\infty}^0 \!\d t \int_{-\infty}^0 \!\d t'\,
    W_{X_1}(0,t) W_{X_2}(0,t')\,
    G^{(F)}_{q_1}(t,t') G^{(F)}_{q_2}(t,t'), \\
    R_{+-} &= \frac{(-ig)(ig)}{\prod_i 2k_i}
    \int_{-\infty}^0 \!\d t \int_{-\infty}^0 \!\d t'\,
    W_{X_1}(0,t') W_{X_2}(t,0)\,
    W_{q_1}(t,t') W_{q_2}(t,t').
\end{align}
Reflection symmetry implies $R_{++}=R_{--}$ and $R_{+-}=R_{-+}$.  
Notice that both contributions carry the same overall sign: the time integrals in $R_{++}$ produce an additional minus sign, so that both terms scale as $+g^2 = (-ig)(ig)$.

We now turn to the triangle diagram. The $D_3$ symmetry allows us to restrict attention to four in-in sectors,
$R_{+++}, R_{---}, R_{+-+}$ and $R_{-+-}$, of which only two are independent by reflection symmetry:
\[
R_{---} = (R_{+++})^*, \qquad
R_{-+-} = (R_{+-+})^*.
\]

The time integrals corresponding to the $+++$ sector involve a triple product of Feynman propagators,
\begin{align}
    R_C^{(+++)} &= (ig)^3
    \int_{-\infty}^0 \!\d t_1
    \int_{-\infty}^0 \!\d t_2
    \int_{-\infty}^0 \!\d t_3\,
    e^{i(X_1 t_1 + X_2 t_2 + X_3 t_3)}
    \frac{1}{8 q_1 q_2 q_3}
    \times \nonumber \\
    &\qquad\qquad\qquad
    \left(
        e^{-iq_1(t_2-t_3)}\Theta(t_2-t_3)
        + e^{-iq_1(t_3-t_2)}\Theta(t_3-t_2)
    \right)\times \nonumber \\
    &\qquad\qquad\qquad\quad 
    \left(
        e^{-iq_2(t_1-t_3)}\Theta(t_1-t_3)
        + e^{-iq_2(t_3-t_1)}\Theta(t_3-t_1)
    \right)\times \nonumber \\
    &\qquad\qquad\qquad\quad \quad
    \left(
        e^{-iq_3(t_1-t_2)}\Theta(t_1-t_2)
        + e^{-iq_3(t_2-t_1)}\Theta(t_2-t_1)
    \right).
\end{align}
This expression can be expanded into six terms, each corresponding to a distinct time ordering. Each term can be put in one-to-one correspondence with a time-ordered graph. Using the tubing rules for time-ordered graphs, the associated rational function can be derived directly at the graphical level.

By contrast, the time integrals for the $+-+$ in-in sector contain only a single Feynman propagator, or equivalently, time-ordered edge:
\begin{align}
    R_C^{(+-+)} &= (-ig)(ig)^2
    \int_{-\infty}^0 \!\d t_1
    \int_{-\infty}^0 \!\d t_2
    \int_{-\infty}^0 \!\d t_3\,
    e^{i(X_1 t_1 + X_2 t_2 + X_3 t_3)}
    \frac{1}{8 q_1 q_2 q_3}
    \times \nonumber \\
    &\quad
    e^{-iq_1(t_2-t_3)} e^{-iq_3(t_2-t_1)}
    \left(
        e^{-iq_2(t_1-t_3)}\Theta(t_1-t_3)
        + e^{-iq_2(t_3-t_1)}\Theta(t_3-t_1)
    \right).
\end{align}
These integrals can be interpreted as the sum of two time-ordered graphs. The two graphs are related by a $D_3$ reflection about the second vertex, corresponding to the exchange $1 \leftrightarrow 3$.


\clearpage
\phantomsection
\addcontentsline{toc}{section}{References}
\bibliographystyle{utphys}
{\linespread{1.075}
\bibliography{Refs}

@misc{Notebook,
title = {Triangle Correlator},
author = {Guilherme L. Pimentel and Tom Westerdijk},
howpublished = {Mathematica notebook available at \url{https://github.com/TomWesterdijk/Cosmology-at-One-Loop}}
}

@article{Guth:1980zm,
    author = "Guth, Alan H.",
    editor = "Fang, Li-Zhi and Ruffini, R.",
    title = "{The Inflationary Universe: A Possible Solution to the Horizon and Flatness Problems}",
    reportNumber = "SLAC-PUB-2576",
    doi = "10.1103/PhysRevD.23.347",
    journal = "Phys. Rev. D",
    volume = "23",
    pages = "347--356",
    year = "1981"
}

@article{Linde:1981mu,
    author = "Linde, Andrei D.",
    editor = "Fang, Li-Zhi and Ruffini, R.",
    title = "{A New Inflationary Universe Scenario: A Possible Solution of the Horizon, Flatness, Homogeneity, Isotropy and Primordial Monopole Problems}",
    reportNumber = "LEBEDEV-81-229",
    doi = "10.1016/0370-2693(82)91219-9",
    journal = "Phys. Lett. B",
    volume = "108",
    pages = "389--393",
    year = "1982"
}

@article{Mukhanov:1981xt,
    author = "Mukhanov, Viatcheslav F. and Chibisov, G. V.",
    title = "{Quantum Fluctuations and a Nonsingular Universe}",
    journal = "JETP Lett.",
    volume = "33",
    pages = "532--535",
    year = "1981"
}

@article{Guth:1982ec,
    author = "Guth, Alan H. and Pi, S. Y.",
    title = "{Fluctuations in the New Inflationary Universe}",
    doi = "10.1103/PhysRevLett.49.1110",
    journal = "Phys. Rev. Lett.",
    volume = "49",
    pages = "1110--1113",
    year = "1982"
}

@article{Starobinsky:1982ee,
    author = "Starobinsky, Alexei A.",
    title = "{Dynamics of Phase Transition in the New Inflationary Universe Scenario and Generation of Perturbations}",
    doi = "10.1016/0370-2693(82)90541-X",
    journal = "Phys. Lett. B",
    volume = "117",
    pages = "175--178",
    year = "1982"
}

@article{Bardeen:1983qw,
    author = "Bardeen, James M. and Steinhardt, Paul J. and Turner, Michael S.",
    title = "{Spontaneous Creation of Almost Scale - Free Density Perturbations in an Inflationary Universe}",
    reportNumber = "UPR-0202T, EFI-83-13-CHICAGO",
    doi = "10.1103/PhysRevD.28.679",
    journal = "Phys. Rev. D",
    volume = "28",
    pages = "679",
    year = "1983"
}

@article{Hawking:1982cz,
    author = "Hawking, S. W.",
    title = "{The Development of Irregularities in a Single Bubble Inflationary Universe}",
    reportNumber = "Print-83-0015 (CAMBRIDGE)",
    doi = "10.1016/0370-2693(82)90373-2",
    journal = "Phys. Lett. B",
    volume = "115",
    pages = "295",
    year = "1982"
}

@misc{BaumannJoyce,
      title={Cosmological Correlations}, 
      author={Baumann, Daniel and Joyce, Austin},
      month={July},
      year={2024},
      url={https://github.com/ddbaumann/cosmo-correlators}, 
}

@misc{goncharov2000,
      title={Multiple zeta-values, Galois groups, and geometry of modular varieties}, 
      author={A. B. Goncharov},
      year={2000},
      eprint={math/0005069},
      archivePrefix={arXiv},
      primaryClass={math.AG},
      url={https://arxiv.org/abs/math/0005069}, 
}

@inproceedings{Duhr:2014woa,
    author = "Duhr, Claude",
    title = "{Mathematical aspects of scattering amplitudes}",
    booktitle = "{Theoretical Advanced Study Institute in Elementary Particle Physics}: {Journeys Through the Precision Frontier: Amplitudes for Colliders}",
    eprint = "1411.7538",
    archivePrefix = "arXiv",
    primaryClass = "hep-ph",
    reportNumber = "CP3-14-70",
    doi = "10.1142/9789814678766_0010",
    pages = "419--476",
    year = "2015"
}

@misc{Brown2009,
      title={Iterated integrals in Quantum Field Theory}, 
      author={Brown, Francis C. S.},
      year={2009},
      publisher={IHES},
      url={https://www.ihes.fr/~brown/ColombiaNotes7.pdf}, 
}

@article{Matsubara-Heo:2023ylc,
    author = "Matsubara-Heo, Saiei-Jaeyeong and Mizera, Sebastian and Telen, Simon",
    title = "{Four lectures on Euler integrals}",
    eprint = "2306.13578",
    archivePrefix = "arXiv",
    primaryClass = "math-ph",
    doi = "10.21468/SciPostPhysLectNotes.75",
    journal = "SciPost Phys. Lect. Notes",
    volume = "75",
    pages = "1",
    year = "2023"
}

@book{Sommerville1958,
  author    = {Sommerville, Duncan M. Y.},
  title     = {An Introduction to the Geometry of n Dimensions},
  publisher = {Dover Publications},
  year      = {1958},
  address   = {New York}
}

@article{Whittaker1902,
    author = {Whittaker, E. T.},
    title = {On the General Solution of Laplace's Equation and the Equation of Wave Motions, and on an undulatory explanation of Gravity},
    journal = {Monthly Notices of the Royal Astronomical Society},
    volume = {62},
    number = {9},
    pages = {617-620},
    year = {1902},
    month = {07},
    issn = {0035-8711},
    doi = {10.1093/mnras/62.9.617},
    url = {https://doi.org/10.1093/mnras/62.9.617},
    eprint = {https://academic.oup.com/mnras/article-pdf/62/9/617/3316471/mnras62-0617.pdf},
}

@misc{Reinke2024,
      title={Hypersurface Arrangements with Generic Hypersurfaces Added}, 
      author={Bernhard Reinke and Kexin Wang},
      year={2024},
      eprint={2412.20869},
      archivePrefix={arXiv},
      primaryClass={math.AG},
      url={https://arxiv.org/abs/2412.20869}, 
}

@book{Weinzierl:2022eaz,
    author = "Weinzierl, Stefan",
    title = "{Feynman Integrals. A Comprehensive Treatment for Students and Researchers}",
    eprint = "2201.03593",
    archivePrefix = "arXiv",
    primaryClass = "hep-th",
    reportNumber = "MITP/22-001",
    doi = "10.1007/978-3-030-99558-4",
    isbn = "978-3-030-99557-7, 978-3-030-99560-7, 978-3-030-99558-4",
    publisher = "Springer",
    series = "UNITEXT for Physics",
    year = "2022"
}

@book{Eden:1966dnq,
    author = "Eden, Richard John and Landshoff, Peter V. and Olive, David I. and Polkinghorne, John Charlton",
    title = "{The analytic S-matrix}",
    isbn = "978-0-521-04869-9",
    publisher = "Cambridge Univ. Press",
    address = "Cambridge",
    year = "1966"
}

@article{Mizera:2023tfe,
    author = "Mizera, Sebastian",
    title = "{Physics of the analytic S-matrix}",
    eprint = "2306.05395",
    archivePrefix = "arXiv",
    primaryClass = "hep-th",
    doi = "10.1016/j.physrep.2023.10.006",
    journal = "Phys. Rept.",
    volume = "1047",
    pages = "1--92",
    year = "2024"
}

@article{Coleman:1965xm,
    author = "Coleman, S. and Norton, R. E.",
    title = "{Singularities in the physical region}",
    doi = "10.1007/BF02750472",
    journal = "Nuovo Cim.",
    volume = "38",
    pages = "438--442",
    year = "1965"
}

@article{Cutkosky:1960sp,
    author = "Cutkosky, R. E.",
    title = "{Singularities and discontinuities of Feynman amplitudes}",
    doi = "10.1063/1.1703676",
    journal = "J. Math. Phys.",
    volume = "1",
    pages = "429--433",
    year = "1960"
}

@incollection{Landau1965,
title = {98 - ON ANALYTIC PROPERTIES OF VERTEX PARTS IN QUANTUM FIELD THEORY},
author = {Landau, L. D.},
editor = {D. {TER HAAR}},
booktitle = {Collected Papers of L.D. Landau},
publisher = {Pergamon},
pages = {787-797},
year = {1965},
isbn = {978-0-08-010586-4},
doi = {10.1016/B978-0-08-010586-4.50103-6},
url = {https://www.sciencedirect.com/science/article/pii/B9780080105864501036}
}

@article{Fairlie2ndType,
    author = {Fairlie, D. B. and Landshoff, P. V. and Nuttall, J. and Polkinghorne, J. C.},
    title = {Singularities of the Second Type},
    journal = {Journal of Mathematical Physics},
    volume = {3},
    number = {4},
    pages = {594-602},
    year = {1962},
    month = {07},
    issn = {0022-2488},
    doi = {10.1063/1.1724262},
    url = {https://doi.org/10.1063/1.1724262},
    eprint = {https://pubs.aip.org/aip/jmp/article-pdf/3/4/594/19167404/594\_1\_online.pdf},
}

@article{Abreu:2017ptx,
    author = "Abreu, Samuel and Britto, Ruth and Duhr, Claude and Gardi, Einan",
    title = "{Cuts from residues: the one-loop case}",
    eprint = "1702.03163",
    archivePrefix = "arXiv",
    primaryClass = "hep-th",
    reportNumber = "CERN-TH-2017-033, CP3-17-05, Edinburgh-2017-05, FR-PHENO-2017-001",
    doi = "10.1007/JHEP06(2017)114",
    journal = "JHEP",
    volume = "06",
    pages = "114",
    year = "2017"
}

@article{Hannesdottir:2022xki,
    author = "Hannesdottir, Holmfridur S. and McLeod, Andrew J. and Schwartz, Matthew D. and Vergu, Cristian",
    title = "{Constraints on sequential discontinuities from the geometry of on-shell spaces}",
    eprint = "2211.07633",
    archivePrefix = "arXiv",
    primaryClass = "hep-th",
    reportNumber = "CERN-TH-2022-189",
    doi = "10.1007/JHEP07(2023)236",
    journal = "JHEP",
    volume = "07",
    pages = "236",
    year = "2023"
}

@inbook{Britto:2024mna,
   title={Cutting-Edge Tools for Cutting Edges},
   ISBN={9780323957069},
   url={http://dx.doi.org/10.1016/B978-0-323-95703-8.00097-5},
   DOI={10.1016/b978-0-323-95703-8.00097-5},
   booktitle={Encyclopedia of Mathematical Physics},
   publisher={Elsevier},
   author={Britto, Ruth and Duhr, Claude and Hannesdottir, Holmfridur S. and Mizera, Sebastian},
   year={2025},
   pages={595–620} 
}

@article{Fevola:2023fzn,
    author = "Fevola, Claudia and Mizera, Sebastian and Telen, Simon",
    title = "{Principal Landau determinants}",
    eprint = "2311.16219",
    archivePrefix = "arXiv",
    primaryClass = "math-ph",
    doi = "10.1016/j.cpc.2024.109278",
    journal = "Comput. Phys. Commun.",
    volume = "303",
    pages = "109278",
    year = "2024"
}

@article{Fevola:2023kaw,
    author = "Fevola, Claudia and Mizera, Sebastian and Telen, Simon",
    title = "{Landau Singularities Revisited: Computational Algebraic Geometry for Feynman Integrals}",
    eprint = "2311.14669",
    archivePrefix = "arXiv",
    primaryClass = "hep-th",
    doi = "10.1103/PhysRevLett.132.101601",
    journal = "Phys. Rev. Lett.",
    volume = "132",
    number = "10",
    pages = "101601",
    year = "2024"
}

@article{Melrose:1965kb,
    author = "Melrose, D. B.",
    title = "{Reduction of Feynman diagrams}",
    doi = "10.1007/BF02832919",
    journal = "Nuovo Cim.",
    volume = "40",
    pages = "181--213",
    year = "1965"
}

@article{vanNeerven:1983vr,
    author = "van Neerven, W. L. and Vermaseren, J. A. M.",
    title = "{LARGE LOOP INTEGRALS}",
    reportNumber = "NIKHEF-H/83-22",
    doi = "10.1016/0370-2693(84)90237-5",
    journal = "Phys. Lett. B",
    volume = "137",
    pages = "241--244",
    year = "1984"
}

@article{Bern:1993kr,
    author = "Bern, Zvi and Dixon, Lance J. and Kosower, David A.",
    title = "{Dimensionally regulated pentagon integrals}",
    eprint = "hep-ph/9306240",
    archivePrefix = "arXiv",
    reportNumber = "SLAC-PUB-5947, SACLAY-SPH-T-92-048, UCLA-92-043",
    doi = "10.1016/0550-3213(94)90398-0",
    journal = "Nucl. Phys. B",
    volume = "412",
    pages = "751--816",
    year = "1994"
}

@article{Binoth:1999sp,
    author = "Binoth, T. and Guillet, J. P. and Heinrich, G.",
    title = "{Reduction formalism for dimensionally regulated one loop N point integrals}",
    eprint = "hep-ph/9911342",
    archivePrefix = "arXiv",
    reportNumber = "LAPTH-759-99",
    doi = "10.1016/S0550-3213(00)00040-7",
    journal = "Nucl. Phys. B",
    volume = "572",
    pages = "361--386",
    year = "2000"
}

@article{Fleischer:1999hq,
    author = "Fleischer, J. and Jegerlehner, F. and Tarasov, O. V.",
    title = "{Algebraic reduction of one loop Feynman graph amplitudes}",
    eprint = "hep-ph/9907327",
    archivePrefix = "arXiv",
    reportNumber = "BUTP-99-11, BI-TP-99-08, DESY-99-086",
    doi = "10.1016/S0550-3213(99)00678-1",
    journal = "Nucl. Phys. B",
    volume = "566",
    pages = "423--440",
    year = "2000"
}

@article{Denner:2002ii,
    author = "Denner, Ansgar and Dittmaier, S.",
    title = "{Reduction of one loop tensor five point integrals}",
    eprint = "hep-ph/0212259",
    archivePrefix = "arXiv",
    reportNumber = "MPI-PHT-2002-63, PSI-PR-02-21",
    doi = "10.1016/S0550-3213(03)00184-6",
    journal = "Nucl. Phys. B",
    volume = "658",
    pages = "175--202",
    year = "2003"
}

@article{Duplancic:2003tv,
    author = "Duplancic, G. and Nizic, B.",
    title = "{Reduction method for dimensionally regulated one loop N point Feynman integrals}",
    eprint = "hep-ph/0303184",
    archivePrefix = "arXiv",
    reportNumber = "IRB-TH-2-03",
    doi = "10.1140/epjc/s2004-01723-7",
    journal = "Eur. Phys. J. C",
    volume = "35",
    pages = "105--118",
    year = "2004"
}

@article{Binoth:2005ff,
    author = "Binoth, T. and Guillet, J. Ph. and Heinrich, G. and Pilon, E. and Schubert, C.",
    title = "{An Algebraic/numerical formalism for one-loop multi-leg amplitudes}",
    eprint = "hep-ph/0504267",
    archivePrefix = "arXiv",
    reportNumber = "LAPTH-1097-05, WUE-ITP-2005-003, ZU-TH-08-05",
    doi = "10.1088/1126-6708/2005/10/015",
    journal = "JHEP",
    volume = "10",
    pages = "015",
    year = "2005"
}

@article{Weinberg:2005vy,
    author = "Weinberg, Steven",
    title = "{Quantum contributions to cosmological correlations}",
    eprint = "hep-th/0506236",
    archivePrefix = "arXiv",
    reportNumber = "UTTG-01-05",
    doi = "10.1103/PhysRevD.72.043514",
    journal = "Phys. Rev. D",
    volume = "72",
    pages = "043514",
    year = "2005"
}

@article{Senatore:2009cf,
    author = "Senatore, Leonardo and Zaldarriaga, Matias",
    title = "{On Loops in Inflation}",
    eprint = "0912.2734",
    archivePrefix = "arXiv",
    primaryClass = "hep-th",
    doi = "10.1007/JHEP12(2010)008",
    journal = "JHEP",
    volume = "12",
    pages = "008",
    year = "2010"
}

@article{Westerdijk:2025ywh,
    author = "Westerdijk, Tom and Yang, Chen",
    title = "{Bananas are unparticles: differential equations and cosmological bootstrap}",
    eprint = "2503.08775",
    archivePrefix = "arXiv",
    primaryClass = "hep-th",
    doi = "10.1007/JHEP09(2025)089",
    journal = "JHEP",
    volume = "09",
    pages = "089",
    year = "2025"
}

@article{Beneke:2023wmt,
    author = "Beneke, Martin and Hager, Patrick and Sanfilippo, Andrea F.",
    title = "{Cosmological correlators in massless {\ensuremath{\phi}}$^{4}$-theory and the method of regions}",
    eprint = "2312.06766",
    archivePrefix = "arXiv",
    primaryClass = "hep-th",
    reportNumber = "TUM-HEP-1485/23, MITP-23-073",
    doi = "10.1007/JHEP04(2024)006",
    journal = "JHEP",
    volume = "04",
    pages = "006",
    year = "2024"
}

@article{delRio:2018vrj,
    author = "del Rio, Adri{\'a}n and Durrer, Ruth and Patil, Subodh P.",
    title = "{Tensor Bounds on the Hidden Universe}",
    eprint = "1808.09282",
    archivePrefix = "arXiv",
    primaryClass = "gr-qc",
    doi = "10.1007/JHEP12(2018)094",
    journal = "JHEP",
    volume = "12",
    pages = "094",
    year = "2018"
}

@article{Chowdhury:2023khl,
    author = "Chowdhury, Chandramouli and Singh, Kajal",
    title = "{Analytic results for loop-level momentum space Witten diagrams}",
    eprint = "2305.18529",
    archivePrefix = "arXiv",
    primaryClass = "hep-th",
    doi = "10.1007/JHEP12(2023)109",
    journal = "JHEP",
    volume = "12",
    pages = "109",
    year = "2023"
}

@article{Chowdhury:2023arc,
    author = "Chowdhury, Chandramouli and Lipstein, Arthur and Mei, Jiajie and Sachs, Ivo and Vanhove, Pierre",
    title = "{The subtle simplicity of cosmological correlators}",
    eprint = "2312.13803",
    archivePrefix = "arXiv",
    primaryClass = "hep-th",
    reportNumber = "LMU-ASC 37/23, IPhT-t23/119",
    doi = "10.1007/JHEP03(2025)007",
    journal = "JHEP",
    volume = "03",
    pages = "007",
    year = "2025"
}

@article{Chowdhury:2025ohm,
    author = "Chowdhury, Chandramouli and Lipstein, Arthur and Marshall, Joe and Mei, Jiajie and Sachs, Ivo",
    title = "{Cosmological dressing rules}",
    eprint = "2503.10598",
    archivePrefix = "arXiv",
    primaryClass = "hep-th",
    reportNumber = "LMU-ASC 03/25",
    doi = "10.1007/JHEP03(2026)076",
    journal = "JHEP",
    volume = "03",
    pages = "076",
    year = "2026"
}

@article{Jain:2025maa,
    author = "Jain, Diksha and Pajer, Enrico and Tong, Xi",
    title = "{Unitary and Analytic Renormalisation of Cosmological Correlators}",
    eprint = "2509.02696",
    archivePrefix = "arXiv",
    primaryClass = "hep-th",
    month = "9",
    year = "2025"
}

@article{Lee:2023jby,
    author = "Lee, Mang Hei Gordon and McCulloch, Ciaran and Pajer, Enrico",
    title = "{Leading loops in cosmological correlators}",
    eprint = "2305.11228",
    archivePrefix = "arXiv",
    primaryClass = "hep-th",
    doi = "10.1007/JHEP11(2023)038",
    journal = "JHEP",
    volume = "11",
    pages = "038",
    year = "2023"
}

@article{Salcedo:2022aal,
    author = "Salcedo, Santiago Agui and Lee, Mang Hei Gordon and Melville, Scott and Pajer, Enrico",
    title = "{The Analytic Wavefunction}",
    eprint = "2212.08009",
    archivePrefix = "arXiv",
    primaryClass = "hep-th",
    doi = "10.1007/JHEP06(2023)020",
    journal = "JHEP",
    volume = "06",
    pages = "020",
    year = "2023"
}

@article{Cacciatori:2023tzp,
    author = "Cacciatori, Sergio Luigi and Epstein, Henri and Moschella, Ugo",
    title = "{Banana integrals in configuration space}",
    eprint = "2304.00624",
    archivePrefix = "arXiv",
    primaryClass = "hep-th",
    doi = "10.1016/j.nuclphysb.2023.116343",
    journal = "Nucl. Phys. B",
    volume = "995",
    pages = "116343",
    year = "2023"
}

@article{Cacciatori:2024zrv,
    author = "Cacciatori, Sergio L. and Epstein, Henri and Moschella, Ugo",
    title = "{Loops in de Sitter space}",
    eprint = "2403.13145",
    archivePrefix = "arXiv",
    primaryClass = "hep-th",
    doi = "10.1007/JHEP07(2024)182",
    journal = "JHEP",
    volume = "07",
    pages = "182",
    year = "2024"
}

@article{Qin:2024gtr,
    author = "Qin, Zhehan",
    title = "{Cosmological correlators at the loop level}",
    eprint = "2411.13636",
    archivePrefix = "arXiv",
    primaryClass = "hep-th",
    doi = "10.1007/JHEP03(2025)051",
    journal = "JHEP",
    volume = "03",
    pages = "051",
    year = "2025"
}

@article{Zhang:2025nzd,
    author = "Zhang, Hongyu",
    title = "{Dimensional Regularization of Bubble Diagrams in de Sitter Spacetime}",
    eprint = "2507.19318",
    archivePrefix = "arXiv",
    primaryClass = "hep-th",
    month = "7",
    year = "2025"
}

@article{Loparco:2023rug,
    author = "Loparco, Manuel and Penedones, Joao and Salehi Vaziri, Kamran and Sun, Zimo",
    title = {{The K{\"a}ll{\'e}n-Lehmann representation in de Sitter spacetime}},
    eprint = "2306.00090",
    archivePrefix = "arXiv",
    primaryClass = "hep-th",
    doi = "10.1007/JHEP12(2023)159",
    journal = "JHEP",
    volume = "12",
    pages = "159",
    year = "2023"
}

@article{Qin:2023nhv,
    author = "Qin, Zhehan and Xianyu, Zhong-Zhi",
    title = "{Nonanalyticity and on-shell factorization of inflation correlators at all loop orders}",
    eprint = "2308.14802",
    archivePrefix = "arXiv",
    primaryClass = "hep-th",
    doi = "10.1007/JHEP01(2024)168",
    journal = "JHEP",
    volume = "01",
    pages = "168",
    year = "2024"
}

@article{Qin:2023bjk,
    author = "Qin, Zhehan and Xianyu, Zhong-Zhi",
    title = "{Inflation correlators at the one-loop order: nonanalyticity, factorization, cutting rule, and OPE}",
    eprint = "2304.13295",
    archivePrefix = "arXiv",
    primaryClass = "hep-th",
    doi = "10.1007/JHEP09(2023)116",
    journal = "JHEP",
    volume = "09",
    pages = "116",
    year = "2023"
}

@article{Baumann:2025qjx,
    author = "Baumann, Daniel and Goodhew, Harry and Joyce, Austin and Lee, Hayden and Pimentel, Guilherme L. and Westerdijk, Tom",
    title = "{Geometry of Kinematic Flow}",
    eprint = "2504.14890",
    archivePrefix = "arXiv",
    primaryClass = "hep-th",
    month = "4",
    year = "2025"
}

@article{Baumann:2024mvm,
    author = "Baumann, Daniel and Goodhew, Harry and Lee, Hayden",
    title = "{Kinematic flow for cosmological loop integrands}",
    eprint = "2410.17994",
    archivePrefix = "arXiv",
    primaryClass = "hep-th",
    doi = "10.1007/JHEP07(2025)131",
    journal = "JHEP",
    volume = "07",
    pages = "131",
    year = "2025"
}

@article{Arkani-Hamed:2023bsv,
    author = "Arkani-Hamed, Nima and Baumann, Daniel and Hillman, Aaron and Joyce, Austin and Lee, Hayden and Pimentel, Guilherme L.",
    title = "{Kinematic Flow and the Emergence of Time}",
    eprint = "2312.05300",
    archivePrefix = "arXiv",
    primaryClass = "hep-th",
    doi = "10.1103/dsjm-tckw",
    journal = "Phys. Rev. Lett.",
    volume = "135",
    number = "3",
    pages = "031602",
    year = "2025"
}

@article{Arkani-Hamed:2023kig,
    author = "Arkani-Hamed, Nima and Baumann, Daniel and Hillman, Aaron and Joyce, Austin and Lee, Hayden and Pimentel, Guilherme L.",
    title = "{Differential equations for cosmological correlators}",
    eprint = "2312.05303",
    archivePrefix = "arXiv",
    primaryClass = "hep-th",
    doi = "10.1007/JHEP09(2025)009",
    journal = "JHEP",
    volume = "09",
    pages = "009",
    year = "2025"
}

@article{Glew:2025ypb,
    author = "Glew, Ross and Pokraka, Andrzej",
    title = "{Kinematic flow from the flow of cuts}",
    eprint = "2508.11568",
    archivePrefix = "arXiv",
    primaryClass = "hep-th",
    month = "8",
    year = "2025"
}

@article{Hang:2024xas,
    author = "Hang, Yanfeng and Shen, Cong",
    title = "{A note on kinematic flow and differential equations for two-site one-loop graph in FRW spacetime}",
    eprint = "2410.17192",
    archivePrefix = "arXiv",
    primaryClass = "hep-th",
    doi = "10.1007/JHEP09(2025)209",
    journal = "JHEP",
    volume = "09",
    pages = "209",
    year = "2025"
}

@article{Chen:2023iix,
    author = "Chen, Jiaqi and Feng, Bo",
    title = "{Towards systematic evaluation of de Sitter correlators via Generalized Integration-By-Parts relations}",
    eprint = "2401.00129",
    archivePrefix = "arXiv",
    primaryClass = "hep-th",
    doi = "10.1007/JHEP06(2024)199",
    journal = "JHEP",
    volume = "06",
    pages = "199",
    year = "2024"
}

@article{Chen:2024glu,
    author = "Chen, Jiaqi and Feng, Bo and Tao, Yi-Xiao",
    title = "{Multivariate hypergeometric solutions of cosmological (dS) correlators by d log-form differential equations}",
    eprint = "2411.03088",
    archivePrefix = "arXiv",
    primaryClass = "hep-th",
    doi = "10.1007/JHEP03(2025)075",
    journal = "JHEP",
    volume = "03",
    pages = "075",
    year = "2025"
}

@article{Mata:2012bx,
    author = "Mata, Ishan and Raju, Suvrat and Trivedi, Sandip",
    title = "{CMB from CFT}",
    eprint = "1211.5482",
    archivePrefix = "arXiv",
    primaryClass = "hep-th",
    reportNumber = "TIFR-TH-12-38, HRI-ST-1210, ICTS-2012-11",
    doi = "10.1007/JHEP07(2013)015",
    journal = "JHEP",
    volume = "07",
    pages = "015",
    year = "2013"
}

@article{Bzowski:2011ab,
    author = "Bzowski, Adam and McFadden, Paul and Skenderis, Kostas",
    title = "{Holographic predictions for cosmological 3-point functions}",
    eprint = "1112.1967",
    archivePrefix = "arXiv",
    primaryClass = "hep-th",
    doi = "10.1007/JHEP03(2012)091",
    journal = "JHEP",
    volume = "03",
    pages = "091",
    year = "2012"
}

@article{Bzowski:2012ih,
    author = "Bzowski, Adam and McFadden, Paul and Skenderis, Kostas",
    title = "{Holography for inflation using conformal perturbation theory}",
    eprint = "1211.4550",
    archivePrefix = "arXiv",
    primaryClass = "hep-th",
    doi = "10.1007/JHEP04(2013)047",
    journal = "JHEP",
    volume = "04",
    pages = "047",
    year = "2013"
}

@article{Bzowski:2013sza,
    author = "Bzowski, Adam and McFadden, Paul and Skenderis, Kostas",
    title = "{Implications of conformal invariance in momentum space}",
    eprint = "1304.7760",
    archivePrefix = "arXiv",
    primaryClass = "hep-th",
    doi = "10.1007/JHEP03(2014)111",
    journal = "JHEP",
    volume = "03",
    pages = "111",
    year = "2014"
}

@article{Bzowski:2019kwd,
    author = "Bzowski, Adam and McFadden, Paul and Skenderis, Kostas",
    title = "{Conformal $n$-point functions in momentum space}",
    eprint = "1910.10162",
    archivePrefix = "arXiv",
    primaryClass = "hep-th",
    doi = "10.1103/PhysRevLett.124.131602",
    journal = "Phys. Rev. Lett.",
    volume = "124",
    number = "13",
    pages = "131602",
    year = "2020"
}

@article{Kundu:2014gxa,
    author = "Kundu, Nilay and Shukla, Ashish and Trivedi, Sandip P.",
    title = "{Constraints from Conformal Symmetry on the Three Point Scalar Correlator in Inflation}",
    eprint = "1410.2606",
    archivePrefix = "arXiv",
    primaryClass = "hep-th",
    reportNumber = "TIFR-TH-14-24, HRI-ST-1414",
    doi = "10.1007/JHEP04(2015)061",
    journal = "JHEP",
    volume = "04",
    pages = "061",
    year = "2015"
}

@article{Kundu:2015xta,
    author = "Kundu, Nilay and Shukla, Ashish and Trivedi, Sandip P.",
    title = "{Ward Identities for Scale and Special Conformal Transformations in Inflation}",
    eprint = "1507.06017",
    archivePrefix = "arXiv",
    primaryClass = "hep-th",
    reportNumber = "TIFR-TH-15-21, HRI-ST-1509",
    doi = "10.1007/JHEP01(2016)046",
    journal = "JHEP",
    volume = "01",
    pages = "046",
    year = "2016"
}

@article{Shukla:2016bnu,
    author = "Shukla, Ashish and Trivedi, Sandip P. and Vishal, V.",
    title = "{Symmetry constraints in inflation, $\alpha$-vacua, and the three point function}",
    eprint = "1607.08636",
    archivePrefix = "arXiv",
    primaryClass = "hep-th",
    reportNumber = "TIFR-TH-16-26",
    doi = "10.1007/JHEP12(2016)102",
    journal = "JHEP",
    volume = "12",
    pages = "102",
    year = "2016"
}

@article{Baumann:2021fxj,
    author = "Baumann, Daniel and Chen, Wei-Ming and Duaso Pueyo, Carlos and Joyce, Austin and Lee, Hayden and Pimentel, Guilherme L.",
    title = "{Linking the singularities of cosmological correlators}",
    eprint = "2106.05294",
    archivePrefix = "arXiv",
    primaryClass = "hep-th",
    doi = "10.1007/JHEP09(2022)010",
    journal = "JHEP",
    volume = "09",
    pages = "010",
    year = "2022"
}

@article{Meltzer:2021zin,
    author = "Meltzer, David",
    title = "{The inflationary wavefunction from analyticity and factorization}",
    eprint = "2107.10266",
    archivePrefix = "arXiv",
    primaryClass = "hep-th",
    reportNumber = "CALT-TH-2021-028",
    doi = "10.1088/1475-7516/2021/12/018",
    journal = "JCAP",
    volume = "12",
    number = "12",
    pages = "018",
    year = "2021"
}

@article{Hogervorst:2021uvp,
    author = "Hogervorst, Matthijs and Penedones, Jo{\~a}o and Vaziri, Kamran Salehi",
    title = "{Towards the non-perturbative cosmological bootstrap}",
    eprint = "2107.13871",
    archivePrefix = "arXiv",
    primaryClass = "hep-th",
    doi = "10.1007/JHEP02(2023)162",
    journal = "JHEP",
    volume = "02",
    pages = "162",
    year = "2023"
}

@article{Goodhew:2021oqg,
    author = "Goodhew, Harry and Jazayeri, Sadra and Lee, Mang Hei Gordon and Pajer, Enrico",
    title = "{Cutting cosmological correlators}",
    eprint = "2104.06587",
    archivePrefix = "arXiv",
    primaryClass = "hep-th",
    doi = "10.1088/1475-7516/2021/08/003",
    journal = "JCAP",
    volume = "08",
    pages = "003",
    year = "2021"
}

@article{Melville:2021lst,
    author = "Melville, Scott and Pajer, Enrico",
    title = "{Cosmological Cutting Rules}",
    eprint = "2103.09832",
    archivePrefix = "arXiv",
    primaryClass = "hep-th",
    doi = "10.1007/JHEP05(2021)249",
    journal = "JHEP",
    volume = "05",
    pages = "249",
    year = "2021"
}

@article{Jazayeri:2021fvk,
    author = "Jazayeri, Sadra and Pajer, Enrico and Stefanyszyn, David",
    title = "{From locality and unitarity to cosmological correlators}",
    eprint = "2103.08649",
    archivePrefix = "arXiv",
    primaryClass = "hep-th",
    doi = "10.1007/JHEP10(2021)065",
    journal = "JHEP",
    volume = "10",
    pages = "065",
    year = "2021"
}

@article{Goodhew:2020hob,
    author = "Goodhew, Harry and Jazayeri, Sadra and Pajer, Enrico",
    title = "{The Cosmological Optical Theorem}",
    eprint = "2009.02898",
    archivePrefix = "arXiv",
    primaryClass = "hep-th",
    doi = "10.1088/1475-7516/2021/04/021",
    journal = "JCAP",
    volume = "04",
    pages = "021",
    year = "2021"
}

@article{Goodhew:2024eup,
    author = "Goodhew, Harry and Thavanesan, Ayngaran and Wall, Aron C.",
    title = "{The Cosmological CPT Theorem}",
    eprint = "2408.17406",
    archivePrefix = "arXiv",
    primaryClass = "hep-th",
    month = "8",
    year = "2024"
}

@article{Thavanesan:2025kyc,
    author = "Thavanesan, Ayngaran",
    title = "{No-go Theorem for Cosmological Parity Violation}",
    eprint = "2501.06383",
    archivePrefix = "arXiv",
    primaryClass = "hep-th",
    month = "1",
    year = "2025"
}

@article{Arkani-Hamed:2018kmz,
    author = "Arkani-Hamed, Nima and Baumann, Daniel and Lee, Hayden and Pimentel, Guilherme L.",
    title = "{The Cosmological Bootstrap: Inflationary Correlators from Symmetries and Singularities}",
    eprint = "1811.00024",
    archivePrefix = "arXiv",
    primaryClass = "hep-th",
    doi = "10.1007/JHEP04(2020)105",
    journal = "JHEP",
    volume = "04",
    pages = "105",
    year = "2020"
}

@article{Baumann:2022jpr,
    author = "Baumann, Daniel and Green, Daniel and Joyce, Austin and Pajer, Enrico and Pimentel, Guilherme L. and Sleight, Charlotte and Taronna, Massimo",
    title = "{Snowmass White Paper: The Cosmological Bootstrap}",
    eprint = "2203.08121",
    archivePrefix = "arXiv",
    primaryClass = "hep-th",
    doi = "10.21468/SciPostPhysCommRep.1",
    journal = "SciPost Phys. Comm. Rep.",
    volume = "2024",
    pages = "1",
    year = "2024"
}

@article{Baumann:2019oyu,
    author = "Baumann, Daniel and Duaso Pueyo, Carlos and Joyce, Austin and Lee, Hayden and Pimentel, Guilherme L.",
    title = "{The cosmological bootstrap: weight-shifting operators and scalar seeds}",
    eprint = "1910.14051",
    archivePrefix = "arXiv",
    primaryClass = "hep-th",
    doi = "10.1007/JHEP12(2020)204",
    journal = "JHEP",
    volume = "12",
    pages = "204",
    year = "2020"
}

@article{Sleight:2019hfp,
    author = "Sleight, Charlotte and Taronna, Massimo",
    title = "{Bootstrapping Inflationary Correlators in Mellin Space}",
    eprint = "1907.01143",
    archivePrefix = "arXiv",
    primaryClass = "hep-th",
    reportNumber = "PUPT-2590",
    doi = "10.1007/JHEP02(2020)098",
    journal = "JHEP",
    volume = "02",
    pages = "098",
    year = "2020"
}

@article{Pajer:2020wxk,
    author = "Pajer, Enrico",
    title = "{Building a Boostless Bootstrap for the Bispectrum}",
    eprint = "2010.12818",
    archivePrefix = "arXiv",
    primaryClass = "hep-th",
    doi = "10.1088/1475-7516/2021/01/023",
    journal = "JCAP",
    volume = "01",
    pages = "023",
    year = "2021"
}

@article{Cabass:2021fnw,
    author = "{Cabass, Giovanni and Pajer, Enrico and Stefanyszyn, David and Supe{\l}, Jakub}",
    title = "{Bootstrapping large graviton non-Gaussianities}",
    eprint = "2109.10189",
    archivePrefix = "arXiv",
    primaryClass = "hep-th",
    doi = "10.1007/JHEP05(2022)077",
    journal = "JHEP",
    volume = "05",
    pages = "077",
    year = "2022"
}

@article{DuasoPueyo:2023kyh,
    author = "Duaso Pueyo, Carlos and Pajer, Enrico",
    title = "{A cosmological bootstrap for resonant non-Gaussianity}",
    eprint = "2311.01395",
    archivePrefix = "arXiv",
    primaryClass = "hep-th",
    doi = "10.1007/JHEP03(2024)098",
    journal = "JHEP",
    volume = "03",
    pages = "098",
    year = "2024"
}

@article{Fevola:2024nzj,
    author = "Fevola, Claudia and Pimentel, Guilherme L. and Sattelberger, Anna-Laura and Westerdijk, Tom",
    title = "{Algebraic approaches to cosmological integrals}",
    eprint = "2410.14757",
    archivePrefix = "arXiv",
    primaryClass = "math.AG",
    doi = "10.4418/2025.80.1.12",
    journal = "Matematiche",
    volume = "80",
    number = "1",
    pages = "303--324",
    year = "2025"
}

@article{Arkani-Hamed:2025mce,
    author = "Arkani-Hamed, Nima and Glew, Ross and Vaz{\~a}o, Francisco",
    title = "{Correlators are simpler than wavefunctions}",
    eprint = "2512.23795",
    archivePrefix = "arXiv",
    primaryClass = "hep-th",
    month = "12",
    year = "2025"
}

@article{Glew:2025mry,
    author = "Glew, Ross",
    title = "{Correlators from graphical amplitudes}",
    doi = "10.1103/cg2r-gtkt",
    journal = "Phys. Rev. D",
    volume = "112",
    number = "6",
    pages = "L061302",
    year = "2025"
}

@article{Frellesvig:2017aai,
    author = "Frellesvig, Hjalte and Papadopoulos, Costas G.",
    title = "{Cuts of Feynman Integrals in Baikov representation}",
    eprint = "1701.07356",
    archivePrefix = "arXiv",
    primaryClass = "hep-ph",
    doi = "10.1007/JHEP04(2017)083",
    journal = "JHEP",
    volume = "04",
    pages = "083",
    year = "2017"
}

@article{Baikov:1996iu,
    author = "Baikov, P. A.",
    editor = "Werlen, M. and Perret-Gallix, D.",
    title = "{Explicit solutions of the multiloop integral recurrence relations and its application}",
    eprint = "hep-ph/9611449",
    archivePrefix = "arXiv",
    reportNumber = "INP-96-42-449",
    doi = "10.1016/S0168-9002(97)00126-5",
    journal = "Nucl. Instrum. Meth. A",
    volume = "389",
    pages = "347--349",
    year = "1997"
}

@article{Wang:2019gbi,
    author = "Wang, Lian-Tao and Xianyu, Zhong-Zhi",
    title = "{In Search of Large Signals at the Cosmological Collider}",
    eprint = "1910.12876",
    archivePrefix = "arXiv",
    primaryClass = "hep-ph",
    doi = "10.1007/JHEP02(2020)044",
    journal = "JHEP",
    volume = "02",
    pages = "044",
    year = "2020"
}

@article{Bodas:2025wuk,
    author = "Bodas, Arushi and Broadberry, Edward and Sundrum, Raman and Xu, Zhaohui",
    title = "{Charged Loops at the Cosmological Collider with Chemical Potential}",
    eprint = "2507.22978",
    archivePrefix = "arXiv",
    primaryClass = "hep-ph",
    reportNumber = "FERMILAB-PUB-25-0519-V",
    month = "7",
    year = "2025"
}

@article{Chen:2018xck,
    author = "Chen, Xingang and Wang, Yi and Xianyu, Zhong-Zhi",
    title = "{Neutrino Signatures in Primordial Non-Gaussianities}",
    eprint = "1805.02656",
    archivePrefix = "arXiv",
    primaryClass = "hep-ph",
    doi = "10.1007/JHEP09(2018)022",
    journal = "JHEP",
    volume = "09",
    pages = "022",
    year = "2018"
}

@article{Wang:2020ioa,
    author = "Wang, Lian-Tao and Xianyu, Zhong-Zhi",
    title = "{Gauge Boson Signals at the Cosmological Collider}",
    eprint = "2004.02887",
    archivePrefix = "arXiv",
    primaryClass = "hep-ph",
    doi = "10.1007/JHEP11(2020)082",
    journal = "JHEP",
    volume = "11",
    pages = "082",
    year = "2020"
}

@article{Bodas:2020yho,
    author = "Bodas, Arushi and Kumar, Soubhik and Sundrum, Raman",
    title = "{The Scalar Chemical Potential in Cosmological Collider Physics}",
    eprint = "2010.04727",
    archivePrefix = "arXiv",
    primaryClass = "hep-ph",
    reportNumber = "UMD-PP-020-09",
    doi = "10.1007/JHEP02(2021)079",
    journal = "JHEP",
    volume = "02",
    pages = "079",
    year = "2021"
}

@article{Chen:2016uwp,
    author = "Chen, Xingang and Wang, Yi and Xianyu, Zhong-Zhi",
    title = "{Standard Model Background of the Cosmological Collider}",
    eprint = "1610.06597",
    archivePrefix = "arXiv",
    primaryClass = "hep-th",
    doi = "10.1103/PhysRevLett.118.261302",
    journal = "Phys. Rev. Lett.",
    volume = "118",
    number = "26",
    pages = "261302",
    year = "2017"
}

@article{Chen:2016hrz,
    author = "Chen, Xingang and Wang, Yi and Xianyu, Zhong-Zhi",
    title = "{Standard Model Mass Spectrum in Inflationary Universe}",
    eprint = "1612.08122",
    archivePrefix = "arXiv",
    primaryClass = "hep-th",
    doi = "10.1007/JHEP04(2017)058",
    journal = "JHEP",
    volume = "04",
    pages = "058",
    year = "2017"
}

@article{Fumagalli:2024jzz,
    author = "Fumagalli, Jacopo",
    title = "{Absence of one-loop effects on large scales from small scales in non-slow-roll dynamics. Part 2. Quartic interactions and consistency relations}",
    eprint = "2408.08296",
    archivePrefix = "arXiv",
    primaryClass = "astro-ph.CO",
    doi = "10.1007/JHEP01(2025)108",
    journal = "JHEP",
    volume = "01",
    pages = "108",
    year = "2025"
}

@article{Cheng:2021lif,
    author = "Cheng, Shu-Lin and Lee, Da-Shin and Ng, Kin-Wang",
    title = "{Power spectrum of primordial perturbations during ultra-slow-roll inflation with back reaction effects}",
    eprint = "2106.09275",
    archivePrefix = "arXiv",
    primaryClass = "astro-ph.CO",
    doi = "10.1016/j.physletb.2022.136956",
    journal = "Phys. Lett. B",
    volume = "827",
    pages = "136956",
    year = "2022"
}

@article{Kawaguchi:2024rsv,
    author = "Kawaguchi, Ryodai and Tsujikawa, Shinji and Yamada, Yusuke",
    title = "{Proving the absence of large one-loop corrections to the power spectrum of curvature perturbations in transient ultra-slow-roll inflation within the path-integral approach}",
    eprint = "2407.19742",
    archivePrefix = "arXiv",
    primaryClass = "hep-th",
    reportNumber = "WUCG-24-07",
    doi = "10.1007/JHEP12(2024)095",
    journal = "JHEP",
    volume = "12",
    pages = "095",
    year = "2024"
}

@article{Ballesteros:2024zdp,
    author = "Ballesteros, Guillermo and Gamb{\'\i}n Egea, Jes{\'u}s",
    title = "{One-loop power spectrum in ultra slow-roll inflation and implications for primordial black hole dark matter}",
    eprint = "2404.07196",
    archivePrefix = "arXiv",
    primaryClass = "astro-ph.CO",
    doi = "10.1088/1475-7516/2024/07/052",
    journal = "JCAP",
    volume = "07",
    pages = "052",
    year = "2024"
}

@article{KristianoYokoyama2211,
    author = "Kristiano, Jason and Yokoyama, Jun'ichi",
    title = "{Constraining Primordial Black Hole Formation from Single-Field Inflation}",
    eprint = "2211.03395",
    archivePrefix = "arXiv",
    primaryClass = "hep-th",
    doi = "10.1103/PhysRevLett.132.221003",
    journal = "Phys. Rev. Lett.",
    volume = "132",
    pages = "221003",
    year = "2024"
}

@article{KristianoYokoyama2303,
    author = "Kristiano, Jason and Yokoyama, Jun'ichi",
    title = "{Note on the bispectrum and one-loop corrections in single-field inflation with primordial black hole formation}",
    eprint = "2303.00341",
    archivePrefix = "arXiv",
    primaryClass = "hep-th",
    doi = "10.1103/PhysRevD.109.103541",
    journal = "Phys. Rev. D",
    volume = "109",
    pages = "103541",
    year = "2024"
}

@article{Riotto2303,
    author = "Riotto, A.",
    title = "{The Primordial Black Hole Formation from Single-Field Inflation is Still Not Ruled Out}",
    eprint = "2303.01727",
    archivePrefix = "arXiv",
    primaryClass = "astro-ph.CO",
    year = "2023"
}

@article{Firouzjahi2303,
    author = "Firouzjahi, Hassan",
    title = "{One-loop Corrections in Power Spectrum in Single Field Inflation}",
    eprint = "2303.12025",
    archivePrefix = "arXiv",
    primaryClass = "astro-ph.CO",
    journal = "JCAP",
    volume = "10",
    pages = "006",
    year = "2023"
}

@article{FirouzjahiRiotto2304,
    author = "Firouzjahi, Hassan and Riotto, Antonio",
    title = "{Primordial Black Holes and loops in single-field inflation}",
    eprint = "2304.07801",
    archivePrefix = "arXiv",
    primaryClass = "astro-ph.CO",
    doi = "10.1088/1475-7516/2024/02/021",
    journal = "JCAP",
    volume = "02",
    pages = "021",
    year = "2024"
}

@article{Firouzjahi2311,
    author = "Firouzjahi, Hassan",
    title = "{Revisiting Loop Corrections in Single Field USR Inflation}",
    eprint = "2311.04080",
    archivePrefix = "arXiv",
    primaryClass = "astro-ph.CO",
    journal = "Phys. Rev. D",
    volume = "109",
    pages = "043514",
    year = "2024",
    doi = "10.1103/PhysRevD.109.043514"
}

@article{ChoudhuryPandaSami2303,
    author = "Choudhury, Sayantan and Panda, Sudhakar and Sami, M.",
    title = "{Quantum loop effects on the power spectrum and constraints on primordial black holes}",
    eprint = "2303.06066",
    archivePrefix = "arXiv",
    primaryClass = "astro-ph.CO",
    journal = "JCAP",
    volume = "11",
    pages = "066",
    year = "2023",
    doi = "10.1088/1475-7516/2023/11/066"
}

@article{MotohashiTada2303,
    author = "Motohashi, Hayato and Tada, Yuichiro",
    title = "{Squeezed bispectrum and one-loop corrections in transient constant-roll inflation}",
    eprint = "2303.16035",
    archivePrefix = "arXiv",
    primaryClass = "astro-ph.CO",
    journal = "JCAP",
    volume = "08",
    pages = "069",
    year = "2023",
    doi = "10.1088/1475-7516/2023/08/069"
}

@article{Tasinato2305,
    author = "Tasinato, Gianmassimo",
    title = "{A large $|\eta|$ approach to single field inflation}",
    eprint = "2305.11568",
    archivePrefix = "arXiv",
    primaryClass = "hep-th",
    journal = "Phys. Rev. D",
    volume = "108",
    pages = "043526",
    year = "2023",
    doi = "10.1103/PhysRevD.108.043526"
}

@article{Fumagalli2305,
    author = "Fumagalli, Jacopo",
    title = "{Absence of one-loop effects on large scales from small scales in non-slow-roll dynamics}",
    eprint = "2305.19263",
    archivePrefix = "arXiv",
    primaryClass = "astro-ph.CO",
    journal = "JHEP",
    volume = "05",
    pages = "162",
    year = "2025",
    doi = "10.1007/JHEP05(2025)162"
}

@article{ChengLee2305,
    author = "Cheng, Shu-Lin and Lee, Da-Shin and Ng, Kin-Wang",
    title = "{Primordial perturbations from ultra-slow-roll single-field inflation with quantum loop effects}",
    eprint = "2305.16810",
    archivePrefix = "arXiv",
    primaryClass = "astro-ph.CO",
    doi = "10.1088/1475-7516/2024/03/008",
    journal = "JCAP",
    volume = "03",
    pages = "008",
    year = "2024"
}

@article{Franciolini2305,
    author = "Franciolini, Gabriele and Iovino, Junior., Antonio and Taoso, Marco and Urbano, Alfredo",
    title = "{Perturbativity in the presence of ultraslow-roll dynamics}",
    eprint = "2305.03491",
    archivePrefix = "arXiv",
    primaryClass = "astro-ph.CO",
    doi = "10.1103/PhysRevD.109.123550",
    journal = "Phys. Rev. D",
    volume = "109",
    number = "12",
    pages = "123550",
    year = "2024"
}

@article{Fumagalli2307,
    author = "Fumagalli, Jacopo and Bhattacharya, Sukannya and Peloso, Marco and Renaux-Petel, Sebastien and Witkowski, Lukas T.",
    title = "{One-loop infrared rescattering by enhanced scalar fluctuations during inflation}",
    eprint = "2307.08358",
    archivePrefix = "arXiv",
    primaryClass = "astro-ph.CO",
    journal = "JCAP",
    volume = "04",
    pages = "029",
    year = "2024",
    doi = "10.1088/1475-7516/2024/04/029"
}

@article{Sheikhahmadi:2024peu,
    author = "Sheikhahmadi, Haidar and Nassiri-Rad, Amin",
    title = "{Renormalized one-Loop Corrections in Power Spectrum in USR Inflation}",
    eprint = "2411.18525",
    archivePrefix = "arXiv",
    primaryClass = "astro-ph.CO",
    month = "11",
    year = "2024"
}

@article{Ford:1984hs,
    author = "Ford, L. H.",
    title = "{Quantum Instability of De Sitter Space-time}",
    reportNumber = "IMPERIAL-TP-83-84-52",
    doi = "10.1103/PhysRevD.31.710",
    journal = "Phys. Rev. D",
    volume = "31",
    pages = "710",
    year = "1985"
}

@article{Antoniadis:1985pj,
    author = "Antoniadis, Ignatios and Iliopoulos, J. and Tomaras, T. N.",
    title = "{Quantum Instability of De Sitter Space}",
    reportNumber = "SLAC-PUB-3812",
    doi = "10.1103/PhysRevLett.56.1319",
    journal = "Phys. Rev. Lett.",
    volume = "56",
    pages = "1319",
    year = "1986"
}

@article{Tsamis:1994ca,
    author = "Tsamis, N. C. and Woodard, R. P.",
    title = "{Strong infrared effects in quantum gravity}",
    reportNumber = "UFIFT-HEP-92-24, CRETE-92-17",
    doi = "10.1006/aphy.1995.1015",
    journal = "Annals Phys.",
    volume = "238",
    pages = "1--82",
    year = "1995"
}

@article{Tsamis:1996qm,
    author = "Tsamis, N. C. and Woodard, R. P.",
    title = "{The Quantum gravitational back reaction on inflation}",
    eprint = "hep-ph/9602316",
    archivePrefix = "arXiv",
    reportNumber = "CPTH-S422-1295, CRETE-96-13, UFIFT-HEP-96-5",
    doi = "10.1006/aphy.1997.5613",
    journal = "Annals Phys.",
    volume = "253",
    pages = "1--54",
    year = "1997"
}

@article{Tsamis:1997za,
    author = "Tsamis, N. C. and Woodard, R. P.",
    title = "{Matter contributions to the expansion rate of the universe}",
    eprint = "hep-ph/9710466",
    archivePrefix = "arXiv",
    reportNumber = "CRETE-97-11, UFIFT-HEP-97-1",
    doi = "10.1016/S0370-2693(98)00159-2",
    journal = "Phys. Lett. B",
    volume = "426",
    pages = "21--28",
    year = "1998"
}

@article{Polyakov:2007mm,
    author = "Polyakov, A. M.",
    title = "{De Sitter space and eternity}",
    eprint = "0709.2899",
    archivePrefix = "arXiv",
    primaryClass = "hep-th",
    reportNumber = "PUPT-2244",
    doi = "10.1016/j.nuclphysb.2008.01.002",
    journal = "Nucl. Phys. B",
    volume = "797",
    pages = "199--217",
    year = "2008"
}

@article{Polyakov:2009nq,
    author = "Polyakov, A. M.",
    title = "{Decay of Vacuum Energy}",
    eprint = "0912.5503",
    archivePrefix = "arXiv",
    primaryClass = "hep-th",
    reportNumber = "PUPT-2320",
    doi = "10.1016/j.nuclphysb.2010.03.021",
    journal = "Nucl. Phys. B",
    volume = "834",
    pages = "316--329",
    year = "2010"
}

@article{Giddings:2010nc,
    author = "Giddings, Steven B. and Sloth, Martin S.",
    title = "{Semiclassical relations and IR effects in de Sitter and slow-roll space-times}",
    eprint = "1005.1056",
    archivePrefix = "arXiv",
    primaryClass = "hep-th",
    reportNumber = "CERN-PH-TH-2010-095",
    doi = "10.1088/1475-7516/2011/01/023",
    journal = "JCAP",
    volume = "01",
    pages = "023",
    year = "2011"
}

@article{Burgess:2010dd,
    author = "Burgess, C. P. and Holman, R. and Leblond, L. and Shandera, S.",
    title = "{Breakdown of Semiclassical Methods in de Sitter Space}",
    eprint = "1005.3551",
    archivePrefix = "arXiv",
    primaryClass = "hep-th",
    doi = "10.1088/1475-7516/2010/10/017",
    journal = "JCAP",
    volume = "10",
    pages = "017",
    year = "2010"
}

@article{Marolf:2010nz,
    author = "Marolf, Donald and Morrison, Ian A.",
    title = "{The IR stability of de Sitter QFT: results at all orders}",
    eprint = "1010.5327",
    archivePrefix = "arXiv",
    primaryClass = "gr-qc",
    doi = "10.1103/PhysRevD.84.044040",
    journal = "Phys. Rev. D",
    volume = "84",
    pages = "044040",
    year = "2011"
}

@article{Krotov:2010ma,
    author = "Krotov, Dmitry and Polyakov, Alexander M.",
    title = "{Infrared Sensitivity of Unstable Vacua}",
    eprint = "1012.2107",
    archivePrefix = "arXiv",
    primaryClass = "hep-th",
    doi = "10.1016/j.nuclphysb.2011.03.025",
    journal = "Nucl. Phys. B",
    volume = "849",
    pages = "410--432",
    year = "2011"
}

@article{Marolf:2010zp,
    author = "Marolf, Donald and Morrison, Ian A.",
    title = "{The IR stability of de Sitter: Loop corrections to scalar propagators}",
    eprint = "1006.0035",
    archivePrefix = "arXiv",
    primaryClass = "gr-qc",
    doi = "10.1103/PhysRevD.82.105032",
    journal = "Phys. Rev. D",
    volume = "82",
    pages = "105032",
    year = "2010"
}

@article{Rajaraman:2010xd,
    author = "Rajaraman, Arvind",
    title = "{On the proper treatment of massless fields in Euclidean de Sitter space}",
    eprint = "1008.1271",
    archivePrefix = "arXiv",
    primaryClass = "hep-th",
    reportNumber = "UCI-TR-2010-14",
    doi = "10.1103/PhysRevD.82.123522",
    journal = "Phys. Rev. D",
    volume = "82",
    pages = "123522",
    year = "2010"
}

@article{Marolf:2011sh,
    author = "Marolf, Donald and Morrison, Ian A.",
    title = "{The IR stability of de Sitter QFT: Physical initial conditions}",
    eprint = "1104.4343",
    archivePrefix = "arXiv",
    primaryClass = "gr-qc",
    doi = "10.1007/s10714-011-1233-3",
    journal = "Gen. Rel. Grav.",
    volume = "43",
    pages = "3497--3530",
    year = "2011"
}

@article{Giddings:2011zd,
    author = "Giddings, Steven B. and Sloth, Martin S.",
    title = "{Cosmological observables, IR growth of fluctuations, and scale-dependent anisotropies}",
    eprint = "1104.0002",
    archivePrefix = "arXiv",
    primaryClass = "hep-th",
    reportNumber = "CERN-PH-TH-2011-070",
    doi = "10.1103/PhysRevD.84.063528",
    journal = "Phys. Rev. D",
    volume = "84",
    pages = "063528",
    year = "2011"
}

@article{Giddings:2011ze,
    author = "Giddings, Steven B. and Sloth, Martin S.",
    title = "{Fluctuating geometries, q-observables, and infrared growth in inflationary spacetimes}",
    eprint = "1109.1000",
    archivePrefix = "arXiv",
    primaryClass = "hep-th",
    reportNumber = "CERN-TH-PH-2011-187, CERN-PH-TH-2011-187",
    doi = "10.1103/PhysRevD.86.083538",
    journal = "Phys. Rev. D",
    volume = "86",
    pages = "083538",
    year = "2012"
}

@article{Senatore:2012nq,
    author = "Senatore, Leonardo and Zaldarriaga, Matias",
    title = "{On Loops in Inflation II: IR Effects in Single Clock Inflation}",
    eprint = "1203.6354",
    archivePrefix = "arXiv",
    primaryClass = "hep-th",
    reportNumber = "SLAC-PUB-15860",
    doi = "10.1007/JHEP01(2013)109",
    journal = "JHEP",
    volume = "01",
    pages = "109",
    year = "2013"
}

@article{Pimentel:2012tw,
    author = "Pimentel, Guilherme L. and Senatore, Leonardo and Zaldarriaga, Matias",
    title = "{On Loops in Inflation III: Time Independence of zeta in Single Clock Inflation}",
    eprint = "1203.6651",
    archivePrefix = "arXiv",
    primaryClass = "hep-th",
    doi = "10.1007/JHEP07(2012)166",
    journal = "JHEP",
    volume = "07",
    pages = "166",
    year = "2012"
}

@article{Senatore:2012ya,
    author = "Senatore, Leonardo and Zaldarriaga, Matias",
    title = "{The constancy of $\zeta$ in single-clock Inflation at all loops}",
    eprint = "1210.6048",
    archivePrefix = "arXiv",
    primaryClass = "hep-th",
    doi = "10.1007/JHEP09(2013)148",
    journal = "JHEP",
    volume = "09",
    pages = "148",
    year = "2013"
}

@article{Polyakov:2012uc,
    author = "Polyakov, A. M.",
    title = "{Infrared instability of the de Sitter space}",
    eprint = "1209.4135",
    archivePrefix = "arXiv",
    primaryClass = "hep-th",
    month = "9",
    year = "2012"
}

@article{Beneke:2012kn,
    author = "Beneke, M. and Moch, P.",
    title = "{On {\textquotedblleft}dynamical mass{\textquotedblright} generation in Euclidean de Sitter space}",
    eprint = "1212.3058",
    archivePrefix = "arXiv",
    primaryClass = "hep-th",
    reportNumber = "TUM-HEP-870-12, TTK-12-49",
    doi = "10.1103/PhysRevD.87.064018",
    journal = "Phys. Rev. D",
    volume = "87",
    pages = "064018",
    year = "2013"
}

@article{Akhmedov:2013vka,
    author = "Akhmedov, E. T.",
    title = "{Lecture notes on interacting quantum fields in de Sitter space}",
    eprint = "1309.2557",
    archivePrefix = "arXiv",
    primaryClass = "hep-th",
    reportNumber = "ITEP-TH-32-13",
    doi = "10.1142/S0218271814300018",
    journal = "Int. J. Mod. Phys. D",
    volume = "23",
    pages = "1430001",
    year = "2014"
}

@article{Anninos:2014lwa,
    author = "Anninos, Dionysios and Anous, Tarek and Freedman, Daniel Z. and Konstantinidis, George",
    title = "{Late-time Structure of the Bunch-Davies De Sitter Wavefunction}",
    eprint = "1406.5490",
    archivePrefix = "arXiv",
    primaryClass = "hep-th",
    reportNumber = "MIT-CTP-4561",
    doi = "10.1088/1475-7516/2015/11/048",
    journal = "JCAP",
    volume = "11",
    pages = "048",
    year = "2015"
}

@article{Akhmedov:2017ooy,
    author = "Akhmedov, E. T. and Moschella, U. and Pavlenko, K. E. and Popov, F. K.",
    title = "{Infrared dynamics of massive scalars from the complementary series in de Sitter space}",
    eprint = "1701.07226",
    archivePrefix = "arXiv",
    primaryClass = "hep-th",
    reportNumber = "ITEP-TH-5-17",
    doi = "10.1103/PhysRevD.96.025002",
    journal = "Phys. Rev. D",
    volume = "96",
    number = "2",
    pages = "025002",
    year = "2017"
}

@article{Hu:2018nxy,
    author = "Hu, Bei-Lok",
    title = "{Infrared Behavior of Quantum Fields in Inflationary Cosmology -- Issues and Approaches: an overview}",
    eprint = "1812.11851",
    archivePrefix = "arXiv",
    primaryClass = "gr-qc",
    month = "12",
    year = "2018"
}

@article{Akhmedov:2019cfd,
    author = "Akhmedov, E. T. and Moschella, U. and Popov, F. K.",
    title = "{Characters of different secular effects in various patches of de Sitter space}",
    eprint = "1901.07293",
    archivePrefix = "arXiv",
    primaryClass = "hep-th",
    doi = "10.1103/PhysRevD.99.086009",
    journal = "Phys. Rev. D",
    volume = "99",
    number = "8",
    pages = "086009",
    year = "2019"
}

@article{Gorbenko:2019rza,
    author = "Gorbenko, Victor and Senatore, Leonardo",
    title = "{$\lambda \phi^4$ in dS}",
    eprint = "1911.00022",
    archivePrefix = "arXiv",
    primaryClass = "hep-th",
    month = "10",
    year = "2019"
}

@article{Baumgart:2019clc,
    author = "Baumgart, Matthew and Sundrum, Raman",
    title = "{De Sitter Diagrammar and the Resummation of Time}",
    eprint = "1912.09502",
    archivePrefix = "arXiv",
    primaryClass = "hep-th",
    doi = "10.1007/JHEP07(2020)119",
    journal = "JHEP",
    volume = "07",
    pages = "119",
    year = "2020"
}

@article{Mirbabayi:2019qtx,
    author = "Mirbabayi, Mehrdad",
    title = "{Infrared dynamics of a light scalar field in de Sitter}",
    eprint = "1911.00564",
    archivePrefix = "arXiv",
    primaryClass = "hep-th",
    doi = "10.1088/1475-7516/2020/12/006",
    journal = "JCAP",
    volume = "12",
    pages = "006",
    year = "2020"
}

@article{Cohen:2020php,
    author = "Cohen, Timothy and Green, Daniel",
    title = "{Soft de Sitter Effective Theory}",
    eprint = "2007.03693",
    archivePrefix = "arXiv",
    primaryClass = "hep-th",
    doi = "10.1007/JHEP12(2020)041",
    journal = "JHEP",
    volume = "12",
    pages = "041",
    year = "2020"
}

@article{Mirbabayi:2020vyt,
    author = "Mirbabayi, Mehrdad",
    title = "{Markovian dynamics in de Sitter}",
    eprint = "2010.06604",
    archivePrefix = "arXiv",
    primaryClass = "hep-th",
    doi = "10.1088/1475-7516/2021/09/038",
    journal = "JCAP",
    volume = "09",
    pages = "038",
    year = "2021"
}

@article{Baumgart:2020oby,
    author = "Baumgart, Matthew and Sundrum, Raman",
    title = "{Manifestly Causal In-In Perturbation Theory about the Interacting Vacuum}",
    eprint = "2010.10785",
    archivePrefix = "arXiv",
    primaryClass = "hep-th",
    doi = "10.1007/JHEP03(2021)080",
    journal = "JHEP",
    volume = "03",
    pages = "080",
    year = "2021"
}

@article{Cohen:2021fzf,
    author = "Cohen, Timothy and Green, Daniel and Premkumar, Akhil and Ridgway, Alexander",
    title = "{Stochastic Inflation at NNLO}",
    eprint = "2106.09728",
    archivePrefix = "arXiv",
    primaryClass = "hep-th",
    doi = "10.1007/JHEP09(2021)159",
    journal = "JHEP",
    volume = "09",
    pages = "159",
    year = "2021"
}

@article{Bzowski:2023nef,
    author = "Bzowski, Adam and McFadden, Paul and Skenderis, Kostas",
    title = "{Renormalisation of IR divergences and holography in de Sitter}",
    eprint = "2312.17316",
    archivePrefix = "arXiv",
    primaryClass = "hep-th",
    doi = "10.1007/JHEP05(2024)053",
    journal = "JHEP",
    volume = "05",
    pages = "053",
    year = "2024"
}

@article{Cespedes:2023aal,
    author = "C{\'e}spedes, Sebasti{\'a}n and Davis, Anne-Christine and Wang, Dong-Gang",
    title = "{On the IR divergences in de Sitter space: loops, resummation and the semi-classical wavefunction}",
    eprint = "2311.17990",
    archivePrefix = "arXiv",
    primaryClass = "hep-th",
    doi = "10.1007/JHEP04(2024)004",
    journal = "JHEP",
    volume = "04",
    pages = "004",
    year = "2024"
}

@article{Benincasa:2024lxe,
    author = "Benincasa, Paolo and Vaz{\~a}o, Francisco",
    title = "{The asymptotic structure of cosmological integrals}",
    eprint = "2402.06558",
    archivePrefix = "arXiv",
    primaryClass = "hep-th",
    doi = "10.21468/SciPostPhys.19.2.029",
    journal = "SciPost Phys.",
    volume = "19",
    number = "2",
    pages = "029",
    year = "2025"
}

@article{Benincasa:2024ptf,
    author = "Benincasa, Paolo and Brunello, Giacomo and Mandal, Manoj K. and Mastrolia, Pierpaolo and Vaz{\~a}o, Francisco",
    title = "{One-loop corrections to the Bunch-Davies wave function of the universe}",
    eprint = "2408.16386",
    archivePrefix = "arXiv",
    primaryClass = "hep-th",
    doi = "10.1103/PhysRevD.111.085016",
    journal = "Phys. Rev. D",
    volume = "111",
    number = "8",
    pages = "085016",
    year = "2025"
}

@article{Bhowmick:2025mxh,
    author = "Bhowmick, Supritha and Lee, Mang Hei Gordon and Ghosh, Diptimoy and Ullah, Farman",
    title = "{Singularities in Cosmological Loop Correlators}",
    eprint = "2503.21880",
    archivePrefix = "arXiv",
    primaryClass = "hep-th",
    month = "3",
    year = "2025"
}

@article{Arkani-Hamed:2013jha,
    author = "Arkani-Hamed, Nima and Trnka, Jaroslav",
    title = "{The Amplituhedron}",
    eprint = "1312.2007",
    archivePrefix = "arXiv",
    primaryClass = "hep-th",
    doi = "10.1007/JHEP10(2014)030",
    journal = "JHEP",
    volume = "10",
    pages = "030",
    year = "2014"
}
}

\end{document}